\documentclass[11pt]{article}
\usepackage{authblk}
\usepackage{footmisc}

\usepackage{adjustbox}
\usepackage{algorithmic}
\usepackage[ruled,linesnumbered]{algorithm2e} 

\usepackage{amsmath,amssymb,amsfonts,amsthm} 

\usepackage[backend=bibtex, style=numeric, natbib=true]{biblatex} 
\usepackage{booktabs} 
\usepackage{bm} 
\usepackage{caption} 
\usepackage{diagbox} 
\usepackage{float} 
\usepackage{graphicx} 

\usepackage[colorlinks=true, allcolors=blue]{hyperref} 

\usepackage{listings} 
\usepackage{mathrsfs} 
\usepackage{mathtools} 
\usepackage{multirow} 

\usepackage{subcaption} 
\usepackage{thmtools} 
\usepackage{thm-restate} 
\usepackage{tikz} 

\usepackage[textsize=normalsize]{todonotes} 
\usepackage{xcolor} 

\usepackage[top=1in, bottom=1in, left=1in, right=1in]{geometry} 

\usepackage[capitalize,noabbrev]{cleveref} 

\graphicspath{{Figures/}} 
\definecolor{pku-red}{RGB}{139,0,18} 

\theoremstyle{plain} 
\newtheorem{theorem}{Theorem}[section] 

\newtheorem{proposition}[theorem]{Proposition} 
\newtheorem{lemma}[theorem]{Lemma} 

\newtheorem{example}[theorem]{Example} 

\theoremstyle{definition}
\newtheorem{assumption}{Assumption}
\newtheorem{definition}[theorem]{Definition}
\theoremstyle{remark}
\newtheorem{remark}[theorem]{Remark}
\crefname{assumption}{Assumption}{Assumptions}
\Crefname{assumption}{Assumption}{Assumptions}

\newcommand{\bbE}{{\mathbb{E}}}

\newcommand{\bbR}{{\mathbb{R}}}

\newcommand{\bbZ}{{\mathbb{Z}}}

\newcommand{\calD}{{\mathcal{D}}}

\newcommand{\calF}{{\mathcal{F}}}
\newcommand{\calG}{{\mathcal{G}}}

\newcommand{\calM}{{\mathcal{M}}}
\newcommand{\calN}{{\mathcal{N}}}

\newcommand{\calT}{{\mathcal{T}}}
\newcommand{\calU}{{\mathcal{U}}}

\newcommand{\calX}{{\mathcal{X}}}

\newcommand{\bma}{{\bm{a}}}

\newcommand{\bmp}{{\bm{p}}}

\newcommand{\bmt}{{\bm{t}}}

\newcommand{\bmx}{{\bm{x}}}
\newcommand{\bmy}{{\bm{y}}}
\newcommand{\bmz}{{\bm{z}}}

\newcommand{\bmbb}{{\bm{\beta}}}

\newcommand{\iinn}{{i\in[n]}} 
\newcommand{\jinn}{{j\in[n]}}

\newcommand{\iinm}{{i\in[m]}}
\newcommand{\jinm}{{j\in[m]}}

\newcommand{\iink}{{i\in[k]}}

\newcommand{\kinM}{{k\in[M]}}

\newcommand{\kinK}{{k\in[K]}}

\newcommand{\jnei}{{j\ne i}}

\newcommand{\ones}{\mathbf{1}} 
\newcommand{\zeros}{\mathbf{0}} 

\newcommand{\argmax}{\mathrm{argmax}}

\newcommand{\dd}{\mathrm{d}} 
\newcommand{\pp}{\partial} 
\newcommand{\Cov}{\mathrm{Cov}} 
\newcommand{\dist}{\mathrm{dist}} 
\newcommand{\diam}{\mathrm{diam}} 

\newcommand{\st}{\mathrm{s.t.}}   
\newcommand{\iid}{\mathrm{i.i.d.}}  
\newcommand{\ie}{\emph{i.e.}}    
\newcommand{\eg}{\emph{e.g.}}    
\newcommand{\wrt}{\emph{w.r.t.}}  
\newcommand{\etc}{\emph{etc.}}    
\newcommand{\wlg}{\emph{w.l.o.g.}}    
\newcommand{\resp}{\emph{resp.}} 

\newcommand{\Wrt}{\wrt~}
\newcommand{\Wlg}{\wlg~}
\newcommand{\Resp}{\resp~}
\newcommand{\widesim}{{\scalebox{1.8}[1]{$\sim$}}} 
\newcommand{\iidd}{\overset{\mathrm{i.d.}}{\widesim}} 
\newcommand{\iidf}{\overset{\mathrm{i.i.d.}}{\widesim}} 

\newcommand{\OBJ}{\mathrm{OBJ}}

\newcommand{\SW}{\mathrm{SW}}

\newcommand{\ASW}{\mathrm{ASW}}

\newcommand{\Ext}{\mathrm{External}}

\newcommand{\MD}{{\calM\calD}}

\newcommand{\method}{{TEDI}}
\newcommand{\Method}{{\method\ }}
\newcommand{\structure}{{outcome separability}}
\newcommand{\Structure}{{\structure\ }}
\newcommand{\network}{{Partial GroupMax Network}}
\newcommand{\Network}{{\network\ }}
\newcommand{\pan}{{Parameterized Affine Network}}
\newcommand{\Pan}{{\pan\ }}
\newcommand{\nbnp}{{no-buy-no-pay}}
\newcommand{\Nbnp}{{\nbnp\ }}
\newcommand{\nbnps}{{\nbnp*}}
\newcommand{\Nbnps}{{\nbnp*\ }}
\newcommand{\ppc}{{partial convexity}}
\newcommand{\Ppc}{{\ppc\ }}
\newcommand{\xlow}{{\underline{x}}}
\newcommand{\xhigh}{{\bar{x}}}
\newcommand{\tlow}{{\underline{t}}}
\newcommand{\thigh}{{\bar{t}}}
\DeclareMathOperator{\GroupMax}{GroupMax}
\DeclareMathOperator{\softplus}{softplus}
\DeclareMathOperator{\EU}{EU}
\DeclareMathOperator{\SEU}{SEU}

\newcommand{\magic}{}
\newcommand{\Magic}{}
\newcommand{\MMagic}{}

\usepackage{graphicx}
\usepackage{color}

\addtocontents{toc}{\protect\setcounter{tocdepth}{2}} 

\begin{document}


\title{Learning Truthful Mechanisms without Discretization}
\date{}
%



\author[1]{Yunxuan Ma}
\author[2]{Siqiang Wang}
\author[1]{Zhijian Duan}
\author[3]{Yukun Cheng}
\author[1,4]{Xiaotie Deng}

\affil[1]{CFCS, School of Computer Science, Peking University \protect\\ \texttt{\{yunxuanma,zjduan,xiaotie\}@pku.edu.cn}}
\affil[2]{School of Mathematical Sciences, Peking University \protect\\ \texttt{2100010867@stu.pku.edu.cn}}
\affil[3]{School of Business, Jiangnan University \protect\\ \texttt{ykcheng@amss.ac.cn}}
\affil[4]{CMAR, Institute for AI, Peking University}

%



\maketitle

\begin{abstract}
\setlength{\parindent}{2em}
\MMagic

This paper introduces TEDI (Truthful, Expressive, and Dimension-Insensitive approach), a discretization-free algorithm to learn truthful and utility-maximizing mechanisms.
Existing learning-based approaches often rely on discretization of outcome spaces to ensure truthfulness, which leads to inefficiency with increasing problem size. To address this limitation, we formalize the concept of \emph{pricing rules}, defined as functions that map outcomes to prices. Based on this concept, we propose a novel menu mechanism, which can be equivalent to a truthful direct mechanism under specific conditions.
The core idea of TEDI lies in its parameterization of pricing rules using Partial GroupMax Network, a new network architecture designed to universally approximate partial convex functions. To learn optimal pricing rules, we develop novel training techniques, including \emph{covariance trick} and \emph{continuous sampling}, to derive unbiased gradient estimators compatible with first-order optimization.
Theoretical analysis establishes that TEDI guarantees truthfulness, full expressiveness, and dimension-insensitivity.
Experimental evaluation in the studied auction setting demonstrates that TEDI achieves strong performance, competitive with or exceeding state-of-the-art methods. 

This work presents the first approaches to learn truthful mechanisms without outcome discretization, thereby enhancing algorithmic efficiency. The proposed concepts, network architecture, and learning techniques might offer potential value and provide new insights for automated mechanism design and differentiable economics.

\end{abstract}
\tableofcontents
\renewcommand{\contentsname}{Contents}
\newpage
\MMagic\Magic
\section{Introduction}
\label{sec:intro}

\magic
Mechanism design is a crucial topic with broad applications in computer science and microeconomics.
It examines how one mechanism designer can incentivize players truthfully reporting their private information, while achieving better resources allocations.
Mechanism design problems have been widely studied in variable contexts, including 
multiple goods monopolists \citep{optimal_auction-special_case:manelli2006bundling,optimal_auction-special_case:pavlov2011optimal,optimal_auction-special_case:giannakopoulos2014duality,optimal_auction-special_case:daskalakis2015strong,optimal_auction-special_case:yao2017dominant},
digital goods auctions \citep{digital-good-auction:goldberg2001competitive,digital-good-auction:goldberg2004lower,digital-good-auction:chen2014optimal},
facility locations \citep{facility-location:barak2024mac,facility-location:gravin2025approximation,facility-location:chan2021mechanism},
mechanism design for LLMs \citep{LLM-aggregation:soumalias2024truthful,LLM-mechanism-design:duetting2024mechanism,LLM-mechanism-design:sun2024mechanism} and many other applications.
Meanwhile, differentiable economics—a technical paradigm developed recently \citep{RegretNet-journal:dutting2024optimal,differentiable-economic:curry2024optimal,differentiable-economic:wang2024deep}—employs differentiable functions (e.g., neural networks) and gradient-based optimization to represent and to learn economic solutions.
Within this framework, notable progress has been made in various mechanism design problems, including auctions \citep{RegretNet-journal:dutting2024optimal,GemNet:wang2024gemnet,lottery-AMA:curry2023differentiable}, 
facility location \citep{AMD-facility-location:golowich2018deep}, social choice \citep{AMD-no_money:narasimhan2016automated}, and stable matching \citep{AMD-matching:ravindranath2021deep}.

However, all existing approaches guaranteeing truthfulness of learned mechanisms (e.g., MenuNet \citep{MenuNet:shen2019automated}, LotteryAMA\citep{lottery-AMA:curry2023differentiable}, GemNet\citep{GemNet:wang2024gemnet}) are typically discretization-based. This discretization step leads to inefficiency as the problem size (e.g., number of goods) increases, a phenomenon often termed the ``curse of dimensionality'' \citep{CoD:bellman1966dynamic,CoD:bengio2005curse}.

\magic

\begin{example}[Curse of Dimensionality \citep{CoD:bellman1966dynamic}]
\label{eg:CoD}
\itshape
Consider an example that a seller wants to sell $m$ goods to one buyer. 
The buyer has linear valuation and her valuation to good $i$, $v_i \iidd Bernoulli(p=0.5)$\footnote{A Bernoulli random variable with $p=0.5$ can be $0$ or $1$ with equal probability.}. 
It's straight-forward that selling each good with price $p_i \equiv 1, \iinm$ is optimal for maixizing the revenue of seller.


The conventional approach is to discretize the outcome space $([0,1]^m)$ with $K$ candidates 
$\{x_1,...,x_K\}\subseteq [0,1]^m$ and prices to each candidates, $\{p_1,...,p_K\}$ ($K$ is a pre-specified hyper-parameter). The $j$'th element in $k$'th candidate, $x_{kj}$, represents the probability of trading good $j$. The candidate-pair set ($M = \{(x_k, p_k): \kinK\}$) is denoted as the menu.
In this problem, there are $2^m$ deterministic outcomes ($\{(a_1,...,a_m): a_i = 0 \text{ or } 1\}$), and all outcomes are indispensable to recover the optimal mechanism. 
Therefore, by conventional menu representation, it requires an exponential number of parameters $(K = \Omega(2^m))$ to recover the optimal solution and therefore suffers from the ``curse of dimensionality''.
However, the optimal mechanism has a precise description, which can be polynomial on problem size. 

\end{example}

\magic
\cref{eg:CoD} suggests that an alternative of discretization step is prone to better performance in moderate-scale scenarios.
This intuition motivates our novel concept of \emph{pricing rules}, as illustrated in \cref{def:pr-informal}.


\begin{definition}[Pricing Rule (Informal)]
\label{def:pr-informal}
A pricing rule is a function $f: \calX \to \bbR$ that assigns a price to each possible outcome. 
If the player selects the outcome $x\in \calX$, then she shall pay $f(x)$. 
\end{definition}




\magic

\begin{example}
\label{eg:PricingRule}
\itshape
Let $X = [0,1]^m$ be the outcome space in \cref{eg:CoD}. 
Given the pricing rule $f$, the rational buyer with valuation $v\in \bbR^m$ will choose the bundle $x\in X$ that maximizes her utility ($\langle v, x \rangle - f(x)$).
In \cref{eg:CoD}, optimal pricing rule has the expression $f(x) = \langle \ones, x \rangle$, which naturally has an $O(m)$ representation.
\end{example}

\begin{figure}[!t]
\centering

\hfill
\begin{subfigure}{0.55\textwidth}
\includegraphics[width=\textwidth]{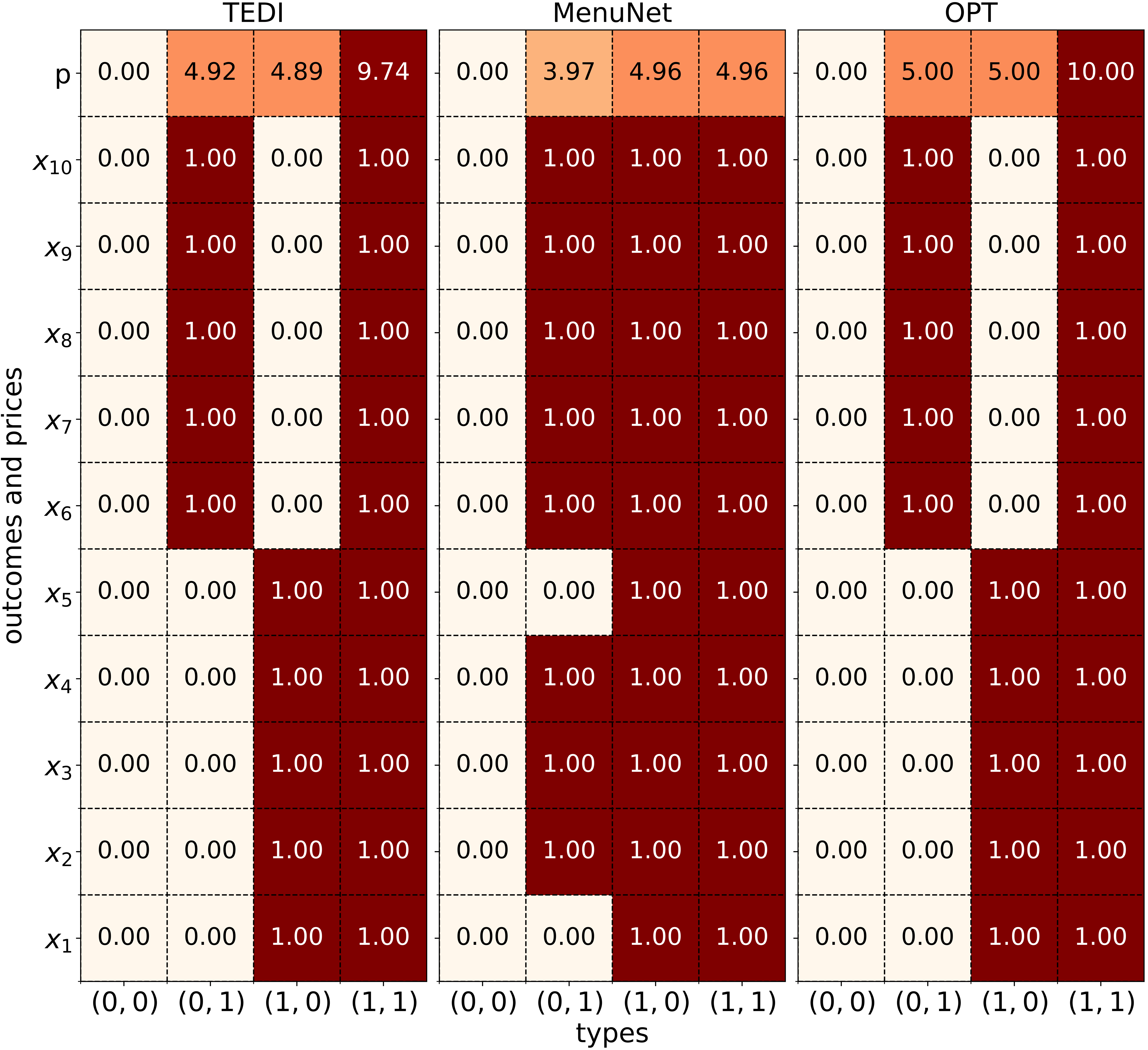}
\caption{}
\label{fig:intro-2-1}
\end{subfigure}
\hfill
\begin{subfigure}{0.165\textwidth}
\includegraphics[width=\textwidth]{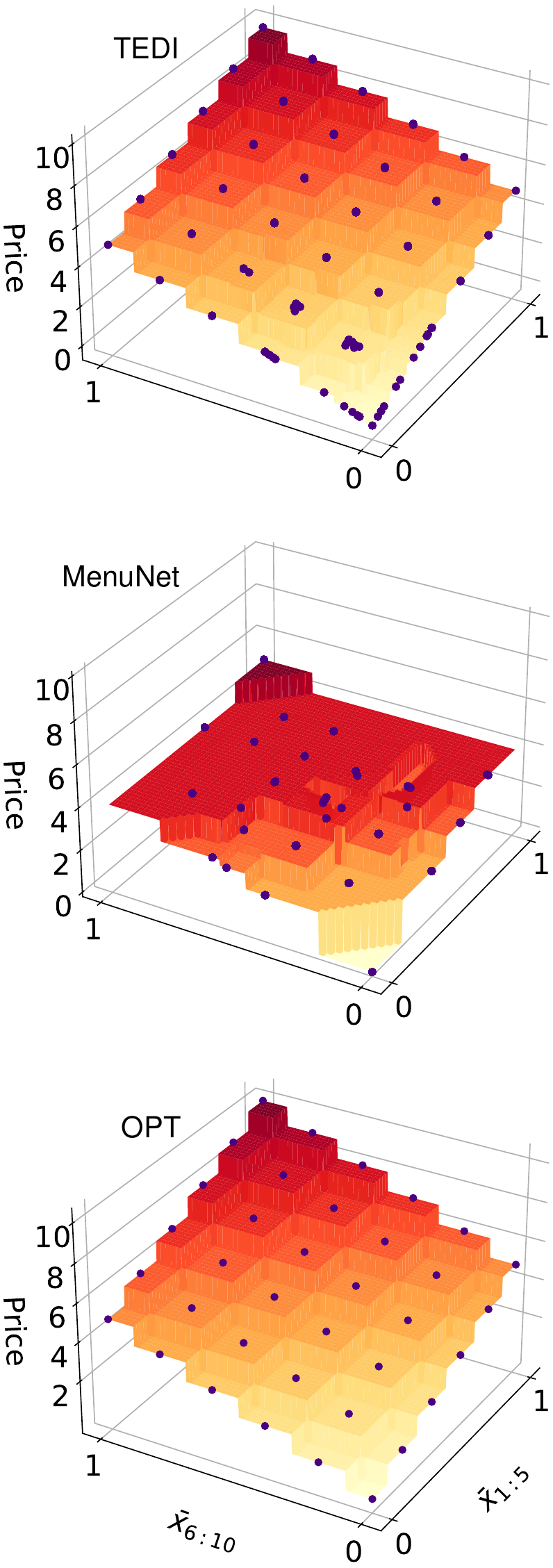}
\caption{}
\label{fig:intro-2-2}
\end{subfigure}
\hfill

\caption{(a): A visualization of the optimal mechanism (denoted as OPT) and the mechanism learned by \Method and MenuNet\citep{MenuNet:shen2019automated}, in the $n=1$ player, $m=10$ goods setting with independent $Bernoulli(p=0.5)$ valuation. 
The horizontal axis, $(x, y)$, represents that the participant's $10$-dimensional valuation has the expression $(x,x,x,x,x,y,y,y,y,y)\in\{0,1\}^{10}$. The $\{x_i\}_{1\le i\le 10}$ (resp. $p$) in the vertical axis represents the amount (probability) of good $i$ sold (resp. the price charged) to the buyer. The color of prices is normalized with a factor of $10$. 
(b): All realized outcomes and corresponding prices of the mechanism given by \method, MenuNet, and OPT. The point $(x,y)$ represents one realized outcome $(a_1,...,a_{10})$ with the mapping $x = \frac{1}{5}\sum_{1\le i\le 5} a_i$ and $y = \frac{1}{5}\sum_{6\le i\le 10} a_i$. The height value represents the price corresponding to each outcome.
(Overall): \Method exhibits $> 98\%$ optimal revenue, MenuNet (similarly, RochetNet, LotteryAMA, GemNet) only exhibits $81.1\%$ optimal revenue. 
}
\label{fig:intro}
\MMagic
\end{figure}

\magic
\subsection{Our Contributions}
The main technical contribution of this paper is to establish an algorithm named \Method (\textbf{T}ruthful, \textbf{E}xpressive, and \textbf{D}imension-\textbf{I}nsensitive approach, detailed in \cref{sec:method}) with a suite of innovative techniques to learn (multi-player) optimal pricing rules.
The core idea is to parameterize pricing rules using neural networks that possess universal approximation properties \Wrt a ``properly defined'' function class. 
We also utilize two novel techniques, namely \emph{covariance trick} and \emph{continuous sampling}, to efficiently sample an unbiased estimator of gradient, which enables the learning process tractable by combining first-order algorithms (\eg, Adam).
\Method exhibits several ideal properties established in \cref{sec:justification}:
\emph{Truthfulness}: Any mechanism learned by \Method is truthful;
\emph{Full-Expressiveness}: \Method can approximate the optimal mechanism with arbitrary precision;
and \emph{Dimension-Insensitivity}: \Method maintains good performance as computational resources grow polynomially with the problem size.
The \emph{dimension-insensitivity} is further demonstrated through experiments in \cref{sec:exp}.
\cref{fig:intro} illustrates a performance comparison among \method, optimal mechanisms, and existing approaches with setting described in \cref{eg:CoD}, which showcases the superior performance of \Method over existing approaches.
For space limits, a detailed discussion of related works is deferred to \cref{subsec:omitted:related}.

\section{Preliminary}
\label{sec:model}

\magic
We consider an auction model consisting of $n$ players (auction participants), one mechanism designer (the auctioneer) and $m$ kinds of goods with sufficient supply.
Each player $i\in [n]$ owns a private type $t_i \in \calT_i \coloneqq [0,1]^m$, where $\calT_i$ is the type space of player $i$ and $t_{ij}$ is the valuation of good $j$ to player $i$.
Denote $\calT = \times_\iinn \calT_i$ as the space of type profile and $\bmt = (t_1,...,t_n) \in \calT$, $\bmt_{-i} = (t_1,...,t_{i-1}, t_{i+1},..., t_n) \in \calT_{-i}$ as the type profiles w/o player $i$'s type, respectively.
Due to the private nature of individual types, the mechanism designer only knows the joint distribution over type profiles, denoted as $\calF \in \Delta(\calT)$.
The mechanism designer aims to design a direct mechanism that uses the players' reported types ${\bmt}'=(t_i')\in \calT$ as input and outputs an outcome $\bmx\in \calX \coloneqq [0,1]^{n\times m}$ as well as a monetary transfer vector $\bmp=(p_1, \dots, p_n)\in \bbR^n$. Here, $\calX$ is the outcome space, $x_{ij}$ is explained as the probability of allocating one-unit good $j$ to player $i$, and $p_i$ is the monetary transfer from player $i$ to the mechanism designer.
Specially, each player receives her local outcome $x_i \in \calX_i\coloneqq [0,1]^m$, and $\bmx = (x_1,...,x_n) \in \calX$ is referred to as a global outcome.
The utility of player $i$, $u_i(x_i,p_i;t_i): \calX_i \times \bbR \times \calT_i \to \bbR$, when her local outcome is $x_i$, the monetary transfer is $p_i$, and her type is $t_i$, is assumed to be quasi-linear and additive among goods as considered in \citep{RegretNet-journal:dutting2024optimal,lottery-AMA:curry2023differentiable,GemNet:wang2024gemnet}, and thus \Wlg represented by the form: $u_i(x_i,p_i;t_i) = \langle x_i, t_i \rangle - p_i$.
Additionally, $u_0(\bmx, \bmp; \bmt): \calX \times \bbR^n \times \calT \to \bbR$ is used to denote the utility of the mechanism designer.
Meanwhile, we require that the mechanism designer is risk-neutral towards players' type profile distribution $\calF$, which is standard in micro-economics \citep{MWG:mas1995microeconomic}.
Given all above, a mechanism design problem $\MD$ (stands for `Mechanism Design') is uniquely determined by a 4-tuple $\MD= (n, m, u_0, \calF)$. 
A list of notations used in this paper is provided in \cref{app:subsec:notations}.

\begin{remark}
\itshape
It is important to note that we do not impose the traditional constraint $\sum_{i\in[n]}x_{ij}\leq 1$ for each good $j\in [m]$.
This is because we consider the setting that the quantity of each good can exceed the number of players, or that goods can be reproduced to satisfy demand. 
Consequently, any solution of the form $(x_1,...,x_n), x_i \in  [0,1]^m$ constitutes a feasible allocation.
This feature is natural in the context of digital goods auctions \citep{digital-good-auction:goldberg2001competitive,digital-good-auction:goldberg2004lower,digital-good-auction:chen2014optimal},
procurements auctions \citep{reverse-auction:smeltzer2003electronic,reverse-auction:1001,reverse-auction:gerstgrasser2019,reverse-auction:jia2021online},
one-player auctions \citep{optimal_auction-special_case:manelli2006bundling,optimal_auction-special_case:pavlov2011optimal,optimal_auction-special_case:giannakopoulos2014duality,optimal_auction-special_case:daskalakis2015strong,optimal_auction-special_case:yao2017dominant},
and auctions with production costs \citep{auction-production:blum2011welfare,auction-production:huang2019welfare,auction-production:jazi2025posted,auction-production:tan2025threshold},
but contrasts with the traditional unit supply setting ($\sum_\iinn x_{ij} \le 1$) in \citep{RegretNet-journal:dutting2024optimal,lottery-AMA:curry2023differentiable,GemNet:wang2024gemnet}.
Furthermore, we impose no special restrictions on the form of the auctioneer's utility function $u_0$, allowing $u_0$ to represent either welfare or revenue functions. 
Specially, if we define $u_0(\bmx, \bmp; \bmt) = \sum_\iinn p_i - c_0(\bmx)$, where $c_0(\bmx)$ is a general regularization function penalizing unfeasible allocations ($\sum_\iinn x_{ij} > 1$), then our model can approximately represent the unit-supply settings in \citep{RegretNet-journal:dutting2024optimal,lottery-AMA:curry2023differentiable,GemNet:wang2024gemnet}.
More discussions about our model are provided in \cref{app:sec-discuss:subsec-boundary}.
\end{remark}

Our goal is to design an algorithm to output a truthful (IC or strategy-proof) and individual rational (IR) direct mechanism that maximizes the mechanism designer's expected utility for each problem $\MD$.


\begin{definition}[Direct Mechanisms]
A direct mechanism $M^d = \left( \bmx^d,\bmp^d \right)$ consists of an allocation rule $\bmx^d: \calT \to \calX$ and a payment rule $\bmp^d: \calT \to \bbR^n$. 

In a direct mechanism $M^d = \left( \bmx^d,\bmp^d \right)$, if players report type profile $\bmt'$, then the outcome $\bmx^d(\bmt') \in \calX$ as well as monetary transfer $\bmp^d(\bmt') \in\bbR^n$ will be realized. Player $i$ receives the local outcome $x_i^d(\bmt')$ and pays $p_i^d(\bmt')$ to the mechanism designer, and thus obtains her utility $u_i(x_i^d(\bmt'),p_i^d(\bmt');t_i)$. 




\end{definition}


\begin{definition}[Truthful and IR Direct Mechanisms]
\label{def:truthful-direct-mechanism}
A direct mechanism $M^d = (\bmx^d,\bmp^d)$ is truthful, if truthfully reporting true type is a dominant strategy for each player, regardless of the reported types $\bm{t}'_{-i}$ of all other players. Mechanism $M^d$ is IR, if each player is happier to participate in the mechanism instead of realizing the outside option ($x_i = \zeros, p_i = 0$, meaning receives nothing without payment) at any case. The properties of truthfulness and IR can be mathematically expressed by following two conditions: 
\begin{align}
    u_i(x^d_i(t_i,\bm{t}'_{-i}),p^d_i(t_i,\bm{t}'_{-i});t_i) \ge& 0, \qquad &\forall& \bm{t}'_{-i}\in\calT, \iinn &\label{eq:model-IR}\tag{IR}
    \\
    u_i(x^d_i(t_i,\bm{t}'_{-i}),p^d_i(t_i,\bm{t}'_{-i});t_i) \ge& u_i(x^d_i(t'_i,\bmt'_{-i}),p^d_i(t'_i,\bmt'_{-i});t_i).\qquad &\forall& \bm{t}'_{-i} \in\calT, t'_i \in \calT_i, \iinn  &\label{eq:model-IC}\tag{IC}
\end{align}
\end{definition}

From now on, we refer truthful and IR direct mechanisms as truthful mechanisms for simplicity when the context allows.
We denote $\calM^d$ as the set of all direct mechanisms.
Formally, the goal \Wrt a mechanism design problem $\MD$ can be stated as solving the Program \eqref{eq:model-6}.

\begin{equation}
\label{eq:model-6}
{\small\max_{(\bmx^d, \bmp^d) = M^d \in \calM^d} \quad
\EU(M^d) \coloneqq \bbE_{\bmt \sim \calF}\left[ u_0(\bmx^d(\bmt),\bmp^d(\bmt);\bmt) \right].
\quad
\st \quad (\ref{eq:model-IC}),}
(\ref{eq:model-IR})
\end{equation}
In \cref{sec:method}, we will introduce the algorithm, referred to as \method, that solves Program \eqref{eq:model-6}.


\Magic
\subsection{Characterization Results}

\magic
Prior to elaborating on \method, we begin with introducing fundamental concepts related to pricing rules and a pivotal characterization result, both of which are essential for understanding \method.

\magic

\begin{definition}[Pricing Rule]
\label{def:pricingrule}
Fix a mechanism design problem $\MD$, a pricing rule \wrt\ player $i$ is a function $p_i: \calX_i \times \calT_{-i} \to \bbR$. 
\end{definition}

\magic
Pricing rules have following economic explanations:
condition on that the type profile of other players $\bmt_{-i}$ is given, if player $i$ selects a local outcome $x_i$, then she shall pay $p_i(x_i;\bmt_{-i})$ to the mechanism designer. 
\magic

\begin{definition}[Menu Mechanisms]
\label{def:menu-mechanism}
A menu mechanism $M^m$ consists of $n$ pricing rules, \ie, $M^m = \{p^m_i\}_\iinn$, where $p^m_i: \calX_i \times \calT_{-i} \to \bbR$ is the pricing rule \Wrt player $i$.



\end{definition}

\magic
Given a type profile $\bmt$, the menu mechanism $M^m$ determines the optimal local outcome $x^*_i \in \calX_i$ for each player $i$ that maximizes her utility, i.e., $x^*_i \in \argmax_{x_i \in \calX_i} u_i(x_i, p^m_i(x_i, \bmt_{-i}); t_i)$\footnote{When there might be multiple maximum points, any tie-breaking rule is feasible.}. 
The corresponding monetary transfer $p^*_i \in\bbR$ for player $i$ is then defined as the price associated with $x^*_i$: $p^*_i = p^m_i(x^*_i;\bmt_{-i})$. 
The resulting global outcome and payment vector are respectively given by, $\bmx^* = (x^*_1,...,x^*_n)$ and $\bmp^* = (p^*_1,...,p^*_n)$.

Denote $\calM^m$ as the set of all menu mechanisms. 
Unlike conventional menu mechanisms (as in \cref{eg:CoD}) that enumerate a set of candidates to discretize the outcome space, the menu mechanisms in this paper are represented by pricing rules, making them discretization-free. This approach allows us to express menu mechanisms more efficiently and to analyze them from a functional perspective.

\magic

\begin{definition}[Equivalence Between Mechanisms]
\label{def:menu-direct-1}
Let $M^m = \{p^m_i\}_\iinn$ and $M^d=(\bmx^d, \bmp^d)$ be a menu mechanism and a direct mechanism, respectively.
We say $M^m$ and $M^d$ are equivalent if following holds,
{\small\begin{align*}
x^d_i(\bmt) \in& \argmax_{x_i \in \calX_i} u_i(x_i, p^m_i(x_i;\bmt_{-i}); t_i), \quad 
p^d_i(\bmt) = p^m_i(x^d_i(\bmt);\bmt_{-i}). \quad \forall \bmt\in\calT
\end{align*}}
\end{definition}

Note that in \cref{def:menu-direct-1}, if $M^d$ is equivalent with $M^m$, then given the same type profile $\bmt$, the outcome as well as the monetary transfer of $M^d$ are exactly those of menu mechanism $M^m$.
Next, we define two necessary conditions for constructing the equivalence between menu mechanisms and direct mechanisms.

\magic

\begin{definition}[No-buy-no-pay]
\label{def:nbnp}
A menu mechanism $M^m = \{p^m_i\}_\iinn$ satisfies \nbnp,
if for all $\iinn, \bmt_{-i} \in \calT_{-i}$, we have $p^m_i(\zeros, \bmt_{-i}) \le 0$.
\end{definition}

\Magic

\begin{definition}[Partial-convexity]
\label{def:ppc}
A menu mechanism $M^m = \{p^m_i\}_\iinn$ satisfies \ppc,
if for all $\iinn$ and all $\bmt_{-i}$, $p^m_i(x_i;\bmt_{-i})$ is convex on $x_i$. 
\end{definition}

\magic
The \Nbnp property can be intuitively interpreted as: when any player selects the outside option in a menu mechanism, they incur no payment to the mechanism designer.
Partial-convexity signifies that the pricing rule exhibits convexity with respect to $x_i$.
The following theorem characterizes the equivalence between menu mechanisms and truthful direct mechanisms, if some specified conditions are satisfied. Due to space constraints, the proof of \cref{thm:direct-menu-equivalence} (as well as other theorems) can be found in \cref{app:omitted-proofs}.

\magic

\begin{restatable}{theorem}{thmDirectMenuEquivalence}
\label{thm:direct-menu-equivalence}
\emph{\bf (Equivalence Between Mechanism Classes)}
Let $\calM^{d,t}$ be the class of all \textbf{truthful} direct mechanisms and $\calM^{m,pn}$ be the class of all menu mechanisms satisfying \textbf{\ppc} and \textbf{\nbnp}.
Then, $\calM^{d,t}$ and $\calM^{m,pn}$ are equivalent in the following sense:

\magic
\begin{itemize}
\item For any truthful direct mechanism $M^d \in \calM^{d,t}$, there is a menu mechanism $M^m \in \calM^{m,pn}$ such that $M^m$ is equivalent with $M^d$.
\item For any menu mechanism $M^m \in \calM^{m,pn}$, and for any direct mechanism $M^d$ that is equivalent with $M^m$, it holds that $M^d \in \calM^{d,t}$ (\ie, $M^d$ is truthful).
\end{itemize}

\end{restatable}

\magic
\Magic
\section{Design of \Method}
\label{sec:method}

\begin{figure}[h]
\centering
\includegraphics[width=0.8\linewidth]{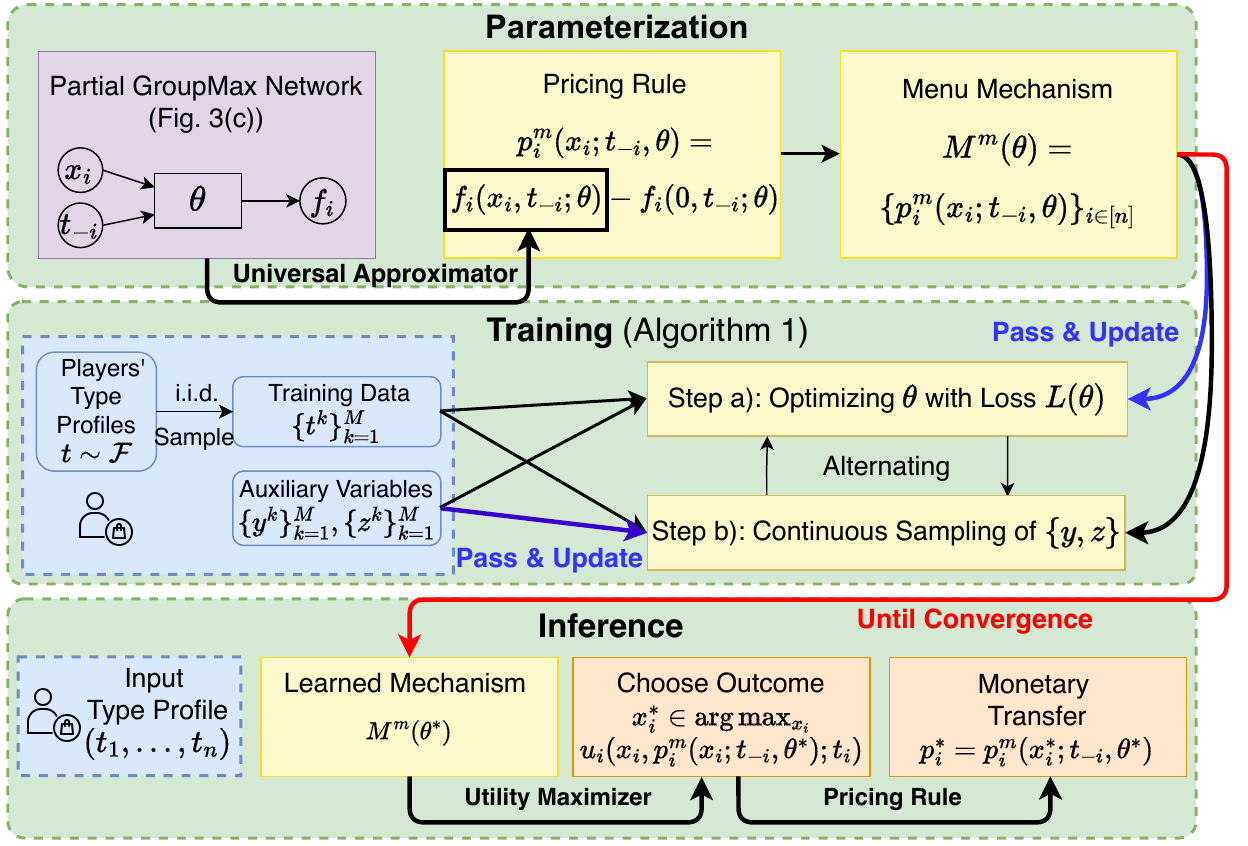}
\caption{An illustration of procedure of \method. Best viewed in color (same for the remaining figures).}
\label{fig:whole-picture}
\MMagic
\end{figure}

\magic
This section presents the design of \method, structured around its three components: parameterization, inference, and training (see \cref{sec:method:para,}, \cref{sec:method:infer}, \cref{sec:method:train}), with whole procedure illustrated at \cref{fig:whole-picture}.
The training procedure's pseudo-codes are provided and explained in the end of \cref{sec:method}.

Prior to delving into the technical details of \method, we first outline the key challenges in its designing~and our high-level ideas to address them.

{\bf Challenge 1: Parameterizing Pricing Rules}.
Theoretical equivalence (\cref{thm:direct-menu-equivalence}) motivates us to parameterize pricing rules using neural networks, explicitly encoding \Ppc and \Nbnp constraints through architectural design.
A critical challenge is that the network should be a universal approximator of partial convex functions in order to capture the (unknown) optimal mechanism.
However, constructing a network-based universal approximator of partial convex functions remains an open problem.

{\bf Challenge 2: Optimization on Pricing Rules}.
Unlike previous works, the inference of menu mechanism here necessitates computing argmax points in continuous spaces ($\calX_i$), rather than discrete ones. This shift leads to conventional "softmax" relaxation methods \citep{MenuNet:shen2019automated,lottery-AMA:curry2023differentiable,GemNet:wang2024gemnet} intractable. 

Our solution leverages the insight that parameter updates do not necessitate direct computation of the objective function. 
Instead, we only require an unbiased gradient estimator and a stream of samples from the underlying distribution. 
For this purpose, we introduce two innovative techniques—\emph{covariance trick} and \emph{continuous sampling}—which together effectively address these challenges.



\Magic
\subsection{Parameterization}
\label{sec:method:para}

Let $f_i(x_i, \bmt_{-i};\theta)$ denote a neural network (specifically, \network) represented by parameters $\theta$, taken $(x_i, \bmt_{-i})$ as input and output $f_i(x_i, \bmt_{-i};\theta)$. The network parameter is shared among all players $\iinn$.
The pricing rule for player $i$, $p^m_i$, is represented in the following expression,
\begin{equation}
\label{eq:pricingrule}
{\small\begin{aligned}
    p^m_i(x_i;\bmt_{-i},\theta) = f_i(x_i,\bmt_{-i};\theta) - f_i(\zeros,\bmt_{-i};\theta).
\end{aligned}}
\end{equation}
Note that \Nbnp is satisfied from \cref{eq:pricingrule} automatically. To ensure the \Ppc of pricing rules, it suffices to enforce \Ppc of $f_i(x_i;\bmt_{-i},\theta)$ solely on $x_i$. 
We subsequently introduce \emph{\network}, a universal approximator of partial convex functions, which is inspired by GroupMax Network \citep{GroupMax:warin2023groupmax} and parameterized max-affine function (PMA) \citep{PMA-universal_approximator:kim2022parameterized}.
Before introducing \network, we formalize two foundational sub-components: GroupMax activation and \Pan (PAN).




\begin{figure}[t]
\centering
\includegraphics[width=0.8\linewidth]{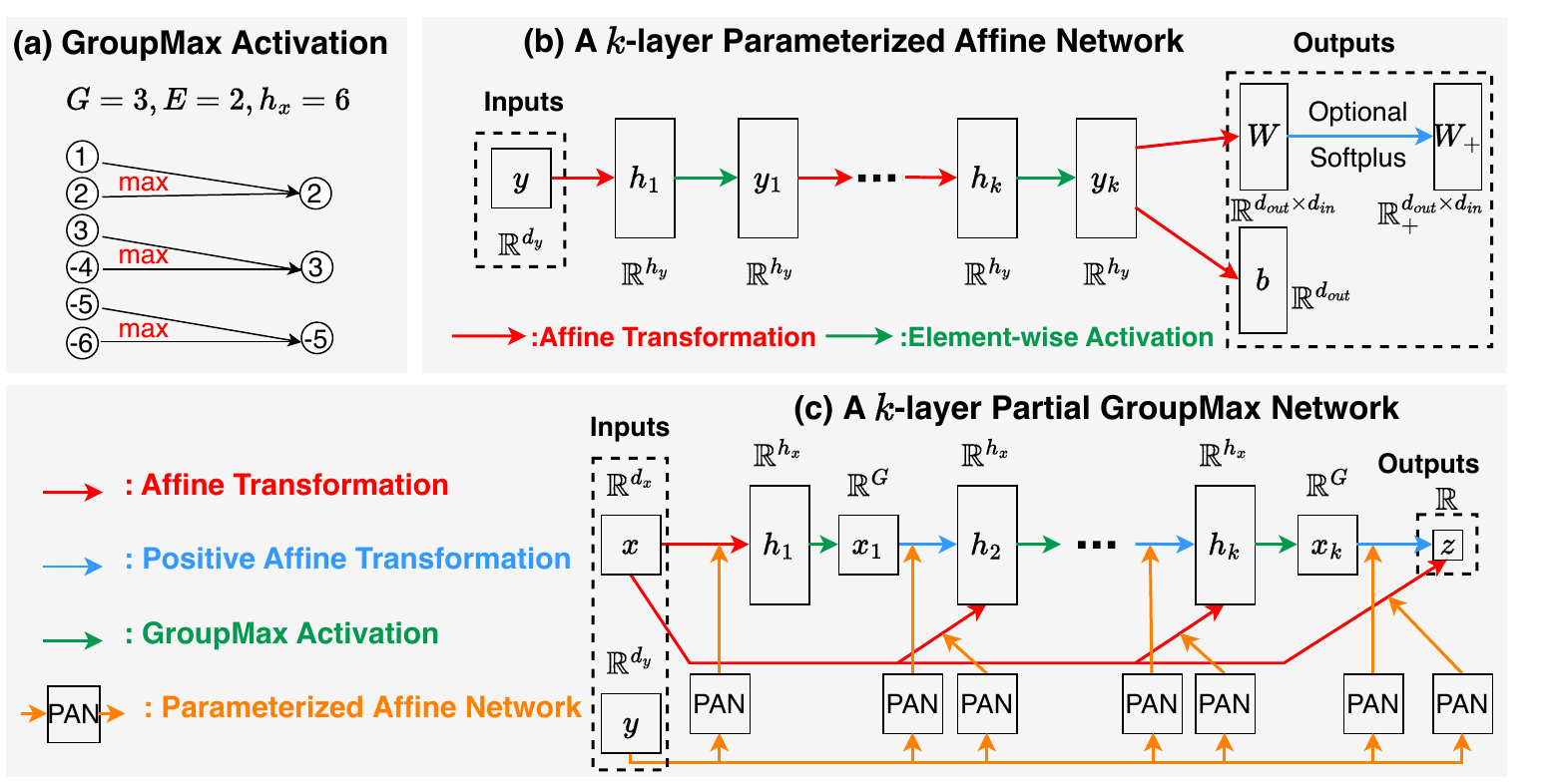}
\caption{The architecture of a $k$-layer Partial GroupMax Network and its sub-components.}
\label{fig:network}
\MMagic
\end{figure}

\magic

\begin{definition}[GroupMax Activation]
\label{def:activation}
The GroupMax activation function has the expression:
\begin{equation}
\label{eq:activation-1}
{\small\begin{aligned}
\GroupMax: \bbR^{G\times E} \to \bbR^{G}\qquad 
\GroupMax(x)_i = \max_{1\le j\le E} x_{i,j}.
\end{aligned}}
\end{equation}
\end{definition}

\magic
The intuition behind \cref{eq:activation-1} is: The GroupMax activation function equally distributes the $G\cdot E$ elements into $G$ groups, with each group containing $E$ elements. It then outputs a vector that collects the largest element from each group.
\cref{fig:network}-(a) illustrates the GroupMax activation function.

\magic

\begin{definition}[\pan]
\label{def:pan}
A \Pan parameterizes functions $W: \bbR^{d_y} \to \bbR^{d^{out}\times d^{in}}$ and $b: \bbR^{d_y} \to \bbR^{d^{out}}$, where $d_y$ is the input dimension of \pan, $d^{in}$ and $d^{out}$ are the dimensions of input vectors and output vectors of an affine transformation.


\end{definition}

\magic
Given an input $y$, a \Pan produces parameters $W(y) \in \bbR^{d^{out}\times d^{in}}$ and $b(y) \in \bbR^{d^{out}}$, which can perform an affine transformation on a vector $z \in \bbR^{d^{in}}$ as: $W(y) \cdot z + b(y)$. 
Sometimes $W(y)$ is required to be positive, then we implement an element-wise softplus function\footnote{A softplus function has the expression: $\softplus(x) = \log (1 + \exp (x))$. It maps a real number to a positive number.} to ensure the positivity.
The architecture of a $k$-layer \Pan is illustrated in \cref{fig:network}-(b).

Next, let us define \Network based on the above components. 

\magic

\begin{definition}[\network]
\label{def:network}
A \Network parameterizes a function $f: \bbR^{d_x} \times \bbR^{d_y} \to \bbR$, such that $f(x,y)$ is convex on $x$ and continuous on $(x,y)$ given inputs $x \in \bbR^{d_x}, y\in \bbR^{d_y}$. 
\end{definition}

\magic
Compared with the conventional neural networks, \Network has two key differences: (1) all activation functions are replaced by GroupMax activation, which is no longer element-wise; and (2) all affine transformations in \Network are generated by a series of Parameterized Affine Networks that take $y$ as input. 

Specifically, a $k$-layer \Network first generates all affine transformations $\{W_k(y), b_k(y)\}_{k=0}^K$ with hardcoded constraints $W_k(y) \ge 0, \forall k\ge 1$, which are output by \Pan on input $y$. Given these affine transformations, the remaining computation process on $x$ is following: $h_k(x,y) = W_{k-1}(y) x_{k-1}(x,y) + b_{k-1}(y)$ and $x_k(x,y) = \GroupMax(h_k(x,y))$ for $k=1,2,\ldots,K$, where $x_0(x,y) = x$. \Network finally outputs $z(x,y) = W_K(y) x_K(x,y) + b_K(y)$.\footnote{The residual connection is omitted for simplicity.}
An illustration of the structure of a $k$-layer \Network is provided in \cref{fig:network}-(c).

We defer more detailed architectures of these networks to \cref{app:omitted:network:architecture} and highlight the following key property of \network, which plays a crucial role in \method.

\magic

\begin{proposition}
\label{prop:network-universal-informal}
\Network is a universal approximator of all functions $f(x,y)$, which are convex on $x$ and continuous on $(x,y)$.
\end{proposition}

\magic
\cref{prop:network-universal-informal}
is formally restated as \cref{thm:network-universal} in \cref{app:omitted:network}. 
\cref{prop:network-universal-informal} has a twofold meaning: 
On one hand, any instance of \network, denoted as $f(x,y)$, is guaranteed to be convex on $x$ and continuous on $(x,y)$. 
On the other hand, any function that is convex on $x$ and continuous on $(x,y)$ can be arbitrarily approximated by \network.

\begin{remark}
\itshape
We briefly delve into why \Network is convex on $x$ as illustrated in \cref{fig:network}, and the insight of universal approximation is left to appendix for space limits.
The convexity is ensured based on the following insights: a positive-weighted sum or point-wise maximum of convex functions is also convex. Specifically, each element in $h_1$, when considered as a function of $x$, is linear in $x$. Each element in $x_1$ is the maximum of some elements in $h_1$, thereby convex in $x$. Each element in $h_2$ is a positive-weighted sum of some elements in $x_1$, which preserves convexity in $x$. This reasoning can be extended to all $x_k$s and $h_k$s, leading to the conclusion that the output $z$ is convex in $x$.
\end{remark}




\Magic
\subsection{Inference}
\label{sec:method:infer}

\magic
The main task of \emph{inference} component of \method, is to efficiently generate the outcome $\bmx^*=(x_1^*,x_2^*,\cdots,x_n^*)$ and monetary transfer $\bmp^*$ given type profile $\bmt$ and parameter $\theta$. To be specific, each local outcome $x_i^*$ is derived by maximizing player $i$'s utility \Wrt pricing rule $p^m_i(x_i;\bmt_{-i},\theta)$, and $p^*_i$ is then determined by $p^*_i = p^m_i(x^*_i;\bmt_{-i},\theta)$. The utility-maximization problem for player $i$ is formalized as follows:

\magic
{\begin{equation}
\label{eq:infer-1}
\begin{aligned}
\max_{x_i\in \calX_i}\quad u_i(x_i,p^m_i(x_i;\bmt_{-i},\theta);t_i) =& \langle t_i, x_i\rangle - p^m_i(x_i;\bmt_{-i},\theta)
= \langle t_i, x_i\rangle - f_i(x_i;\bmt_{-i},\theta) + f_i(\zeros;\bmt_{-i},\theta).
\end{aligned}
\end{equation}}

\magic
Note that the first term, $\langle t_i, x_i\rangle$, is linear on $x_i$; the second term, $- f_i(x_i;\bmt_{-i},\theta)$, is concave on $x_i$; and the third term, $f_i(\zeros;\bmt_{-i},\theta)$, is constant \Wrt $x_i$.
Overall, the objective in \eqref{eq:infer-1} is concave on $x_i$. 
Finding the maximum point $x^*_i$ of a concave function with convex constraints $x_i \in \calX_i$ is tractable through many convex optimization approaches \citep{convex:boyd2004convex}, and the corresponding $p^*_i$ can be derived naturally.
\cref{alg:infer} (left to \cref{app:omitted:inference}) presents the pseudo-codes of the inference procedure.

\Magic
\subsection{Training}
\label{sec:method:train}

\magic
\emph{Training} is a key component of \method, and this subsection provides a detailed introduction to it at length. 
For simplicity, we use $u_0(\bmx,\theta;\bmt)$ and $u_i(x_i;\bmt,\theta)$ to represent $u_0(\bmx, \{p_i(x_i;\bmt_{-i},\theta)\}_\iinn;\bmt)$ and $u_i(x_i,p_i(x_i;\bmt_{-i},\theta);t_i)$, 
as the utilities of mechanism designer and players depend on outcome $\bmx$, type profile $\bmt$ and menu mechanism parameterized by $\theta$. (Note that the prices $\{p_i(x_i;\bmt_{-i},\theta)\}_\iinn$ are then uniquely determined by $\theta$, $\bmx$, and $\bmt$.) 

Denote $\bmx(\bmt)$ as the outcome induced by type profile $\bmt$. Given $\theta$ fixed, the corresponding local outcome $x_i(\bmt)$ must be subject to the players' utility maximization constraints, \ie, $\forall \iinn,\ x_i(\bmt) \in \argmax_{a_i\in \calX_i} u_i(a_i;\bmt,\theta)$.
The mechanism designer's utility is then determined by $u_0(\bmx(\bmt),\theta;\bmt)$.
Because the type profiles are drawn from $\calF$, the mechanism designer would like to maximize the expected utility, shown as \eqref{eq:original-program}:

\magic
\begin{equation}
\label{eq:original-program}
\begin{aligned}
\max_{\theta, \bmx(\bmt)} \quad \bbE_{\bmt\sim \calF} \left[ u_0(\bmx(\bmt),\theta;\bmt) \right]
\quad
\st\quad x_i(\bmt) \in \argmax_{a_i\in \calX_i} u_i(a_i;\bmt,\theta)\quad \forall \iinn, \forall \bmt\in \calT
\end{aligned}
\end{equation}
\magic

In the remainder of \cref{sec:method:train}, we will explain how \Method solves Program \eqref{eq:original-program}
step-by-step.

\Magic
\subsubsection{Simplifying the Program.}

The first difficulty is that Program \eqref{eq:original-program}  
consists of infinite constraints and the objective has an expectation form. 
To address this issue, we begin by sampling $M$ $\iid$ type profiles from the distribution $\calF$, denoted as $\calT^M = \{\bmt^1,...,\bmt^M\}$. 
Thus, we can simplify the objective and constraints by using a sample-average approximation on these samples. 
Here, Program \eqref{eq:original-program} becomes
\magic
\begin{equation}
\label{eq:program-1}
{\small\begin{aligned}
\max_{\theta, \{\bmx^{k}\}_\kinM} \quad \frac{1}{M}\sum_{k=1}^M \left[ u_0(\bmx^k,\theta;\bmt^k) \right]
\quad
\st\quad x^k_i \in \argmax_{a_i\in \calX_i} u_i(a_i;\bmt^k,\theta), \quad \forall \iinn, \forall \kinM
\end{aligned}}
\end{equation}
\magic
When $M$ goes to infinity, Program \eqref{eq:program-1} approaches the original form \eqref{eq:original-program}.

\magic
\subsubsection{Softening the Constraints.}

The second difficulty is that the computation process of the $\argmax$ operator in Program \eqref{eq:program-1} is non-differentiable. What's worse, the $\argmax$ takes on $\calX_i$, which is a continuous space, leading the conventional softmax relaxation \citep{MenuNet:shen2019automated,lottery-AMA:curry2023differentiable,GemNet:wang2024gemnet} impossible.
To resolve this difficulty, we consider relaxing the $\argmax$ operator as a ``softmax-like'' distribution.
Specifically, given type profile $\bmt$ and mechanism parameter $\theta$, we construct the (energy-based) distribution $x_i \sim q^{\beta_i}_i(\cdot;\bmt,\theta)$, where
\magic
{\small\begin{align}
\label{eq:algo-1}
q_i^\beta(a_i;\bmt,\theta) = \frac{\exp\left(\beta\cdot u_i(a_i;\bmt,\theta) \right)}{Z_i^\beta(\bmt,\theta)},\quad 
Z_i^\beta(\bmt,\theta) = \int_{a_i \in \calX_i} \exp\left(\beta \cdot u_i(a_i;\bmt,\theta) \right) \dd a_i,
\magic
\end{align}}
$\beta > 0$ is the temperature and $Z_i^\beta(\bmt,\theta) $ is the normalizing constant (also called partition function \wrt\ $a_i$). 
It's clear to see that $q^\beta_i(\cdot;\bmt,\theta)$ is indeed a distribution since $\int_{\calX_i} q^\beta_i(a_i;\bmt,\theta) \dd a_i \equiv 1$.
Program \eqref{eq:program-1} then becomes

\magic
\begin{equation}
\label{eq:program-2}
{\small\begin{aligned}
\max_{\theta} \quad& \OBJ^\bmbb(\theta) = \frac{1}{M}\sum_{k=1}^M 
\bbE_{x^k_i \iidd q^{\beta_i}_i(\cdot;\bmt^k,\theta)} \left[ u_0(\bmx^k,\theta;\bmt^k) \right].
\end{aligned}}
\end{equation}
Note that when $\beta_i \to \infty$, the distribution $q_i^{\beta_i}(\cdot;\bmt,\theta)$ approximates the point mass at $\{x_i: x_i \in \argmax_{a_i\in \calX_i} u_i(a_i;\bmt,\theta)\}$. Therefore, Program \eqref{eq:program-2} approaches Program \eqref{eq:program-1} as all $\beta_i \to \infty$.

\Magic
\subsubsection{Differentiating over $\theta$ with \emph{Covariance Trick}.}

From now on, we focus on the latest Program \eqref{eq:program-2}, which takes the form of an unconstrained optimization Program on $\theta$ with objective function $\OBJ^\bmbb(\theta)$.
Following conventions in optimization and machine learning \citep{SGD:bottou2010large}, what we need is an unbiased estimator of $\nabla_\theta \OBJ^\bmbb(\theta)$, namely the first-order gradient of the objective.
Given this estimator, $\theta$ can be optimized by the downstream first-order algorithm.

One might think $\nabla_\theta \OBJ^\bmbb(\theta)$ is intractable, since $\theta$ appears not only in the objective function but also under the expectation operator ($\bbE_{x^k_i \iidd q_i^{\beta_i}(\cdot;\bmt^k,{\color{red}\theta}) }[\cdot]$), making $\nabla_\theta$ and $\bbE[\cdot]$ non-interchangeable.
Surprisingly, the identity in \cref{prop:cov-trick} holds universally, enabling the interchange of $\nabla_\theta$ with the expectation operator.


\magic

\begin{restatable}{proposition}{propAlgoOne}
\emph{\bf (Covariance Trick)}
\label{prop:cov-trick}
For any $\bmt$, the following identity holds.
\begin{equation}
\label{eq:algo-13-1}
\begin{aligned}
\nabla_\theta \bbE_{x_i} \left[ u_0(\bmx,\theta;\bmt) \right] 
=& \bbE_{x_i} [\nabla_\theta u_0(\bmx,\theta;\bmt)] 
+ \Cov_{x_i} \left[ u_0(\bmx,\theta;\bmt), \nabla_\theta \ASW^\bmbb(\bmx,\bmt;\theta) \right],
\end{aligned}
\end{equation}
where the expectation is taken on $x_i \iidd q^{\beta_i}_i(\cdot;\bmt,\theta)$,
$\ASW^\bmbb (\bma,\bmt;\theta) = \sum_\iinn \beta_i u_i(a_i,\bmt;\theta)$ is the affine social welfare of players under outcome $\bma \in \calX$, weighted by $\bmbb$. $\Cov[X, Y]=\bbE[XY]-\bbE[X]\bbE[Y]$ is the covariance of random variables $(X,Y)$.
\end{restatable}


\magic
The proof of \cref{prop:cov-trick} is left to \cref{app:omitted-proofs}.
The mathematical insight of \cref{prop:cov-trick} is as follows: since the underlying distribution $q_i^{\beta_i}$ is not arbitrary but has a specific energy-based form, exchanging differential operator $\nabla_\theta$ and expectation operator $\bbE_{x_i}$ will exactly induce an additional covariance term.
We then replace covariance $\Cov[\cdot]$ with expectation $\bbE[\cdot]$ and transform \cref{eq:algo-13-1} to following form:
{\small
\begin{equation}
\label{eq:algo-13-2}
{\small\begin{aligned}
\frac{1}{2} \bbE_{\{y_i,z_i\}_\iinn} \left[ \nabla_\theta \left( u_0(\bmy,\theta;\bmt) + u_0(\bmz,\theta;\bmt) \right)
+
\left( u_0(\bmy,\theta;\bmt) - u_0(\bmz,\theta;\bmt) \right) \cdot \nabla_\theta \left( \ASW^\bmbb(\bmy;\bmt,\theta) - \ASW^\bmbb(\bmz;\bmt,\theta) \right) \right],
\end{aligned}}
\end{equation}
}
where the expectation is taken on $\{y_i, z_i\} \iidf q_i^{\beta_i} (\cdot;\bmt,\theta)$ and $\{y_i, z_i\}$s are independent among $i \in [n]$.

\Magic
\subsubsection{Deriving the Loss Function.}
The next step is to construct a computable loss function $L$ such that, by differentiating $L$ over $\theta$, we can exactly obtain \cref{eq:algo-13-2}. Actually, the loss function in \cref{eq:algo-14} below satisfies our requirement.

\Magic
{\small
\begin{equation}
\label{eq:algo-14}
\begin{aligned}
L^\bmbb(\theta;\bmt) = \frac{1}{2}\bbE \left[ \left( u_0(\bmy,\theta;\bmt) + u_0(\bmz,\theta;\bmt) \right)
+ \left( \tilde{u}_0(\bmy,\theta;\bmt) - \tilde{u}_0(\bmz,\theta;\bmt) \right) \left( \ASW^\bmbb(\bmy;\bmt,\theta) - \ASW^\bmbb(\bmz;\bmt,\theta) \right) \right],
\end{aligned}
\end{equation}
}
where $\tilde{u}_0$ indicates that this term does not participate in gradient computation. It can be implemented by `.detach()' module in PyTorch for the term $u_0(\bmy,\theta;\bmt) - u_0(\bmz,\theta;\bmt)$.

Until now, the only task left for obtaining an unbiased estimator of $\nabla_\theta \OBJ^\bmbb(\theta)$ is to draw $y^k_i, z^k_i \sim q_i^{\beta_i} (\cdot;\bmt^k,\theta)$ for each sample index $\kinM$, player index $\iinn$ and time-contingent parameter $\theta$. 
One would imagine drawing samples after each network iteration using existing techniques
\citep{importance-sampling:kahn1951estimation,importance-sampling:kloek1978bayesian,MCMC:metropolis1953equation,Gibbs:geman1984stochastic}.
However, re-drawing samples from scratch for every iteration is costly.
The next paragraph introduces how we utilize the ``continuous sampling'' technique to draw these samples efficiently during full training period.

\Magic
\subsubsection{\emph{Continuous Sampling} of \texorpdfstring{$y_i \sim q^{\beta_i}_i(\cdot;\bmt,\theta)$ for time-contingent $\theta$}{Continuous Sampling}.}
\label{subsubsec:method:sample}
Without loss of generality, we focus on continuous sampling of $y_i \sim q^{\beta_i}_i(\cdot;\bmt,\theta)$ for single player $i$ and one type profile $\bmt$.
Denote $\theta^{last}$ and $\theta^{new}$ as the last-iteration and current network parameters, respectively.
Assume that we already have an approximate sample $y_i^{last}$ from $q^{\beta_i}_i(\cdot;\bmt,\theta^{last})$.
We construct a Euler–Maruyama discretization of Langevin dynamic \citep{Langevin:roberts1996exponential} with initial point $y^0_i = y_i^{last}$ as follows:

Fix the iteration time $T$, the random variables $y_i^1,...,y_i^T$ are constructed sequentially and finally output $y_i^T$. The generating process of $y_i^t$ given $y_i^{t-1}$ is,
\begin{equation}
\label{eq:algo-15}
{\small\begin{aligned}
y_i^t = y_i^{t-1} + \eta_i \cdot \nabla_{x_i} u_i(y_i^{t-1};\bmt,\theta^{new}) + \sqrt{2 \eta_i {\beta_i}^{-1}} \epsilon^t_i, \quad
\epsilon^t_i \iidf \calN(0,1),
\end{aligned}}
\end{equation}
where $\eta_i>0$ is the step size or learning rate. 
Notice that the distribution of $y_i^{last}$ is close to $q^{\beta_i}_i(\cdot;\bmt,\theta^{new})$ (since $\theta^{new}$ is close to $\theta^{last}$), and the distribution $q^{\beta_i}_i(\cdot;\bmt,\theta^{new})$ is log-concave. 
Thus, by nature of Langevin dynamics, the distribution of $\{y_i^t\}$ converges quickly to $q^{\beta_i}_i(\cdot;\bmt,\theta^{new})$ \citep{Langevin:roberts1996exponential} and it's natural to take $y^T_i$ as an approximate sample from $q^{\beta_i}_i(\cdot;\bmt,\theta^{new})$. 

\Magic
\subsubsection{Training Procedure.}
\label{subsubsec:method:train-pseudo}
\begin{algorithm}[t]
\caption{Training of \Method}
\label{alg:train}
\KwIn{$\MD = (n, m, u_0, \calF)$: a mechanism design problem}
\KwOut{$\theta$: parameter of menu mechanisms}
\textbf{parameter}: $M$: number of samples, $S$: iteration in Langevin dynamics, $T$: total training iteration, $\beta_0$: initial temperature, $\eta_0$: initial learning rate in Langevin dynamics, $\eta_\theta$: initial learning rate for $\theta$\\
Sample $M$ type profiles $\{\bmt^k\}_\kinM \iidd \calF(\calT)$, initialize $\{\bmy^k, \bmz^k \}_\kinM \iidd \calU(\calX)$\;
Initialize $\beta_i \gets \beta_0, \eta_i \gets \eta_0, \forall i$, and Adam Optimizer on $\theta$ with learning rate $\eta_\theta$\;
\tcc{Training Loop}
\For{$t=1,2,...,T$}
{
    \tcc{a) Updating mechanism $\theta$ by an approximate unbiased-estimator of true loss}
    Compute $L^\bmbb(\theta;\bmt^k)$ in \cref{eq:algo-14} for all $\kinM$, with $\bmy, \bmz$ in \eqref{eq:algo-14} replaced by $\bmy^k, \bmz^k$\;
    $L(\theta) \gets - \frac{1}{M}\sum_\kinM L^\bmbb(\theta;\bmt^k)$\;
    Achieve $\nabla_\theta L(\theta)$ by calling .backward() on $L(\theta)$, optimize $\theta$ with Adam optimizer\;

    \tcc{b) Resampling $\{\bmy^k, \bmz^k \}_\kinM$ by Langevin Dynamic}
    
    \For{$k=1,2,...,M; i=1,2,...,n$}
    {
        $y^{k,0}_i \gets y^k_i$, $z^{k,0}_i \gets z^k_i$\;
        \For{$s = 1,2,...,S$}
        {
        $y^{k,s}_i = y^{k,s-1}_i + \eta_i \cdot \nabla_{x_i} u_i(y^{k,s-1}_i;\bmt,\theta) + \sqrt{2\eta_i \beta_i^{-1}} \cdot \varepsilon^{k,s}_i$ where $\varepsilon^{k,s}_i \iidf \calN(0,1)$, do the same on $z^{k,s}_i$.
        }
        $y^k_i \gets y^{k,S}_i$, $z^k_i \gets z^{k,S}_i$\;
    }
    Increase $\{\beta_i\}_\iinn$ as well as decrease $\{\eta_i\}_\iinn$ and $\eta_\theta$ when appropriate\;
}
\textbf{return}: Learned menu mechanism $\theta$\;
\end{algorithm}

\cref{alg:train} presents the pseudo-codes of our training procedure.
The training procedure begins with sampling $M$ $\iid$ type profiles ($\{\bmt^k\}_\kinM$) as well as initializing auxiliary variables ($\{\bmy^k, \bmz^k\}_\kinM$) \Wrt type profiles.
In the training process, the algorithm alternates between (a) optimizing mechanism parameter $\theta$ with samples $\{\bmy^k, \bmz^k\}_\kinM$ (from line 5-7) and (b) continuous sampling of $\{\bmy^k, \bmz^k\}_\kinM$ as $\theta$ varies (from line 8-14). 

To be specific, (a) In line 5-7, we regard $\bmy^k_i$ and $\bmz^k_i$ as approximate samples from $q_i^{\beta_i}(\cdot;\bmt^k,\theta)$. 
If these samples are exactly drawn from $q_i^{\beta_i}(\cdot;\bmt^k,\theta)$, then the expectation of $\nabla_\theta L(\theta;\bmt^k)$ in line 6 is exactly the true gradient. 
Therefore, we can regard $\nabla_\theta L(\theta;\bmt^k)$ taken on samples $\{\bmy^k,\bmz^k\}_\kinM$ as an approximate unbiased-gradient of the objective function in \cref{eq:program-2}.
(b) In line 8-14, since we require the distribution of $\bmy^k_i$ and $\bmz^k_i$ close to $q_i^{\beta_i}(\cdot;\bmt^k,\theta)$, yet $\theta$ is continuously updated in step (a). 
Therefore, we also need to update $\{\bmy^k, \bmz^k\}_\kinM$ after each iteration of $\theta$, to guarantee our requirements throughout the full training process.




\Magic
\subsection{Theoretical Justifications of \method}
\label{sec:justification}

\magic
\Method preserves several ideal properties, respectively: truthfulness, full expressiveness, and dimension-insensitivity. Specially, truthfulness of \Method is a direct corollary of \cref{thm:direct-menu-equivalence}, as stated in \cref{cor:tedi-truthful}.



\magic

\begin{restatable}{corollary}{corTediTruthful}
\emph{\bf (\Method is Truthful)}
\label{cor:tedi-truthful}
For any menu mechanism $M^m$ that can be expressed by \method, and any direct mechanism $M^d$ that is equivalent with $M^m$, it holds that $M^d$ is truthful.
\end{restatable}


\magic
Given the results established in \cref{thm:direct-menu-equivalence} and \cref{prop:network-universal-informal},
the insight of the full-expressiveness of \Method then becomes clear:
On one hand, truthful mechanisms are equivalent to the menu mechanisms with partial convex pricing rules. Subsequently, partial convex pricing rules can be arbitrarily approximated by \network.
Following above insight, we (informally) state the \emph{full-expressiveness} in \cref{prop:method-expressive-informal}.
The key point is showing that the optimal mechanism over all truthful mechanisms would also be expressed by \method.
\cref{prop:method-expressive-informal} is formally restated as \cref{thm:tedi-expressive} in \cref{app:omitted:justification}. 

\magic
\begin{proposition}[\Method is Full-Expressive (Informal)]
\label{prop:method-expressive-informal}
Given a mechanism design problem $\MD$. 
Denote $\SEU^*$ $(\SEU^{TEDI})$ as the designer's supreme expected utility over all truthful mechanisms (mechanisms expressed by \method) \wrt\ $\MD$, then, $\SEU^* = \SEU^{TEDI}$.
\end{proposition}





\magic
Lastly, we provide a \emph{dimension-insensitivity} justification. Following the machine learning convention, it's widely believed that neural networks are ideal approaches to overcome the ``curse of dimensionality'', because network architecture has the potential to discover simple structures \citep[\S 6.4]{book-deep-learning:Goodfellow-et-al-2016}.
In \cref{app:omitted:expressive} we designate the parameters of a \Network to represent the optimal mechanism in \cref{eg:CoD} with $O(m)$ parameters, as a contrast with $\Omega(2^m)$ parameters for discretization-based approaches. 
As \Network utilizes neural network structure, we believe that \Network is an ideal dimension-insensitive approach to learn optimal menu mechanisms. 
\Magic
\section{Experiments}
\label{sec:exp}

\magic
In this section, we conduct thorough experiments to show the competitive performance of \Method in a reproducible goods auctions setting. 
We leave the detailed implementations of all approaches to \cref{app:subsec:exp:implementation-detail}.

\Magic
\subsection{Experimental Settings}
\label{sec:exp:setting}

\magic
In our experiments, the problem $\MD$ is economically interpreted as reproducible goods auctions \citep{reproducible-goods,digital-good-auction:chen2014optimal,auction-production:huang2019welfare}, where
mechanism designer has commitment power over outcomes \citep{borgers-mechanism-design} thus commits to produce goods with one-time production cost $c^p\in \bbR_+$ and duplication cost per-unit $c^d \in \bbR_+$.
Mechanism designer's utility is $u_0(\bmx,\bmp,\bmt) = \sum_\iinn p_i - c_0(\bmx)$, where $c_0(\bmx)=\sum_\jinm \left( c^p \cdot \max_\iinn x_{ij} + c^d \cdot \sum_\iinn x_{ij} \right)$ is the cost of realizing the outcome $\bmx$ through production.
The formulation of $u_0(\bmx,\bmp;\bmt)$ is only for experimental purpose, and TEDI can be easily applied to settings with general differentiable $u_0(\bmx,\bmp;\bmt)$.
In \cref{app:subsec:exp:more-results} we provide another choice of $u_0$ where the economic explanations inherit the unit-supply auctions in \citep{RegretNet-journal:dutting2024optimal,lottery-AMA:curry2023differentiable,GemNet:wang2024gemnet}.

$c_0(\bmx)$ has following explanations:
For each good $j$, since player $i$ buys the good $j$ with probability $x_{ij}$, good $j$ should be produced with probability at least $x_{ij}$. This argument is universal for $i$, thus good $j$ should be produced with probability at least $\max_\iinn x_{ij}$.
A rational mechanism designer will exert her effort to fully decrease the cost, thus she will produce good $j$ with probability exactly $\max_\iinn x_{ij}$.
Therefore, the expected one-time production cost for good $j$ is $c^p \cdot \max_\iinn x_{ij}$.
Similarly, good $j$ should be duplicated with $\sum_\iinn x_{ij}$ times in expectation, thus $c^d \cdot \sum_\iinn x_{ij}$ is the expected duplication cost for good $j$. 

We refer $F^{c^p, c^d}_{n,m}$ as the setting with $n$ players, $m$ goods, production cost $c^p$ and duplication cost $c^d$ for each good, and type distribution $F$. We omit superscripts $c^p, c^d$ when both are $0$.
When there is $n=1$ player and $c^p = c^d = 0$, the problem degenerates to a well-studied problem called \emph{multiple goods monopolist} \citep{optimal_auction-special_case:giannakopoulos2014duality,optimal_auction-special_case:daskalakis2015strong}.
The type distribution is from $F\in\{U, B, C\}$, representing that the type distribution is $\iid$ from $U([0,1])$, $Ber(p=0.5)$, or correlated, respectively.
For correlated distribution, we first generate $t_j \iidf U([0.25, 0.75])$ for each good $j$ and then generate $t_{ij} \iidf U([t_j-0.25, t_j+0.25])$ for each player $i$.

\Magic
\subsection{Baselines}
\label{sec:exp:baseline}

\magic
\subsubsection{Learning-Based Approaches.}

We choose the representative approach from the differentiable economics literature as baselines, respectively: RegretNet \citep{RegretNet:dutting2019optimal}, Lottery-AMA \citep{lottery-AMA:curry2023differentiable} and GemNet \citep{GemNet:wang2024gemnet}. 
These approaches were originally designated for unit-supply auctions, and we minimally modify these approaches to fit our setting and denote them as MD-RegretNet, MD-LotteryAMA, and MD-GemNet\footnote{MD stands for ``Mechanism Design''.}, respectively. 

\magic
\begin{itemize}
\item MD-RegretNet \citep{RegretNet-journal:dutting2024optimal}: Parameterize the allocation rule and payment rule directly with neural networks. The measure of untruthfulness is penalized during training. 
\item MD-LotteryAMA \citep{lottery-AMA:curry2023differentiable}: Learn an AMA mechanism with the discretization of the global outcome space and shift values.
\item MD-GemNet \citep{GemNet:wang2024gemnet}: Learn a discrete menu for each player, where the menus are output by a neural network taking others' types as input. 
\end{itemize}

\magic
For these baselines, we align the hyper-parameters with \Method to make a fair comparison, including batch size, network iterations, temperature, \etc~ (See \cref{app:subsec:exp:implementation-detail} for more details.)
We note that when there is a single player, GemNet and Lottery-AMA reduce to MenuNet \citep{MenuNet:shen2019automated}. Besides, RochetNet \citep{RegretNet-journal:dutting2024optimal} is also identical with MenuNet. Therefore, we merge them as MD-MenuNet when $n=1$.

\subsubsection{Traditional Approaches.} We consider four traditional baselines as follows.

\magic
\begin{itemize}
\item VCG \citep{VCG-V:vickrey1961counterspeculation}: The most classical mechanism with strong versatility.
\item SJA \citep{optimal_auction-special_case:giannakopoulos2014duality}: A mechanism that is proven to be optimal in multiple goods monopolists with $m\le 6$ goods and uniform distribution. SJA is also conjectured to be optimal when $m>6$.
\item Bundle-OPT: Bundling all goods together at a specific reserve price.
The reserve price is parameterized, and optimized through grid search. 
This baseline is also used in LotteryAMA \citep{lottery-AMA:curry2023differentiable}.
\item Separable-OPT: Modified from Item-wise Myerson \citep{RegretNet:dutting2019optimal}, that runs optimal mechanism on each good separately.
\end{itemize}

\Magic
\subsubsection{Ablation Studies.}
We conduct four ablation studies to show how each component in \Method will affect its performance. 
The first operates on learning algorithm and the remaining operates on network architectures.

\magic
\begin{itemize}
\item \method-Discrete: Discretize the local outcome space as MD-GemNet does. The payments are output by \network.
\item \method-PICNN: Replace \Network with PICNN (Partial Input Convex Neural Network) \citep{PICNN:amos2017input}.
PICNN also parameterizes partial convex functions, but it has no universal approximation guarantee.
\item \method-PMA: Replace \Network with PMA (Parameterized Max-Affine) \citep{PMA-universal_approximator:kim2022parameterized}.
PMA parameterizes partial convex functions, but it has incomplete network architecture.
\item \method-MLP: Replace \Network with MLP (Multi-Layer Perceptron).
MLP has no convexity guarantee on its (partial) inputs.
\end{itemize}


\Magic
\subsection{Experimental Results and Analysis}
\label{sec:exp:result}


\begin{table}[t]
\centering
\caption{Experimental results with diverse approaches (including ablations studies) and diverse settings. 
`=0' ('<0') represents the value is exactly $0$ (smaller than $0$); `x' represents the approach fails to execute the setting due to its large memory demand; `-' represents that the approach (row) does not apply to the setting (column).
The \emph{GPU Time} and \emph{TFLOPS} results are measured when the model has $n=3$ players and $m=10$ goods.
For the last two columns of \method, the first number (resp. the second number) represents the GPU time (TFLOPS) \Wrt network updating (resp. continuous sampling).
All approaches have consistent hyper-parameters in different settings except for MD-GemNet and MD-RegretNet, with fine-tuned hyper-parameters in bracketed values (\eg, (x.xxxx)) and details described in \cref{app:subsec:exp:implementation-detail}.
}
\adjustbox{max width=\textwidth}{
\begin{tabular}{c|cccccc|cccccc|cc}
\toprule
Methods\textbackslash Settings & $U_{1,2}$  & $U_{1,10}$ & $U_{1,20}$ & $B_{1,10}$ & $B_{1,20}$ & $B_{1,40}$ & $U^{1,0}_{3,2}$ & $U^{1,0}_{3,10}$ & $B^{1,0}_{3,10}$ & $C^{1,0}_{3,10}$ & $U^{0,0.5}_{3,10}$ & $C^{0,0.5}_{3,10}$ & GPU Time (s) & TFLOPS \\
\midrule
\method & \underline{0.5482} & \underline{3.4620} & 7.5245 & 4.9128 & \underline{9.7927} & \underline{19.643} & 0.2678 & \underline{1.7285} & \underline{5.8090} & \underline{2.2400} & \textbf{\underline{2.1989}} & \textbf{\underline{1.4101}} & 670 \& 1,061 & 311 \& 79 \\
\midrule
\method-Discrete & 0.5443 & 3.4292 & 7.5177 & 3.4932 & 7.0206 & x & 0.2463 & 1.4221 & 4.2375 & 1.7460 & =0 & =0 & 739 & 103 \\
\method-PICNN & 0.5442 & 3.4527 & 7.5121 & 4.7256 & 9.5346 & 19.0270 & 0.1288 & 0.9798 & 5.7548 & 1.5281 & 2.0337 & 1.2192 & 1,425 & 340 \\
\method-PMA & 0.4960 & 2.4867 & 4.9732 & \underline{4.9619} & 9.5789 & x & 0.2297 & 0.8976 & 5.7047 & 1.0720 & 1.8424 & 1.1570 & 2,211 & 467\\
\method-MLP & =0 & =0 & =0 & =0 & =0 & =0 & =0 & =0 & 0.2474 & =0 & =0 & =0 & 505 & 170\\
\midrule
MD-MenuNet & 0.5475 & 3.4472 & \underline{7.5261} & 4.0565 & 6.5714 & 13.703 & - & - & - & - & - & - & - & - \\
MD-GemNet & - & - & - & - & - & - & \underline{0.2748} & 1.4814 & (1.3309) & 1.7893 & (1.2848) & (0.6350) & 1,025 & 467 \\
MD-LotteryAMA & - & - & - & - & - & - & 0.2212 & 1.2050 & 1.8118 & 1.7017 & 0.3089 & 0.2286 & 708 & 106 \\
\midrule
MD-RegretNet & \textbf{0.5530} & \textbf{3.5186} & \textbf{7.6565} & 4.9444 & 9.9663 & 10.0198 & (\textbf{0.4053}) & \textbf{1.9053} & 5.6432 & \textbf{2.9853} & (2.1900) & (1.0874) & 8,793 & 552  \\
(regret) & 0.0003 & 0.0031 & 0.0168 & 0.0141 & 0.0005 & 0.0245 & (0.0507) & 0.0409 & 0.0171 & 0.0756 & (0.2852) & (0.1537) & - & -  \\
\midrule
SJA & 0.5491* & 3.4748 & 7.5132 & 1.9882 & 1.9999 & 3.0000 & - & - & - & - & - & - & - & - \\
VCG & =0 & =0 & =0 & =0 & =0 & =0 & <0 & <0 & <0 & <0 & =0 & =0 & - & - \\
Separable-OPT & 0.5000 & 2.5000 & 5.0000 & \textbf{5.0000}* & \textbf{10.000}* & \textbf{20.000}* & 0.2601 & 1.3019 & \textbf{6.2507} & 1.4981 & 1.8752 & 1.1894 & - & - \\
Bundle-OPT & 0.5441 & 3.4477 & 7.5252 & 3.3059 & 6.9424 & 14.762 & =0 & 0.3616 & =0 & 1.4033 & 0.4709 & 0.3319 & - & - \\
\bottomrule
\end{tabular}
}
\label{tab:exp}
\MMagic
\end{table}

\cref{tab:exp} presents all experimental results with $n=1$ or $n=3$ players, respectively.
The optimal values among all approaches under comparison are highlighted in bold. 
The optimal values among all truthful, learning-based approaches are underlined (this excludes MD-RegretNet into comparison). 
The superscript (*) represents the optimal value known for each setting. 

\Magic
\subsubsection{Comparison between \Method and Baselines.}
In the $1$-player experiments, \Method performs optimally $5$ out of $6$ times compared with MD-MenuNet. 
The performance is comparable in $U_{1,20}$, since the Bundle-OPT, which serves as a single-menu baseline, is already strong.
Surprisingly, \Method outperforms MD-MenuNet, SJA, and Bundle-OPT with a significant gap in the case of $B_{1,m}$. \cref{fig:intro} in \cref{sec:intro} presents an illustration of \Method and MD-MenuNet in the setting of $B_{1,10}$, showing that while \Method successfully learns the optimal scheme of selling goods independently, MD-MenuNet fails.
We speculate that such a phenomenon is mainly caused by the curse-of-dimensionality issue formerly explained in \cref{eg:CoD}.

For the multiple-players settings, \Method performs optimally $4$ out of $6$ times among all truthful approaches and $5$ out of $6$ times among all truthful, learning-based approaches.
A notable comparison is the settings of $U^{1,0}_{3,2}$ and $U^{1,0}_{3,10}$, where MD-GemNet outperforms \Method in the former but \Method outperforms MD-GemNet in the latter.
This comparison demonstrates that \Method can result in more robust performance even the problem size becomes large.
The other $4$ settings ($B^{1,0}_{3,10}$, $C^{1,0}_{3,10}$, $U^{0,0.5}_{3,10}$, $C^{0,0.5}_{3,10}$) demonstrate that \Method can consistently result in good performance in diverse settings, even if the types are correlated among players.

We also point out some (seemingly surprising) phenomena about baselines:
1) MD-LotteryAMA performs extremely worse in some settings ($U^{0,0.5}_{3,10}$, $C^{0,0.5}_{3,10}$) and even worse than Separable-OPT in the $2$-goods settings, $U^{1,0}_{3,2}$. This result discloses the limited expressiveness of affine maximizer mechanisms \citep{AMA:milgrom1982theory}, which is consistent with existing results \citep{AMA-limitation:carbajal2013truthful,AMD-CA:sandholm2015automated};
2) MD-GemNet (as well as \method-Discrete) has $0$ value in some random seeds.
We speculate that this phenomenon is a joint impact of following two factors:
a) the naive, allocate-nothing mechanism is a local optimum (since we introduce the cost function), and
b) discretization-based approaches are sensitive to parameter initialization \citep{lottery-AMA:curry2023differentiable,discretization:dulac2015deep}, thus is prone to stuck in local optimum.
3) MD-RegretNet has extremely large regret and smaller value in some settings (\eg, $U^{0,0.5}_{3,10}, C^{0,0.5}_{3,10}$).
We speculate that this happens because MD-RegretNet needs to compute regret in high-dimensional type space ($\calT_i$). However, computing the maximum point \Wrt a non-concave function in high-dimensional space is highly unstable \citep[\S 8]{nonconvex-optimization:dauphin2014identifying,book-deep-learning:Goodfellow-et-al-2016}.
Due to space limits, we defer more analysis (including the above phenomena) of baselines to \cref{app:subsec:exp:more-analysis}.


\Magic
\subsubsection{Analysis of Ablation Studies.}

Comparing with \method-Discrete, \Method generalizes the finite candidates to full spaces ($\calX_i$).
We note that \Method outperforms \method-Discrete with a significant gap.
This comparison suggests that replacing the discretization step plays an important role on improving the performance.


The performance of \method-PICNN, \method-PMA and \method-MLP have many take-away messages:
1) The performance gap between \Method and \method-PICNN suggests that PICNN \citep{PICNN:amos2017input} is highly unlikely to be universal approximator of partial convex functions, although the existence of counter-examples are still open.
2) The performance of \method-PMA is close to and slightly worse than Separable-OPT, especially in $1$-player settings.
We speculate that this arises from following reasons:
a) Since \method-PMA exhibits incomplete network architecture, it is likely to simply learn a linear pricing rule, and
b) Separable-OPT always gives the optimal linear pricing rule.
3) The extremely-poor performance of \method-MLP suggests that guaranteeing the partial convexity of pricing rule is of vital importance. The most important reason is below: The fast convergence of Langevin dynamics requires that the distribution $q_i^{\beta_i}(\cdot;\bmt_{-i},\theta)$ is log-concave, and this requirement gets satisfied only when pricing rule is partially convex.


\Magic
\section{Conclusions and Future Works}
\label{sec:conclusion}

To conclude this paper, we propose \method, an discretization-free algorithm that can efficiently learn truthful mechanisms.
We conduct thorough theoretical and empirical justifications to demonstrate the \emph{truthfulness}, \emph{full-expressiveness}, and \emph{dimension-insensitivity} of \method.
This paper not only preserves technical contributions by proposing \Network as well as an innovative training algorithm, which altogether resolve the ``curse-of-dimensionality'' faced in existing discretization-based approaches.
More importantly, by refining the concept of pricing rules and menu mechanisms, this paper might provide new insights on learning truthful mechanisms in more general scenarios.

We note that \Method can not be directly applied to learn feasible unit-supply auctions studied in \cite{RegretNet-journal:dutting2024optimal,lottery-AMA:curry2023differentiable,GemNet:wang2024gemnet} by nature of menu mechanisms, and we leave the discussion of the algorithmic boundary of \Method to \cref{app:sec-discuss:subsec-boundary}.
For future works, it is promising to extend the proposed technique to learn large-scale unit-supply auctions by parameterizing the boosting function in AMA mechanisms \citep{AMA:milgrom1982theory} with universally convex neural networks, thus improving the direct discretization as in \citep{lottery-AMA:curry2023differentiable,AMenuNet:duan2024scalable}.
Besides, all experiments share the ``permutation equivariance property'' \citep{permutation-equivariance:qin2022benefits,permutation-equivariance:zaheer2017deep,permutation-equivariance:lee2019set,permutation-equivariance:han2022universal}.
A permutation-equivariant pricing rule may likely help the training process of \Method and decrease the risk of over-fitting. 
For this reason, we hope that this study will motivate future explorations on the universal approximators of the convex and permutation-equivariant functions.

\nocite{*}
\newpage
\printbibliography
\newpage
\appendix

\newpage
\section{Preface to the Appendix}
\label{app:sec-preface}

Before delving into appendix, we first make some changes in notations, which makes the technical results in this paper more generalizable and more consistent with the algorithmic boundary discussed in \cref{app:sec-discuss:subsec-boundary}.

The first change is that we do not assume $\calT_i = [0,1]^m$ and $\calX_i = [0,1]^m$ any more. We only require $\calT_i$ and $\calX_i$ be convex and compact subset of Euclidean space with same dimensionality.
They need not be the same among player index $i$.
This change generalizes the original models in \cref{sec:model}.

The second change is about players' utilities. We generalize player $i$'s utility as follows.
\begin{align*}
    u_i(x_i,p_i;t_i) = \langle x_i, t_i \rangle + c_i(x_i) - p_i
\end{align*}
where we add a new term $c_i(x_i)$ on player $i$'s utility function. $c_i(x_i)$ is referred as the cost function of player $i$.
We do not pose additional requirements on $c_i(x_i)$ except normalization to $c_i(\zeros) = 0$. As a consequence, setting $c_i(x_i) \equiv 0$ could represent the original model in \cref{sec:model}.

The introduction of $c_i(x_i)$ also changes the statement of \Ppc and parameterization of pricing rules.
Firstly, \Ppc is restated as follows:
$p_i(x_i; \bmt_{-i}) - c_i(x_i)$ is convex on $x_i$ for all $i$.
The pricing rule is parameterized in following expression:
\begin{align*}
    p^m_i(x_i;\bmt_{-i},\theta) = f_i(x_i,\bmt_{-i};\theta) - f_i(\zeros,\bmt_{-i};\theta) + c_i(x_i)
\end{align*}

When $c_i(x_i) \equiv 0$, the property of \Ppc and the parameterization of pricing rule are identical with those introduced in \cref{sec:model}. Therefore, introducing $c_i(x_i)$ generalizes all results in main body.

Above all, a mechanism design problem is formally described by a $6$-tuple: $\MD = (n, \calT, \calX, \{u_i\}_\iinn, u_0, \calF)$.

Readers may feel the introduction of $c_i(x_i)$ nitpicking. However, introducing this term has strong economic insights, which will be explained in \cref{app:sec-discuss:subsec-boundary} but omitted here.

\newpage
\section{Omitted Contents}
\label{app:sec:omitted-contents}

\subsection{Notations in This Paper}
\label{app:subsec:notations}

\cref{tab:notation} shows the notations that appear in this paper.

\begin{table}[H]
\centering
\caption{Notations, and their meanings and structures appeared in this paper.}
\adjustbox{max width=\textwidth}{
\begin{tabular}{ccc}
\toprule
Notations & Meanings & Structures \\
\midrule
$n$ & number of players & - \\
$m$ & number of goods & - \\
$[n]$ & the set of players & - \\
$\bbR$ & the set of real numbers & - \\
$\bbR_+$ & the set of non-negative numbers & - \\
$\ones$ ($\zeros$) & a vector with all elements equal to $1$ ($0$) & - \\
$\calT_i$ & type space of player $i$ & $\calT_i = [0,1]^m$ \\
$t_i$ & player $i$'s type & $t_i \in \calT_i$ \\
$\calT$ ($\calT_{-i}$) & type profile space & $\calT = \times_\iinn \calT_i$ ($\calT_{-i} = \times_\jnei \calT_j$) \\
$\bmt$ ($\bmt_{-i}$) & type profile (w/o player $i$) & $\bmt = \{t_1,...,t_n\}\in \calT$ ($\bmt_{-i} = \bmt \backslash \{t_i\} \in \calT_i$) \\
$\calF$ & the distribution of type profiles & $\calF \in \Delta(\calT)$ \\
$\calX_i$ & local outcome space of player $i$ & $\calX_i = [0,1]^m$ \\
$x_i$ & a local outcome of player $i$ & $x_i \in \calX_i$ \\
$\calX$ ($\calX_{-i}$) & global outcome space & $\calX = \times_\iinn \calX_i$ ($\calX_{-i} = \times_\jnei \calX_j$) \\
$\bmx$ ($\bmx_{-i}$) &  a global outcome & $\bmx = (x_1,...,x_n)\in\calX$ ($\bmx_{-i} = \bmx \backslash \{x_i\} \in \calX_{-i}$)\\
\hline
\multirow{2}{*}{$u_i(x_i, p_i; t_i)$} & \multirow{2}{*}{utility of player $i$} & $u_i: \calX_i \times \bbR \times \calT_i \to \bbR$ with the form \\
& & $u_i(x_i, p_i; t_i) = \langle x_i, t_i \rangle - p_i$ \\
\hline
$u_0(\bmx,\bmp;\bmt)$ & utility of mechanism designer & $u_0: \calX \times \bbR^n \times \calT \to \bbR$, differentiable \\
$\MD$ & a mechanism design problem & $\MD = (n, m, \calF, u_0)$ \\
$M^d$ & a direct mechanism & $M^d = (x^d, p^d)$ \\
$x^d$ & allocation rule in direct mechanism & $x^d: \calT \to \calX$ \\
$p^d$ & payment rule in direct mechanism & $p^d: \calT \to \bbR^n$ \\
$M^m$ & a menu mechanism & $M^m = \{p^m_i\}_\iinn$ \\
$p^m_i$ & pricing rule for player $i$ in menu mechanism & $p^m_i: \calX_i \times \calT_{-i} \to \bbR$ \\
$\theta$ & parameter of \network & - \\
$x \sim q(\cdot;...)$ & random variable $x$ with pdf $q(x_0;...)$ at point $x_0$ & - \\
$x_i \iidd q_i(\cdot;...)$ & $\{x_i\}$ are independently distributed from $q_i(\cdot;...)$ & - \\
$x_i \iidf q(\cdot;...)$ & identically and independently distributed from $q(\cdot;...)$ & - \\
\bottomrule
\end{tabular}
}
\label{tab:notation}
\end{table}

\subsection{Related Works}
\label{subsec:omitted:related}

\paragraph{Automated Mechanism Design.}
The paradigm of automated mechanism design (AMD) was initiated by \citet{AMD-initial:sandholm2003automated}, aiming at finding the optimal mechanism with computational approaches rather than analytical ones.
Following this pioneering work, automated mechanism design began to prosper with diverse applications including ad auctions \citep{AMD-ad:benisch2009methodology}, combinatorial auctions \citep{AMD-CA:sandholm2015automated}, social choice, stable matching \citep{AMD-no_money:narasimhan2016automated} and facility locations \citep{AMD-facility-location:sui2013analysis}. 
Subsequent research has extended the study of AMD from multiple perspectives, including sample complexity, \citep{AMD-sample_complexity:balcan2016sample}, robustness \citep{AMD-robust:albert2017automated}, online settings \citep{AMD-online:hajiaghayi2007automated}, and dynamic settings \citep{AMD-dynamic:zhang2021automated}.
This paper builds on the idea of AMD.

\paragraph{Differentiable Economics.}
Differentiable economics aims at utilizing differential functions (\eg, neural networks) to parameterize economic solutions, followed by gradient computation for finding optimal solutions.
It inherits a similar spirit with automated mechanism design, yet emphasizes the differentiability of the objective functions for ease of optimization.
The applications include contract design \citep{differentiable-economic:wang2024deep}, multi-facility location \citep{AMD-facility-location:golowich2018deep}, market making \citep{differentiable-economic:curry2024optimal} and others.
In addition, auction design is one of the mainstream applications of differential economics with relatively mature solutions.
These solutions can be roughly partitioned into three categories: regret-based approaches~\citep{RegretNet:dutting2019optimal, curry2020certifying, peri2021preferencenet, rahme2021auction, rahme2021permutation, curry2022learning, duan2022context, regret-former:ivanov2022optimal}, VCG-based approaches (\eg, Lottery-AMA \citep{lottery-AMA:curry2023differentiable}) and menu-based approaches (\eg, MenuNet \citep{MenuNet:shen2019automated}, GemNet \citep{GemNet:wang2024gemnet}). 

This paper falls within the realm of differentiable economics. Although our setting differs from conventional auctions, the approaches mentioned above for conventional auctions are representative, and we adapt them to our setting to serve as baselines.

\paragraph{Characterization of Optimal Mechanisms.}
VCG mechanism is proposed by \citet{VCG-V:vickrey1961counterspeculation}, \citet{VCG-C:clarke1971multipart} and \citet{VCG-G:groves1973incentives} for mechanism design in the quasi-linear contexts. For social welfare maximizing designer, VCG mechanism is always optimal. 
In many cases mechanism designer has a different objective (\eg, revenue). 
For such cases, 
there has no unified solution between Bayesian settings (mechanism designer holds a prior over type profiles) and prior-free settings.

For Bayesian setting, the cornerstone result of
\citet{Myerson:myerson1981optimal} characterizes the optimal mechanism when a revenue-maximizing mechanism designer sells a single item to multiple buyers with independent valuation. They show that truthfulness can be transformed into monotonicity condition, and under truthfulness condition, revenue equals to virtual welfare. Later it has been shown that Myerson's results also applies to single-dimensional settings, as long as players’ have independent valuations.
Towards multi-dimensional revenue-maximizing settings, \citet{optimal_auction-special_case:manelli2006bundling} studies the optimal mechanism of selling $m=2,3$ goods to single buyer, and \citet{optimal_auction-special_case:giannakopoulos2014duality} extends this results up to $m=6$ goods.
These results rely on distribution assumptions such as the $\iid, U[0,1]$ valuations.
To our knowledge, there lacks a general theoretical solution for the general problem of multi-dimensional mechanism design and non-social-welfare-maximizing designer.

For prior-free setting, \citet{digital-good-auction:goldberg2001competitive} initializes the research on competitive ratio of truthful digital goods auctions in prior-free and identical-items settings. 
\citet{digital-good-auction:goldberg2004lower} later finds a competitive ratio lower bound $\lambda_n$ for $n$-player digital goods auctions with $\lim_{n\to\infty} \lambda_n \approx 2.42$. 
Based on prior works, \citet{digital-good-auction:chen2014optimal} shows that the bounds proposed in \citet{digital-good-auction:goldberg2004lower} are indeed tights. 
The results of \citet{digital-good-auction:chen2014optimal} are finished by proving the existence of a truthful mechanism that meets the competitive ratio requirement, yet do not provide a computational approach to find such a mechanism in polynomial time.
Towards prior-free truthful auctions with heterogeneous players and single-item, \citet{auction-no-prior:guruganesh2024prior} shows that prior-free auctions exhibit a significant gap with Myerson's revenue and can never perform consistently better than second-price auctions.

\paragraph{Characterization of Truthful Mechanisms.}
Comparing our equivalence results between truthful direct mechanisms and menu mechanisms in \cref{thm:direct-menu-equivalence}, 
\citet{menu-IC:hammond1979straightforward} showed that menu-based mechanism is a sufficient condition for IC, and IC mechanism is in some sense a menu mechanism.
\citet{utility-convex:rochet1987necessary} showed that with a truthful mechanism, the utilities of players are convex on their individual types.
Comparing with these results, our characterization in \cref{thm:direct-menu-equivalence} is more general than \citet{menu-IC:hammond1979straightforward} and has different perspective with \citet{utility-convex:rochet1987necessary}.

There are also plenty of works characterizing the relation between truthful mechanisms and VCG-based mechanisms. \citet{Robert:roberts1979characterization} shows that if the valuation spaces are full domain and there are at least $3$ possible outcomes, then any truthful mechanism must be an AMA. 
A simplified proof is later provided by \citet{Robert:lavi2009two}.
In a more general setting of combinatorial auctions, \citet{almost_AMA:lavi2003towards} proves that any truthful mechanism should be ``almost'' AMA.
\citet{AMA-continuous-domain:nath2015affine} extends the AMA characterization to the continuous domain, in which valuation spaces are whole continuous functions.
All these works focus on the case of deterministic mechanism.
However, there is no full characterization about the class of truthful randomized mechanisms in these settings.

Beyond quasi-linearity contexts, the Gibbard-Satterthwaite Theorem \citep{GS-G:gibbard1973manipulation,GS-S:satterthwaite1975strategy} shows that an incentive-compatible social choice function must be a dictatorship with more than $3$ candidates. Later \citet{median-rule:moulin1980strategy,median-rule:ching1997strategy} extends to the setting with single-peak preference and shows that only generalized median-rule is incentive-compatible among all social choice functions. The incentive-compatibility results have also been characterized for domains of house-allocation \citep{house-allocation:shapley1974cores} and stable matching \citep{stable-marriage:gale1962college}.

\paragraph{Representation of Convex Functions.}
Max-of-Affine (MoA) functions and Log-sum-exp (LSE) functions are well-known convex functions by design. 
\citet{LSE-universal_approximator:calafiore2019log} demonstrates that both the maximum-of-affine and log-sum-exp functions are universal approximators for the class of convex functions under the $L_\infty$ norm. 
Later, \citet{PMA-universal_approximator:kim2022parameterized} further shows that parameterized maximum-of-affine (PMA) and parameterized log-sum-exp (PLSE) are also universal approximators for the class of continuous functions that exhibit convexity over partial inputs. 
However, MoA and LSE suffer from curse-of-dimensionality since it requires an exponential number ($O(K^{d/2})$) of affine functions to represent an $d$-dimension, $K$-point polytope in the worst case \citep{CoD-PMA:polytope:mcmullen1970maximum}, while the polytope exhibits a natural, polynomial description.

\citet{PICNN:amos2017input} proposes Partial Input Convex Neural Network (PICNN) to represent partial convex functions with networks. However, until now, there has been no established result about whether PICNN is a universal approximator of the partial convex function class.
On another direction, \citet{GroupMax:warin2023groupmax} proposes the GroupMax Network and proves that GroupMax can represent the maximum-of-affine function, establishing the universal approximation property of GroupMax towards the class of convex functions. \citet{GroupMax:warin2023groupmax} also proposes another GroupMax architecture (without adequately naming it) to express partial convex functions, yet still unable to show whether such network is a universal approximator.

\Network draws ideas from results above while making improvements, being the first proven network-based universal approximator of partial convex functions. 
We also tried to incorporate PICNN and the GroupMax Network (partial convex version) into our pricing rule representation, but we found the performances of these attempts are sub-optimal when there are multiple players. 
Such suboptimal results provide evidence that PICNN and GroupMax Network (partial convex version) are \textbf{probably not} universal approximators.



\subsection{Details of \network}
\label{app:omitted:network}


\subsubsection{Architecture of \network.}
\label{app:omitted:network:architecture}

The detailed architectures of \Network and \Pan are presented in \cref{alg:PAN-initialization,alg:PAN-computation,alg:network-initialization,alg:network-computation}, demonstrating the corresponding dimensions or properties of their parameters, as well as the computation flows on given inputs.

\begin{algorithm}[t]
\caption{Initialization of \network}
\label{alg:network-initialization}
\KwIn{$d_x, d_y$: input dimension;}
\textbf{hyper-parameters}: $k$: hidden layer; $h_x$: hidden dimension; $G$: the number of groups; $E$: the number of elements in each group\\
\textbf{constraints}: $h_x = G \cdot E$\\
\tcc{Initialize Parameterized Affine Networks}
$PAN^x_0 \gets Parameterized Affine Network(d_y = d_y, d^{in} = d_x, d^{out} = h_x, positive=false)$\;
\For{$i=1,...k-1$}
{
    $PAN^x_i \gets Parameterized Affine Network(d_y = d_y, d^{in} = G, d^{out} = h_x, positive=true)$\;
    $PAN^r_i \gets Parameterized Affine Network(d_y = d_y, d^{in} = d_x, d^{out} = h_x, positive=false)$\;
}
$PAN^x_k \gets Parameterized Affine Network(d_y = d_y, d^{in} = G, d^{out} = 1, positive=true)$\;
$PAN^r_k \gets Parameterized Affine Network(d_y = d_y, d^{in} = d_x, d^{out} = 1, positive=false)$\;
\end{algorithm}

\begin{algorithm}[t]
\caption{Computation of \network}
\label{alg:network-computation}
\KwIn{$x\in\bbR^{d_x}, y \in \bbR^{d_y}$}
\KwOut{$z\in \bbR$}
\textbf{parameters}: $PAN^x_i$ for $0\le i \le k$, $PAN^r_i$ for $1\le i\le k$\;
\textbf{constraints}: $W^x_i \ge 0$, for $i=1,...,k-1$, $w^{x} \ge 0$\\
\tcc{Generate all affine transformations by calling Parameterized Affine Network on input $y$}
$W^x_0, b_0 \gets PAN^x_i(y)$, $W^r_0 \gets 0$\;
\For{$1\le i\le k-1$}
{
    $W^x_i, b_i \gets PAN^x_i(y)$\;
    $W^r_i, * \gets PAN^r_i(y)$\;
}
$w^x, b \gets PAN^x_k(y)$\;
$w^r, * \gets PAN^r_k(y)$\;
\tcc{The computation flow of \network.}
$x_0 \gets x$\;
\For{$i=1,2,...,k$}
{
    $h^x_{i} = W^x_{i-1} x_{i-1} + W^{r}_{i-1} x_0 + b_{i-1}$\\
    $x_i = GroupMax(h^x_{i})$\\
}
$z = \langle w^{x}, x_k\rangle + \langle w^{r}, x_0\rangle + b$\\
\textbf{return}: $z$
\end{algorithm}

\begin{algorithm}[t]
\caption{Initialization of Parameterized Affine Network}
\label{alg:PAN-initialization}
\KwIn{$d_y\in \bbZ_+$: the dimension of input $y$; $d^{in}, d^{out}\in \bbZ_+$ the dimension of affine transformation output; $positive\in \{true, false\}$: whether the affine transformation is hard-coded positive}
\textbf{hyper-parameters}: $k$: the hidden layer; $h_y$: hidden dimension\\
\tcc{Initialize the network parameters}
Initialize  $W^y_0 \gets LinearLayer(d_y, h_y)$\;
\For{$i=1,,,k-1$}
{
    Initialize $W^y_i \gets LinearLayer(h_y, h_y)$\;
}
Initialize $H_W \gets LinearLayer(h_y, d^{in}\cdot d^{out}), H_b \gets Linear Layer(h_y, d^{out})$\;
\end{algorithm}

\begin{algorithm}[t]
\caption{Computation of Parameterized Affine Network}
\label{alg:PAN-computation}
\KwIn{$y\in \bbR^{d_y}$}
\KwOut{$W\in \bbR^{d^{out}\times d^{in}}, b\in \bbR^{d^{out}}$}
\textbf{parameters}: $W^y_i$ for $0\le i\le k-1$, $H_W, H_b$, $positive \in \{true, false\}$\\
\textbf{constraints}: $W \ge 0$, if $positive = true$\\
$y_0 \gets y$\;
\For{$i=1,2,...,k$}
{
    $h^y_i \gets W^y_{i-1} (y_{i-1})$\;
    $y_i \gets \sigma(h^y_i)$\qquad \text{element-wise activation function}\; 
}
$W \gets H_W (y_k)$\;
Transform $W$ to a matrix with dimension $d^{out}\times d^{in}$\;
$b \gets H_b (y_k) $\;
\If{$positive = true$}
{
    $W \gets \softplus(W)$\;
}
\textbf{return}: $(W, b)$
\end{algorithm}

\begin{algorithm}[t]
\caption{Inference of \Method}
\label{alg:infer}
\KwIn{$\bmt\in\calT$: type profile}
\KwOut{$\bmx \in \calX$: outcome, $\bmp\in\bbR^n$: money transfer}
\textbf{parameter}: parameterized menu mechanism $\theta$\\
Initialize $\bmx$\;
\Repeat{convergence}
{
\For{i=1,...,n}
{
    $p_i \leftarrow p^m_i(x_i;\bmt_{-i},\theta)$\;
    $u_i \leftarrow \langle x_i,t_i \rangle + c_i(x_i) - p_i$\;
    \tcc{Call .backward() on $u_i$ to achieve the gradient}
    Compute $\nabla_{x_i} u_i$\; 
    Optimize $x_i$ with $\nabla_{x_i} u_i$ while hard-coding $x_i \in \calX_i$\;
}
}
Let $\bmp = (p_1,...,p_n)$\\
\textbf{return}: $(\bmx, \bmp)$\\
\end{algorithm}

\subsubsection{Universal Approximation Property.}
\label{app:omitted:network:universal}

We formally states the universal approximation property of \Network in this section.
To begin with, we give the formal definitions of universal approximators.


\begin{definition}[Universal Approximators]
\label{def:universal}
Let $\calF_1, \calF_2$ be two class of functions, if following holds,
\begin{itemize}
\item $\calF_1 \subseteq \calF_2$
\item $\forall f_2 \in \calF_2$, $\forall \varepsilon > 0$, there is $f_1 \in \calF_1$ such that $l_\infty(f_1,f_2) < \varepsilon$.
\end{itemize}

Then we say $\calF_1$ is a universal approximator of $\calF_2$.
\end{definition}


\cref{thm:network-universal} below shows the formal statement of universal approximation property of \network. \cref{thm:network-universal} is a formal version of \cref{prop:network-universal-informal}.


\begin{restatable}{theorem}{thmNetworkUniversal}
\label{thm:network-universal}
Fix $d_x, d_y \in \bbZ_+$, $B > 0$ and $X\subseteq \bbR^{d_x}$ is a compact, convex subset of $\bbR^{d_x}$ and $Y\subseteq \bbR^{d_y}$ is a compact set of $\bbR^{d_y}$.
Define $\calF = \{f: X \times Y \to \bbR |\text{$f(x,y)$ is continuous on $(x,y)$, convex on $x$}\}$ and
$\calG = \{g: X \times Y \to \bbR | g \text{ can be expressed by a Partial GroupMax Network} \}$. 
Then $\calG$ is a universal approximator of $\calF$.
\end{restatable}

\subsection{Inference Procedure of \method}
\label{app:omitted:inference}

The inference procedure of \Method is provided in \cref{alg:infer}, showing how \Method can output the outcome $\bmx$ and monetary transfer $\bmp$ given any type profile $\bmt$ efficiently.

\subsection{Rigorous Justifications of Full-Expressiveness}
\label{app:omitted:justification}

In this section, we provide the rigorous justification of \method's full-expressiveness. We begin with defining the distance between menu mechanisms in \cref{def:dist-menu}.


\begin{definition}[Distance between menu mechanisms]
\label{def:dist-menu}
Let $M^{m,1} = \{p^{m,1}_i\}_\iinn$ and $M^{m,2} = \{p^{m,2}_i\}_\iinn$ be two menu mechanisms. We define the distance between $M^{m,1}$ and $M^{m,2}$ as follows,
\begin{equation}
\label{eq:dist-menu}
\dist(M^{m,1}, M^{m,2}) = \max_\iinn\ l_\infty (p^{m,1}_i, p^{m,2}_i)
\end{equation}

We also introduce the $l_\infty$-norm as follows,
\begin{align*}
    l_\infty(p^{m,1}_i, p^{m,2}_i) = \max_{x_i \in \calX_i, \bmt_{-i} \in \calT_{-i}} | p^{m,1}_i(x_i,\bmt_{-i}) - p^{m,2}_i(x_i,\bmt_{-i}) |
\end{align*}

\end{definition}

\cref{def:dist-menu} is intuitive: the distance between menu mechanisms is exactly the maximum $l_\infty$-norm between pricing rules.

We also require a technical assumption on the mechanism design model, which is fairly weak.


\begin{assumption}
\label{asp:designer-utility}
The mechanism designer's utility $u_0(\bmx, \bmp; \bmt)$ is non-decreasing on $p_i$.    
\end{assumption}

\cref{asp:designer-utility} assumes that the mechanism designer does not hate more money. The technical intuition of this assumption lies in that there is an optimal mechanism that drains players' utilities to $0$ sometimes; otherwise, increasing the monetary transfer by a minimal constant would neither affect the properties of truthfulness and IR or decrease the mechanism designer's utility, thus is also optimal.

We also note that \cref{asp:designer-utility} can be moved if we impose an additional term on the pricing rule representations as follows,
\begin{align*}
    p^m_i(x_i;\bmt_{-i};\theta) = c_i(x_i) + f_i(x_i;\bmt_{-i};\theta) - f_i(\zeros;\bmt_{-i};\theta) + g_i(\bmt_{-i};\theta)
\end{align*}
where $f_i(x_i;\bmt_{-i};\theta)$ is an instance of \Network and $g_i(\bmt_{-i};\theta)$ is an instance of any universal neural networks with hard-coded constraint $g_i(\bmt_{-i};\theta) \le 0$.

Next, we define the \emph{optimality} \Wrt \emph{menu mechanism}. 
Given menu mechanism $M^m = \{p^m_i\}_\iinn$, we define the expected utility of the designer by committing menu mechanism $M^m$ as the maximum value by committing an equivalent direct mechanism $M^d$, with the definition of expected utilities of direct mechanisms in \cref{eq:model-6}. 
Formally,
\begin{align*}
\EU(M^m) =& \sup_{M^d \in \calM^d}\ \EU(M^d) \\
\st\quad& \text{$M^d$ is equivalent with $M^m$}
\end{align*}

The supreme expected utility of the mechanism design problem $\MD$, $\SEU$, is defined by the supreme value of $\EU(M^d)$ on all truthful direct mechanisms $M^d$.
Or equivalently, the supreme value of $\EU(M^m)$ on all menu mechanisms with \Ppc and \Nbnp by \cref{thm:direct-menu-equivalence}. 
Recall that $\calM^{m,pn}$ and $\calM^{d,t}$ represent the class of menu mechanisms that satisfy \Ppc and \nbnp, and the class of truthful direct mechanisms, respectively. Then, by \cref{thm:direct-menu-equivalence} we have
\begin{align*}
    \SEU \coloneqq \sup_{\substack{M^d \in \calM^{d,t}}} \EU(M^d) = \sup_{M^m \in \calM^{m,pn}} \EU(M^m)
\end{align*}

The following lemma shows that focusing on the menu mechanisms with $p^m_i(0,\bmt_{-i}) \equiv 0$ is without loss of generality.


\begin{restatable}{lemma}{lemNbnpWlog}
\label{lem:nbnp-wlog}
Under \cref{asp:designer-utility},
for any $\varepsilon > 0$, there exists a menu mechanism $M^m$ satisfying \Ppc and the follow identity, called \nbnps,
\begin{align*}
    p^m_i(0,\bmt_{-i}) \equiv 0, \forall \iinn, \bmt_{-i} \in \calT_{-i}
\end{align*}
that satisfies $\EU(M^m) \ge \SEU - \varepsilon$.
\end{restatable}

We denote $\calM^{m,pn*} \subseteq \calM^{m,pn}$ as the menu mechanisms satisfying \Ppc and \nbnps. 
\cref{lem:nbnp-wlog} indicates that $\SEU = \sup_{M^m \in \calM^{m,pn*}} \EU(M^m)$. Therefore, from now on we care about the mechanism class $\calM^{m,pn*}$.
The expressive power of \Method can be shown in the following proposition. 


\begin{restatable}{proposition}{propMethodExpressive}
\label{prop:method-expressive}
For any menu mechanism $M^m \in \calM^{m,pn*}$ and any $\varepsilon > 0$, there is another menu mechanism $M^m_0$ that can be expressed by \method, such that $\dist(M^m_0, M^m) \le \varepsilon$.
\end{restatable}

\cref{prop:method-expressive} shows that \Method can approximate all menu mechanisms with \Ppc and \nbnps, including the optimal menu mechanism, within an arbitrarily small distance.

Until now, the full-expressiveness justification is finished from the menu mechanism perspective. 
Yet, there is another question that states as follows: Does the full-expressiveness also hold in the perspective of the mechanism designer's expected utility?
Intuitively, one may think that an arbitrarily small difference in menu mechanisms will not explode to a large difference in the corresponding utilities. 
Therefore, it's imaginable that \Method also ensures the same $\SEU$.
We will positively answer this question in the rest of this section, with a technical assumption of the non-degeneration of the distribution on type profiles.


\begin{definition}
\label{def:non-degenerate-distribution}
We call a distribution $\calF$ on $X$ is non-degenerate, if for any $\mu$-measurable subset $X_0 \subseteq X$, we have $\Pr_{x\sim \calF}[ x \in X_0] > 0$ indicates that $\mu(X_0) > 0$, where $\mu$ is Lebesgue measure.
\end{definition}


\begin{assumption}
\label{asp:type-non-degenerate}
The distribution $\calF$ on type profile $\calT$ is non-degenerate.    
\end{assumption}

We note that \cref{asp:type-non-degenerate} is fairly weak, because a distribution that preserves a probability density function is always non-degenerate. 

Next, we require some technical definitions and lemmas, that bridge the gap between menu mechanisms and the corresponding designer's utility.
We begin by introducing the distance between direct mechanisms.


\begin{definition}[Distance between direct mechanisms]
\label{def:dist-direct}
Let $M^{d,1} = (\bmx^{d,1}, \bmp^{d,1})$ and $M^{m,2} = (\bmx^{d,2}, \bmp^{d,2})$ be two direct mechanisms. We define the distance between $M^{d,1}$ and $M^{d,2}$ as follows,
\begin{equation}
\label{eq:dist-direct}
\dist(M^{d,1}, M^{d,2}) = l_1 (\bmx^{d,1}, \bmx^{d,2}) + l_1 (\bmp^{d,1}, \bmp^{d,2})
\end{equation}

We also introduce the $l_1$-norm as follows.
\begin{align*}
l_1(\bmx^{d,1}, \bmx^{d,2}) = \sum_\iinn \int_{\bmt \in \calT} \| x^{d,1}_i(\bmt) - x^{d,2}_i(\bmt) \|_1 \dd \mu(\bmt)
\\
l_1(\bmp^{d,1}, \bmp^{d,2}) = \sum_\iinn \int_{\bmt \in \calT} | p^{d,1}_i(\bmt) - p^{d,2}_i(\bmt) | \dd \mu(\bmt)
\end{align*}

\end{definition}

\cref{def:dist-direct} is intuitive: the distance between direct mechanisms is exactly the summation of $l_1$-norm between allocation rules and payment rules.

The following lemma shows that, the distance measure between direct mechanisms can be arbitrarily small, if the distance measure between their equivalent menu mechanisms is arbitrarily small.


\begin{restatable}{lemma}{lemDistancePreserve}
\label{lem:distance-preserve}
Let $M^{m,1}$ and $M^{m,2}$ be two menu mechanisms and $M^{d,i}$ is equivalent with $M^{m,i}$ for $i=1,2$. 
For any $\varepsilon$, there is $\delta$ relies only on $\varepsilon$ and problem instance $\MD$ (which actually says that $\delta$ is independent of $M^{m,i}$ and $M^{d,i}$) such that, 
if $\dist(M^{m,1},M^{m,2}) \le \delta$, then $\dist(M^{d,1}, M^{d,2}) \le \varepsilon$.
\end{restatable}

On the other direction, we also show that the difference between the mechanism designer's expected utilities \Wrt direct mechanisms can be arbitrarily small, if the distance measure between direct mechanisms is arbitrarily small.
Specifically,



\begin{restatable}{lemma}{lemDirectEU}
\label{lem:direct-EU}
Under \cref{asp:type-non-degenerate},
for any $\varepsilon>0$ and bound $B > 0$, there is $\delta>0$ that relies only on $\varepsilon, B$ and problem instance $\MD$ such that, if $M^{d,i} = (\bmx^{d,i}, \bmp^{d,i}), i=1,2$ holds that $\dist(M^{d,1}, M^{d,2})<\delta$, and $\bmp^{d,i}(\bmt)$ is bounded by $\pm B$, then $|\EU(M^{d,1}) - \EU(M^{d,2})| < \varepsilon$.
\end{restatable}

Together with \cref{lem:distance-preserve,lem:direct-EU}, we can derive the final result, demonstrating the full-expressiveness of \method.


\begin{restatable}{theorem}{thmTediExpressive}
\emph{(\Method is full-expressive)}
\label{thm:tedi-expressive}
Under \cref{asp:designer-utility} and \cref{asp:type-non-degenerate},
denote $\calM^{tedi}$ as the menu mechanism class that can be expressed by \method, then 
\begin{align*}
\SEU = \sup_{M^m \in \calM^{tedi}} \EU(M^m)
\end{align*}

In other words, for any problem instance $\MD$, \Method can always recover the optimal mechanism within arbitrarily small difference on the designer's welfare.

\end{restatable}

\subsection{Dimension-Insensitivity of \Method}
\label{app:omitted:expressive}

In \cref{sec:justification}, we show that \Method can express optimal mechanism with polynomial parameters. 
Recall that in \cref{eg:CoD}, the optimal mechanism has pricing rule $p(x) = \sum_\jinm x_j p_j$, where $p_j \equiv 1$ is the price of good $j$.
The pricing rule representation in \Method is $p^m(x) = c(x) + f(x) - f(0)$, where $f(x)$ is an instance of \network.



\begin{restatable}{example}{egExpressOptimalMechanism}
\emph{(Express Optimal Pricing Rule with \network)}
\label{eg:express-optimal-mechanism}
In this case, $c(x) \equiv 0$ and we guarantee $f(0) = 0$ by construction. Since there is only one player, and then $t_{-i}$ does not make any sense thus the \Pan has empty input and its output can be seen as constant. 
We let \Network has no hidden layer and only one final affine transform, \ie, let $\{w_j\}_\jinm \in \bbR^m_+, b \in \bbR$ be the parameters, then the network output is exactly $\sum_\jinm x_j w_j + b$ given input $x\in [0,1]^m$. By designating $w_j = p_j$ and $b=0$, the network exactly expresses the optimal mechanism. 
Note that this network has only $m+1 = O(m)$ parameters, far more efficient than the conventional menu mechanism that requires $(m+1)\cdot 2^m$ parameters.
\end{restatable}

\newpage
\section{Omitted Proofs}
\label{app:omitted-proofs}

\subsection{Derivation of \texorpdfstring{\cref{prop:cov-trick}}{}}

\propAlgoOne*

\begin{proof}
\label{prf:prop:algo-1}
We abbreviate $\nabla_\theta$ as $\nabla$ when context is clear.
To begin with, we have
\begin{equation}
\label{eq:algo-3}
\begin{aligned}
\bbE_{x_i \iidd q^{\beta_i}_i(y;\bmt,\theta)} \left[ u_0(\bmx,\theta;\bmt) \right] 
= \int_{\bmy \in \calX} u_0(\bmy,\theta;\bmt) \cdot \prod_\iinn q^{\beta_i}_i(y_i;\bmt,\theta) \cdot \prod_\iinn (\dd y_i)
\end{aligned}
\end{equation}

By taking differential operator $\nabla$ \wrt\ $\theta$, we achieve,
\begin{equation}
\label{eq:algo-4}
\begin{aligned}
& \nabla \bbE_{x_i \iidd q^{\beta_i}_i(y;\bmt,\theta)} \left[ u_0(\bmx,\theta;\bmt) \right] 
= \nabla \left[ \int_{\bmy \in \calX} u_0(\bmy,\theta;\bmt) \cdot \prod_\iinn q^{\beta_i}_i(y_i;\bmt,\theta) \cdot \prod_\iinn (\dd y_i) \right]
\\
=& \int_{\bmy \in \calX} \nabla u_0(\bmy,\theta;\bmt) \cdot \prod_\iinn q^{\beta_i}_i(y_i;\bmt,\theta) \cdot \prod_\iinn (\dd y_i)
\\
+& \sum_\iinn \int_{\bmy \in \calX} u_0(\bmy,\theta;\bmt) \cdot \prod_\jnei q^{\beta_j}_j(y_j;\bmt,\theta) \cdot \nabla q^{\beta_i}_i(y_i;\bmt,\theta) \cdot \prod_\iinn (\dd y_i)
\\
=& H + \sum_\iinn I_i
\end{aligned}
\end{equation}
where 
\begin{align*}
    H \coloneqq& \int_{\bmy \in \calX} \nabla u_0(\bmy,\theta;\bmt) \cdot \prod_\iinn q^{\beta_i}_i(y_i;\bmt,\theta) \cdot \prod_\iinn (\dd y_i)
    \\
    I_i \coloneqq& \int_{\bmy \in \calX} u_0(\bmy,\theta;\bmt) \cdot \prod_\jnei q^{\beta_j}_j(y_j;\bmt,\theta) \cdot \nabla q^{\beta_i}_i(y_i;\bmt,\theta) \cdot \prod_\iinn (\dd y_i)
\end{align*}

We focus on analyzing the unbiased estimator of $H$ and $I_i$, respectively. We begin with analyzing $H$ since this part is straight-forward. However, analyzing $I_i$ is much more tricky.

\paragraph{Unbiased estimator of $H$.} 

Observe that 
\begin{equation}
\label{eq:algo-5}
\begin{aligned}
H = \bbE_{y_i \iidd q^{\beta_i}_i(y;\bmt,\theta)} \left[ \nabla u_0(\bmy,\theta;\bmt) \right]
\end{aligned}
\end{equation}

Therefore, as long as we can sample $y_i \iidd q^{\beta_i}_i (y_i; \bmt, \theta)$, then $\nabla u_0(\bmy,\theta;\bmt)$ forms an unbiased estimator. 
$\nabla u_0(\bmy,\theta;\bmt)$ can be computed by directly computing $u_0(\bmy,\theta;\bmt)$, followed by auto-grad module in PyTorch.

\paragraph{Unbiased estimator of $I_i$.}

We first analyze the trickiest term $\nabla q^{\beta_i}_i (y_i;\bmt,\theta)$ in $I_i$.

\begin{equation}
\label{eq:algo-6}
\begin{aligned}
& \nabla q^{\beta_i}_i (y_i;\bmt,\theta) 
= \nabla \left[ \frac{\exp\left( {\beta_i} \cdot u_i(y_i;\bmt,\theta) \right)}{Z(\bmt,\theta)} \right]
\\
=& \frac{1}{Z^2(\bmt,\theta)} \left[ \nabla \exp\left( {\beta_i} \cdot u_i(y_i;\bmt,\theta) \right) \cdot Z(\bmt,\theta) 
- \exp\left( {\beta_i} \cdot u_i(y_i;\bmt,\theta) \right) \cdot \nabla Z(\bmt,\theta) \right]
\\
=& J^1_i - J^2_i
\end{aligned}
\end{equation}
where
\begin{equation}
\label{eq:algo-7}
\begin{aligned}
    J^1_i \coloneqq& \frac{\nabla \exp\left( {\beta_i} \cdot u_i(y_i;\bmt,\theta) \right)}{Z(\bmt,\theta)}
    \\
    J^2_i \coloneqq& \frac{\exp\left( {\beta_i} \cdot u_i(y_i;\bmt,\theta) \right) \cdot \nabla Z(\bmt,\theta)}{Z^2(\bmt,\theta)}
\end{aligned}
\end{equation}

We next analyze $J^1_i$ and $J^2_i$ respectively.

\textbf{Analyze of $J^1_i$:}

Notice that $\nabla \exp\left( {\beta_i} \cdot u_i(y_i;\bmt,\theta) \right) = {\beta_i} \cdot \nabla u_i(y_i;\bmt,\theta) \cdot \exp\left( {\beta_i} \cdot u_i(y_i;\bmt,\theta) \right)$ and $\frac{\exp\left( {\beta_i} \cdot u_i(y_i;\bmt,\theta) \right)}{Z(\bmt,\theta)} = q^{\beta_i}_i(y_i;\bmt,\theta)$. Therefore, we have

\begin{equation}
\label{eq:algo-8}
J^1_i = {\beta_i} \cdot \nabla u_i(y_i;\bmt,\theta) \cdot q^{\beta_i}_i(y_i;\bmt,\theta)
\end{equation}

\textbf{Analyze of $J^2_i$:}

Observe that
\begin{equation}
\label{eq:algo-9}
\begin{aligned}
J^2_i =& q^{\beta_i}_i(y_i;\bmt,\theta) \cdot \frac{\nabla Z_i(\bmt,\theta)}{Z_i(\bmt,\theta)}
\\
\nabla Z_i(\bmt,\theta) =& \nabla \left[ \int_{z_i \in X_i} \exp({\beta_i} \cdot u_i(z_i;\bmt,\theta)) \dd z_i \right]
\\
=& \int_{z_i \in X_i} {\beta_i} \cdot \nabla u_i(z_i;\bmt,\theta) \cdot \exp({\beta_i} \cdot u_i(z_i;\bmt,\theta)) \dd z_i
\\
=& \int_{z_i \in X_i} {\beta_i} \cdot \nabla u_i(z_i;\bmt,\theta) \cdot q^{\beta_i}_i(z_i;\bmt,\theta) \cdot Z_i(\bmt,\theta) \dd z_i
\\
=& Z_i(\bmt,\theta) \cdot \int_{z_i \in X_i} {\beta_i} \cdot \nabla u_i(z_i;\bmt,\theta) \cdot q^{\beta_i}_i(z_i;\bmt,\theta) \dd z_i
\end{aligned}
\end{equation}

Therefore, we have
\begin{equation}
\label{eq:algo-10}
\begin{aligned}
J^2_i =& q^{\beta_i}_i(y_i;\bmt,\theta) \cdot \int_{z_i \in X_i} {\beta_i} \cdot \nabla u_i(z_i;\bmt,\theta) \cdot q^{\beta_i}_i(z_i;\bmt,\theta) \dd z_i
\end{aligned}
\end{equation}

Combining \cref{eq:algo-8} and \cref{eq:algo-10} and notice that $\int_{z_i \in X_i} q^{\beta_i}_i(z_i;\bmt,\theta) \dd z_i = 1$, we achieve that,
\begin{equation}
\label{eq:algo-11}
\begin{aligned}
\nabla q^{\beta_i}_i (y_i;\bmt,\theta) = J^1_i - J^2_i = q^{\beta_i}_i(y_i;\bmt,\theta) \cdot \int_{z_i\in X_i} {\beta_i} \cdot \left[ \nabla u_i(y_i;\bmt,\theta) - \nabla u_i(z_i;\bmt,\theta) \right] q^{\beta_i}_i(z_i;\bmt,\theta) \dd z_i
\end{aligned}
\end{equation}

Taking \cref{eq:algo-11} into the expression of $I_i$, we have
\begin{equation}
\label{eq:algo-12}
\begin{aligned}
I_i =& \int_{\bmy \in \calX} u_0(\bmy,\theta;\bmt) \cdot \prod_\jnei q^{\beta_j}_j(y_j;\bmt,\theta) \cdot \nabla q^{\beta_i}_i(y_i;\bmt,\theta) \cdot \prod_\iinn (\dd y_i)
\\
=& \int_{\bmy \in \calX} u_0(\bmy,\theta;\bmt) \cdot \prod_\jinn q^{\beta_j}_j(y_j;\bmt,\theta) \cdot \int_{z_i\in \calX_i} {\beta_i} \cdot \left[ \nabla u_i(y_i;\bmt,\theta) - \nabla u_i(z_i;\bmt,\theta) \right] q^{\beta_i}_i(z_i;\bmt,\theta) \dd z_i \cdot \prod_\iinn (\dd y_i)
\\
=& \bbE_{\{y_i,z_i\}_\iinn} \left[ {\beta_i} \cdot u_0(\bmy,\theta;\bmt) \cdot \nabla \left( u_i(y_i;\bmt,\theta) - u_i(z_i;\bmt,\theta) \right) \right]
\end{aligned}
\end{equation}
where on the expectation on $\{y_i,z_i\}_\iinn$, we omit $\{y_i, z_i\}_\iinn \iidd \{ q^{\beta_i}_i(y_i;\bmt,\theta), q^{\beta_i}_i(z_i;\bmt,\theta) \}_\iinn$ for space limitation, when the context is clear.

Since \cref{eq:algo-12} may has a large variance (because $u_0$ might be large), we further utilize a variance reduction technique. Notice that
\begin{equation}
\label{eq:algo-12-1}
\bbE_{\{y_i, z_i\}_\iinn}\left[ \frac{{\beta_i}}{2} 
\cdot \left( u_0(\bmz, \theta; \bmt) + u_0(\bmy, \theta; \bmt)\right) 
\cdot \nabla \left( u_i(y_i;\bmt,\theta) - u_i(z_i;\bmt,\theta) \right) \right] 
= 0
\end{equation}

This equation holds because LHS is anti-symmetric if we exchange $\bmy$ and $\bmz$. Combining \cref{eq:algo-12} and \cref{eq:algo-12-1}, we have
\begin{equation}
\label{eq:algo-12-2}
I_i = \bbE_{\{y_i,z_i\}_\iinn} \left[ \frac{{\beta_i}}{2} \cdot \left( u_0(\bmy,\theta;\bmt) - u_0(\bmz,\theta;\bmt) \right) \cdot \nabla \left( u_i(y_i;\bmt,\theta) - u_i(z_i;\bmt,\theta) \right) \right]
\end{equation}

Taking \cref{eq:algo-5} and \cref{eq:algo-12-2} into \cref{eq:algo-4}, we obtain that,
\begin{equation}
\label{eq:algo-13}
\begin{aligned}
& \nabla \bbE_{x_i \iidd q^{\beta_i}_i(y;\bmt,\theta)} \left[ u_0(\bmx,\theta;\bmt) \right] = H + \sum_\iinn I_i
\\
=& \frac{1}{2} \bbE_{\{y_i,z_i\}_\iinn} \left[ \nabla \left( u_0(\bmy,\theta;\bmt) + u_0(\bmz,\theta;\bmt) \right)
\right.
\\
+& \sum_\iinn \left.
{\beta_i} \cdot \left( u_0(\bmy,\theta;\bmt) - u_0(\bmz,\theta;\bmt) \right) \cdot \nabla \left( u_i(y_i;\bmt,\theta) - u_i(z_i;\bmt,\theta) \right) \right]
\\
=& \frac{1}{2} \bbE_{\{y_i,z_i\}_\iinn} \left[ \nabla \left( u_0(\bmy,\theta;\bmt) + u_0(\bmz,\theta;\bmt) \right)
\right.
\\
+& \left.
\left( u_0(\bmy,\theta;\bmt) - u_0(\bmz,\theta;\bmt) \right) \cdot \nabla \left( \ASW(\bmy;\bmt,\theta) - \ASW(\bmz;\bmt,\theta) \right) \right]
\end{aligned}
\end{equation}

\cref{eq:algo-13} is equivalent to the form in \cref{prop:cov-trick}.

\end{proof}

\subsection{Proof of \texorpdfstring{\cref{thm:direct-menu-equivalence}}{}}

\thmDirectMenuEquivalence*

\begin{proof}
\label{prf:thm:direct-menu-equivalence}
We prove \cref{thm:direct-menu-equivalence} by two separate sub-theorems.
\begin{theorem}[Menu mechanism leads to truthfulness]
\label{thm:menu-to-direct}
Let $M^m$ be a menu mechanism satisfying \nbnp, and $M^d$ is a direct mechanism that is equivalent to $M^m$, then $M^d$ is truthful.
\end{theorem}

\begin{theorem}[Truthful direct mechanism is in some sense a menu mechanism]
\label{thm:direct-to-menu}
Let $M^d$ be a truthful direct mechanism, then there is a menu mechanism $M^m$ satisfying \Nbnp and \Ppc such that $M^d$ is equivalent to $M^m$.
\end{theorem}

\begin{proof}[Proof of \cref{thm:menu-to-direct}]
\label{prf:thm:menu-to-direct}

Let $M^m = \{p^m_i(x_i;\bmt_{-i}\}_\iinn$ be a menu mechanism satisfies \nbnp, and $M^d = (\bmx^d, \bmp^d)$ is a equivalent direct mechanism of $M^m$.

We begin with showing $M^d$ satisfies IR.
Fix player $i$, player $i$'s type $t_i$ and other players' types $\bmt_{-i}$, mechanism $M^d$ should satisfy:
\begin{align*}
    x^d_i(\bmt) \in \argmax_{x_i \in \calX_i} u_i(x_i,p^m_i(x_i,\bmt_{-i});t_i)
\end{align*}

Since $0 \in \calX_i$, we should have 
\begin{align*}
0 \le& - p^m_i(0,\bmt_{-i}) = u_i(0,p^m_i(0,\bmt_{-i});t_i) 
\\
\le& u_i(x^d_i(\bmt),p^m_i(x^d_i(\bmt),\bmt_{-i});t_i) = u_i(x^d_i(\bmt),p^d_i(\bmt);t_i)
\end{align*}
The first inequality arises from \nbnp, the second equality arises from the definition of players' utility, the third inequality arises from the property of $\argmax$, the last inequality arises from the definition of equivalent direct mechanism.
This derivation means that $M^d$ satisfies IR.

Next we show that $M^d$ satisfies IC. Fix $\iinn, t_i, t'_i, \bmt_{-i}$, meaning that a $t_i$-type player $i$ wants to deviate to $t'_i$, then it should satisfy:
\begin{align*}
    x^d_i(t'_i,\bmt_{-i}) \in \argmax_{x_i \in \calX_i} u_i(x_i,p^m_i(x_i,\bmt_{-i});t'_i)
\end{align*}

We have
\begin{align*}
& u_i(x^d_i(\bmt),p^d_i(\bmt);t_i) 
= u_i(x^d_i(\bmt),p^m_i(x^d_i(\bmt),\bmt_{-i});t_i)
\\
\ge& u_i(x^d_i(t'_i,\bmt_{-i}),p^m_i(x^d_i(t'_i,\bmt_{-i}),\bmt_{-i});t_i)
= u_i(x^d_i(t'_i,\bmt_{-i}),p^d_i(t'_i,\bmt_{-i});t_i)
\end{align*}
The first and third equality arises from definition, and the second inequality arises from the $\argmax$ property and $x^d_i(t'_i, \bmt_{-i}) \in \calX_i$. This means that $M^d$ satisfies IC.

\end{proof}

\begin{proof}[Proof of \cref{thm:direct-to-menu}]
\label{prf:thm:direct-to-menu}
Let $M^d = (\bmx^d, \bmp^d)$ be a truthful direct mechanism. We will construct pricing rules $p^m_i: \calX_i \times \calT_{-i} \to \bbR$ such that, $M^d$ is equivalent to the menu mechanism $M^m = \{p^m_i\}_\iinn$, and show that $M^m$ satisfies \Ppc and \nbnp.

We first shed light on how to construct $p^m_i$ without much rigorous. 
We introduce some notations here. Denote $\Tilde{u}^d_i(\bmt)$ as the utility of player $i$ under direct mechanism $M^d$, when the type profile is $\bmt$ and the players are honest. 
Denote $u^d_i(t'_i;\bmt) $ as the utility of player $i$ under mechanism $M^d$, when the reported type profile is $\bmt$ and player $i$'s true type is $t'_i$. We have the expression
\begin{align*}
u^d_i(t'_i; \bmt) = u_i(x^d_i(\bmt), p^d_i(\bmt), t'_i)
\\
\Tilde{u}^d_i(\bmt) = u^d_i(t_i;\bmt)
\end{align*}

If player $i$ wants to deviate to $t'_i$, truthfulness requires that
\begin{equation}
\label{eq:direct-to-menu:1}
\begin{aligned}
\langle t_i, x_i^d(t'_i;\bmt_{-i}) \rangle + c_i(x_i^d(t'_i;\bmt_{-i})) - p_i^d(t'_i;\bmt_{-i}) 
\le& \langle t_i, x_i^d(t_i;\bmt_{-i}) \rangle + c_i(x_i^d(t_i;\bmt_{-i})) - p_i^d(t_i;\bmt_{-i}) 
\\
=& \Tilde{u}^d_i(\bmt)
\end{aligned}
\end{equation}

We temporarily assume that player $i$ can manipulate $x^d_i$ and $p^d_i$ by misreporting $t'_i$. Then we introduce free variables $x_i$ and $p_i$ replacing $x_i^d(t'_i;\bmt_{-i})$ and $p_i^d(t'_i;\bmt_{-i})$ in \cref{eq:direct-to-menu:1}. Then it becomes
\begin{equation}
\label{eq:direct-to-menu:2}
\begin{aligned}
\langle t_i, x_i \rangle + c_i(x_i) - p_i 
\le& \Tilde{u}^d_i(\bmt)
\end{aligned}
\end{equation}

\cref{eq:direct-to-menu:2} is equivalent to
\begin{equation}
\label{eq:direct-to-menu:3}
\begin{aligned}
p_i \ge& - \tilde{u}^d_i(\bmt) + c_i(x_i) + \langle t_i, x_i\rangle
\end{aligned}
\end{equation}

An observation is that, if we take $p^m_i(x_i;\bmt_{-i})$ into $p_i$ in \cref{eq:direct-to-menu:3}, then \cref{eq:direct-to-menu:3} should always hold, otherwise it can not induce a truthful equivalent direct mechanism. Besides, the inequality should becomes equality at some point, because the inequality in \cref{eq:direct-to-menu:1} becomes equality when $t_i = t'_i$. Therefore, by this insight, the only possible choice of $p^m_i(x_i;\bmt_{-i})$ should have the expression,
\begin{equation}
\label{eq:direct-to-menu:4}
\begin{aligned}
p^m_i(x_i;\bmt_{-i}) =& \sup_{t_i \in \calT_i}  \left[ - \tilde{u}^d_i(\bmt) + c_i(x_i) + \langle t_i, x_i\rangle \right]
\\
=& c_i(x_i) + \sup_{t_i \in \calT_i} \left[ \langle t_i, x_i\rangle - \tilde{u}^d_i(\bmt) \right]
\end{aligned}
\end{equation}

Next we focus on rigorous analysis on $p^m_i(x_i;\bmt_{-i})$. We will show that the menu mechanism $M^m = \{p^m_i\}_\iinn$ with \cref{eq:direct-to-menu:4} indeed satisfies our requirement. We concludes our initial requirements to following points and prove these points separately.
\begin{enumerate}
\item $M^m$ satisfies \ppc.
\item $M^m$ satisfies \nbnp.
\item $x^d_i(\bmt)\in\argmax_{x_i\in \calX_i} v_i(x_i;t_i) - p^m_i(x_i;\bmt_{-i}) $.
\item $p^d_i(\bmt) = p^m_i(x^d_i(\bmt);\bmt_{-i})$.
\end{enumerate}

\begin{proof}[Proof of (1)]
Note that $p^m_i(x_i;\bmt_{-i}) - c_i(x_i) = \sup_{t_i \in \calT_i} \langle t_i, x_i\rangle - \tilde{u}^d_i(\bmt)$ is exactly the Fenchel conjugate of $-\tilde{u}^d_i(\bmt)$ \wrt\ $t_i$. Therefore, it's convex by nature of Fenchel conjugate \citep{convex:boyd2004convex}. This completes the \ppc.
\end{proof}

\begin{proof}[Proof of (2)]
Note that $p^m_i(0;\bmt_{-i}) = c_i(0) + \sup_{t_i \in \calT_i} [- \tilde{u}^d_i(\bmt)] = - \inf_{t_i \in \calT_i} \tilde{u}^d_i(\bmt)$. By IR of $M^d$, we know that $\tilde{u}^d_i(\bmt)\ge 0$ for all $\bmt$. Then $- \inf_{t_i \in \calT_i} \tilde{u}^d_i(\bmt) \le 0$. This completes the \nbnp.
\end{proof}

\begin{proof}[Proof of (4)]
By definition,
\begin{equation}
\label{eq:direct-to-menu:5}
\begin{aligned}
    p^m_i(x^d_i(\bmt);\bmt_{-i}) =& \sup_{t'_i \in\calT_i} p^d_i(t'_i;\bmt_{-i}) + v_i(x^d_i(\bmt);t'_i) - v_i(x^d_i(t'_i;\bmt_{-i});t'_i)
    \\
    \ge& p^d_i(t_i;\bmt_{-i}),\qquad\text{by letting $t'_i = t_i$}
\end{aligned}
\end{equation}
This completes one direction of (4).

In order to prove the other direction, we first observe that, by IC of $M^d$, 
\begin{align*}
    \tilde{u}^d_i(\bmt) \ge& u_i^d(t_i;t'_i, \bmt_{-i})
    \\
    \Leftrightarrow v_i(x^d_i(\bmt);t_i) - p^d_i(\bmt) \ge& v_i(x^d_i(t'_i,\bmt_{-i});t_i) - p^d_i(t'_i,\bmt_{-i})
\end{align*}

By switching $t_i$ and $t'_i$ we get,
\begin{align*}
    v_i(x^d_i(t'_i,\bmt_{-i});t'_i) - p^d_i(t'_i,\bmt_{-i}) \ge& v_i(x^d_i(t_i,\bmt_{-i});t'_i) - p^d_i(t_i,\bmt_{-i})
    \\
    \Leftrightarrow v_i(x^d_i(t_i,\bmt_{-i});t'_i) - v_i(x^d_i(t'_i,\bmt_{-i});t'_i) \le& p^d_i(t_i,\bmt_{-i}) - p^d_i(t'_i,\bmt_{-i})
\end{align*}

Taking it into the first line of \cref{eq:direct-to-menu:5}, we derive,
\begin{equation}
\label{eq:direct-to-menu:6}
\begin{aligned}
    p^m_i(x^d_i(\bmt);\bmt_{-i}) =& \sup_{t'_i \in\calT_i} p^d_i(t'_i;\bmt_{-i}) 
    + v_i(x^d_i(\bmt);t'_i) 
    - v_i(x^d_i(t'_i;\bmt_{-i});t'_i)
    \\
    \le& \sup_{t'_i \in\calT_i} p^d_i(t'_i;\bmt_{-i})
    + p^d_i(t_i,\bmt_{-i})
    - p^d_i(t'_i,\bmt_{-i})
    \\
    =& p^d_i(t_i,\bmt_{-i})
\end{aligned}
\end{equation}

This completes the other direction of (4).
Together with \cref{eq:direct-to-menu:5} and \cref{eq:direct-to-menu:6}, we complete the proof of (4).
\end{proof}

\begin{proof}[Proof of (3)]
We have 
\begin{align*}
& v_i(x^d_i(\bmt);t_i) - p^m_i(x^d_i(\bmt);\bmt_{-i})
\\
=& v_i(x^d_i(\bmt);t_i) - p^d_i(\bmt)
= \tilde{u}^d_i(\bmt)
\end{align*}

We need to prove $v_i(x_i;t_i) - p^m_i(x_i;\bmt_{-i}) \le \tilde{u}^d_i(\bmt)$ for all $x_i \in \calX_i$.

Notice that
\begin{align*}
\mathrm{LHS} =& v_i(x_i;t_i) - c_i(x_i) - \left( \sup_{t'_i\in\calT_i} \tilde{u}^d_i(t'_i,\bmt_{-i}) + \langle t'_i,x_i\rangle \right)
\\
\le& v_i(x_i;t_i) - c_i(x_i) - \tilde{u}^d_i(\bmt) - \langle t_i, x_i \rangle, \qquad\text{let $t'_i = t_i$}
\\
=& \tilde{u}^d_i(\bmt)
\end{align*}

Hence we complete the proof of (3).
\end{proof}

\end{proof}

Combining \cref{thm:direct-to-menu} and \cref{thm:menu-to-direct}, we complete the proof of \cref{thm:direct-menu-equivalence}.

\end{proof}

\subsection{Proof of \texorpdfstring{\cref{cor:tedi-truthful}}{}}

\corTediTruthful*

\begin{proof}
\label{prf:cor:tedi-truthful}
Denote $M^m = \{p^m_i\}_\iinn$ be a menu mechanism in \method. By \cref{eq:pricingrule} we have that
\begin{align*}
    p^m_i (x_i;\bmt_{-i}) = c_i(x_i) + f_i(x_i;\bmt_{-i}) - f_i(0;\bmt_{-i})
\end{align*}
Easy to see that $\forall \iinn, \bmt_{-i}$, we have $p^m_i(0;\bmt_{-i}) = c_i(0) = 0$ and $p^m_i(x_i;\bmt_{-i}) - c_i(x_i) = f_i(x_i;\bmt_{-i}) - f_i(0;\bmt_{-i})$. Therefore, $M^m$ satisfies \Nbnp and \Ppc. By \cref{thm:direct-menu-equivalence}, we know that if $M^d$ is equivalent to $M^m$ then $M^d$ is truthful. 
\end{proof}

\subsection{Proof of \texorpdfstring{\cref{thm:network-universal}}{}}

\thmNetworkUniversal*

\begin{proof}[Proof of \cref{thm:network-universal}]
\label{prf:thm:network-universal}

We prove this theorem by two separate theorems.

\begin{theorem}[\Network is partial convex]
\label{thm:network-to-convex}
Let $g(x,y): X \times Y \to \bbR$ be a function expressed by \Network, then $g(x,y)$ is convex on $x$ and continuous on $(x,y)$.
\end{theorem}

\begin{theorem}[Partial convex functions can be approximated by \network]
\label{thm:convex-to-network}
Let $f(x,y): X \times Y \to \bbR$ be a function continuous on $(x,y)$ and convex on $x$, then, for any $\varepsilon > 0$ there is a $g(x,y) \in \calG$ such that $l_\infty (f,g) \le \varepsilon$.
\end{theorem}

\begin{proof}[Proof of \cref{thm:network-to-convex}]
\label{prf:thm:network-to-convex}
We show the continuity and convexity separately.

\emph{Continuity on $(x,y)$.}
Given $x,y$, each operation within \Network (including the operations in \pan) is continuous operation. Therefore, the output of \Network is continuous on $(x,y)$.

\emph{Convexity on $x$.} \citet{GroupMax:warin2023groupmax} shows that, for an instance of GroupMax Network, as long as all affine transformations except the one in first layer are positive transformation, then such instance is bound to be convex. The convexity of \Network comes from this insight: Our construction of affine transformation is guaranteed to be positive except the one in first layer. These affine transformations are operated on the input $x$ side, thus \Network is convex on $x$. 

However, we also introduce residual connections between hidden layers of \Network as well as an addition positive affine transformation in the final layer. Thus our construction differs from the construction in \citet{GroupMax:warin2023groupmax}. 
However, it's not hard to imagine and prove that these slight modifications from the original GroupMax Network do not affect the core property of the convexity.

For the sake of rigor, we also present a self-contained proof of the convexity on $x$ in \Network in \cref{prf:thm:network-to-convex-self}.

\end{proof}

\begin{proof}[Proof of \cref{thm:convex-to-network}]
\label{prf:thm:convex-to-network}

We begin with definitions of parameterized max-affine functions (PMA) and then introduce a lemma from \citet{PMA-universal_approximator:kim2022parameterized} as follows.

\begin{definition}[Parameterized Max-Affine function (PMA)]
\label{def:PMA}
A parameterized max-affine function $f^{PMA}: X \times Y \to \bbR$ is represented as follows.
\begin{equation}
\label{eq:PMA}
f^{PMA}(x,y) = \max_{k\in [I]}\ \langle a^k(y), x \rangle + b^k(y)
\end{equation}
where $X$ is a convex, compact subspace of $\bbR^{d_x}$, $Y$ is a compact subspace of $\bbR^{d_y}$, $I \in \bbZ_+$ and $a^k: Y \to \bbR^{d_x}, b^k: T \to \bbR$ are continuous functions for $k\in [I]$.
\end{definition}

\begin{lemma}[Theorem 4 in \citet{PMA-universal_approximator:kim2022parameterized}]
\label{lem:PMA-universal}
Denote $\calF(A_1, A_2) = \{f: A_1 \to A_2|\text{$f(a)$ is}$ $\text{continuous on $a$}\}$ and $\calF^a$ is a universal approximator of $\calF(Y, \bbR^{d_x})$, $\calF^b$ is a universal approximator of $\calF(Y, \bbR)$. 
Denote $\calF^{PMA}(\calF^a, \calF^b)$ as the function class that can be represented by PMA under $\calF^a$ and $\calF^b$. 
Specifically, we have $\calF^{PMA} = \{f^{PMA}: X \times Y \to \bbR | \text{$f^{PMA}(x,y)$ has the form in \cref{eq:PMA}, with}\ $
$a^k \in \calF^a, b^k \in \calF^b$ for each $\ k \in [I]\}$. 
Then, $\calF^{PMA}$ is a universal approximator of $\calF$.
\end{lemma}

The key point is to show that the \Network can express a PMA with universal approximation on $a^k(\cdot)$ and $b^k(\cdot)$. 
To see this, we construct the 1-layer \Network with the residual affine transformation to be zero transform. This can be done by hard-coding that each \pan always outputs zero-matrix in the network outputs. Besides, we set the number of group, $G=1$, and the \pan in the second layer always outputs an identity transformation. By this way, we can see that the computation of \Network then has the form of PMA, where the first affine transformation serves as $a^k(x) + b^k$ and the following GroupMax activation becomes the max operation. 
See \cref{fig:network-PMA} for an illustration.


\begin{figure}[H]
\centering
\includegraphics[width=0.8\linewidth]{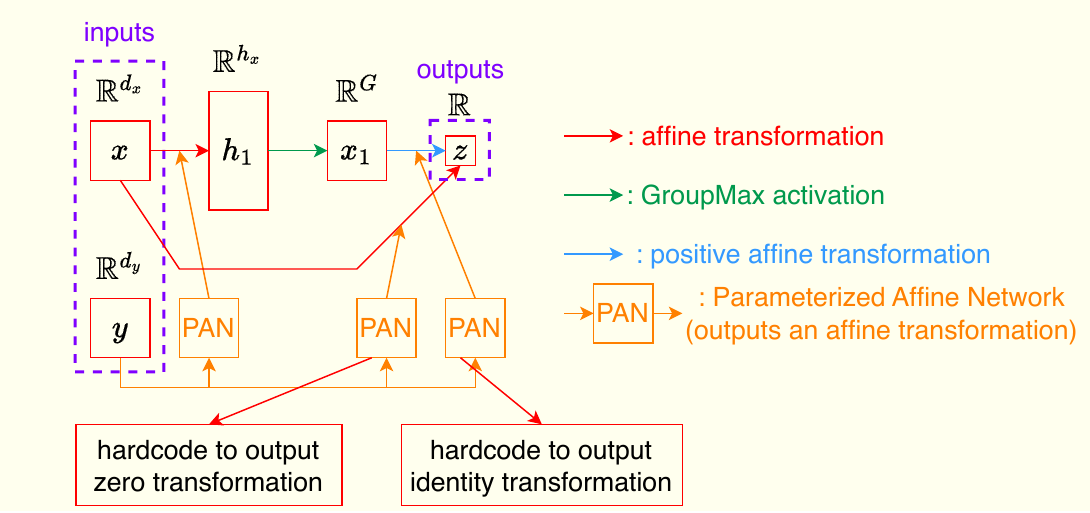}
\caption{Using \Network to express a PMA.}
\label{fig:network-PMA}
\end{figure}

Note that \Pan utilizes neural networks as architecture, and ordinary neural networks are universal approximators of all continuous functions on compact space \citep{universal-neural-network:cybenko1989approximation}.
Therefore, let $\calF^a, \calF^b$ be the class of parameterized affine transformation, then $\calF^a$ is a universal approximator of $\calF(Y, \bbR^{d_x})$, $\calF^b$ is a universal approximator of $\calF(Y, \bbR)$. 
Besides, one neural network can always represent multiple neural networks by stacking hidden neurons. Therefore, any $\{a^k, b^k\}_{k\in [I]}$ with $a_k \in \calF^a, b_k \in \calF^b$ can be represented by a \pan. 
It then follows by \cref{lem:PMA-universal} that such construction of \Network (a subset of $\calG$) is a universal approximator of $\calF$. The same argument then holds for $\calG$.

\end{proof}

Together with \cref{thm:convex-to-network} and \cref{thm:network-to-convex}, we complete the proof of \cref{thm:network-universal}.
\end{proof}

\subsection{Proof of \texorpdfstring{\cref{lem:nbnp-wlog}}{}}

\lemNbnpWlog*

\begin{proof}[Proof of \cref{lem:nbnp-wlog}]
\label{prf:lem:nbnp-wlog}

By definition of $\SEU$, there is a menu mechanism $M^{m,1} = \{p^{m,1}_i\}$ satisfying \Ppc and \nbnp, such that $\EU(M^{m,1}) \ge \SEU - \varepsilon$. We will construct another menu mechanism $M^{m,2}$ satisfying \Ppc and \Nbnps and show that $\EU(M^{m,2}) \ge \EU(M^{m,1})$, then we completes the proof.

We construct $M^{m,2} = \{p^{m,2}_i\}_\iinn$ and $p^{m,2}_i$ has following expression:

\begin{align*}
p^{m,2}_i(x_i;\bmt_{-i}) = p^{m,1}_i(x_i;\bmt_{-i}) - p^{m,1}_i(0;\bmt_{-i})
\end{align*}

It's easy to show that $M^{m,2}$ satisfies \Ppc and \Nbnps (and thus \nbnp). Next we show $\EU(M^{m,2}) \ge \EU(M^{m,1})$.

Let $M^{d,1} = (\bmx^{d,1}, \bmp^{d,1})$ be an equivalent direct mechanism of $M^{m,1}$ such that $\EU(M^{d,1}) > \EU(M^{m,1}) - \varepsilon_0$, we will construct $M^{d,2} = (\bmx^{d,2}, \bmp^{d,2})$ such that $\EU(M^{d,2}) \ge \EU(M^{d,1})$ and $M^{d,2}$ is equivalent to $M^{m,2}$. The construction is follows:
\begin{align*}
x^{d,2}_i(\bmt) =& x^{d,1}_i(\bmt)
\\
p^{d,2}_i(\bmt) =& p^{d,1}_i(\bmt) - p^{m,1}_i(0;\bmt_{-i})
\end{align*}

Not hard to verify that $M^{d,2}$ is indeed equivalent to $M^{m,2}$. Then, 
\begin{align*}
\EU(M^{d,2}) =& \bbE_{\bmt\sim\calF}\left[ u_0(\bmx^{d,2}_i(\bmt),\bmp^{d,2}(\bmt);\bmt) \right]
\\
=& \bbE_{\bmt\sim\calF}\left[ u_0(\bmx^{d,1}_i(\bmt),\bmp^{d,2}(\bmt);\bmt) \right]
\\
\ge& \bbE_{\bmt\sim\calF}\left[ u_0(\bmx^{d,1}_i(\bmt),\bmp^{d,1}(\bmt);\bmt) \right]
\\
=& \EU(M^{d,1})
\end{align*}
where the inequality arises from the fact that $u_0$ in non-decreasing in $p_i$ and $p^{d,2}_i(\bmt) \ge p^{d,1}_i(\bmt)$.

Therefore, $\EU(M^{m,2}) \ge \EU(M^{d,2}) \ge \EU(M^{d,1}) > \EU(M^{m,1}) - \varepsilon_0$. By arbitrariness of $\varepsilon_0$, we have that $\EU(M^{m,2})\ge \EU(M^{m,1})$, that completes the proof.

\end{proof}

\subsection{Proof of \texorpdfstring{\cref{prop:method-expressive}}{}}

\propMethodExpressive*

\begin{proof}[Proof of \cref{prop:method-expressive}]
\label{prf:prop:method-expressive}

Let $M^m = \{p^m_i\} \in \calM^m_2$. We will construct a \method-representable mechanism $M^\method = \{p^\method_i\}$ and show that $\dist(M^m, M^\method) \le \varepsilon$.

Specifically, since $p^m_i(x_i;\bmt_{-i}) - c_i(x_i)$ is convex on $x_i$ and continuous on $(x_i,\bmt_{-i})$ in a compact domain ($\calX_i \times \calT_{-i}$), by universal approximation property of \network, there is a \network-representable function $f_i(x_i;\bmt_{-i})$ such that,
\begin{align*}
l_\infty(p^m_i - c_i, f_i) \le \frac{\varepsilon}{2}.
\end{align*}

We construct $p^\method_i = c_i(x_i) + f_i(x_i;\bmt_{-i}) - f_i(0;\bmt_{-i})$, then
\begin{align*}
l_\infty(p^m_i, p^\method_i) =& l_\infty (p^m_i - c_i, p^\method_i - c_i)
\\ 
=& l_\infty(p^m_i(x_i;\bmt_{-i}) - c_i(x_i), f_i(x_i;\bmt_{-i}) - f_i(0;\bmt_{-i}))
\\
\le& l_\infty(p^m_i(x_i;\bmt_{-i}) - c_i(x_i), f_i(x_i;\bmt_{-i}))
+ l_\infty(0, f_i(0;\bmt_{-i}))
\\
=& l_\infty(p^m_i(x_i;\bmt_{-i}) - c_i(x_i), f_i(x_i;\bmt_{-i}))
+ l_\infty(p_i(0;\bmt_{-i}) - c_i(0), f_i(0;\bmt_{-i}))
\\
\le& 2 l_\infty(p^m_i(x_i;\bmt_{-i}) - c_i(x_i), f_i(x_i;\bmt_{-i}))
\\
\le& \varepsilon
\end{align*}
where the first inequality is the triangle inequality of $l_\infty$, the second inequality is because $l_\infty(p^m_i(x_i;\bmt_{-i}) - c_i(x_i), f_i(x_i;\bmt_{-i}))$ is an upper bound of fixing $x_i=0$, which is $l_\infty(p_i(0;\bmt_{-i}) - c_i(0), f_i(0;\bmt_{-i}))$.
This completes the proof.

\end{proof}

\subsection{Proof of \texorpdfstring{\cref{lem:distance-preserve}}{}}

\lemDistancePreserve*

\begin{proof}[Proof of \cref{lem:distance-preserve}]
\label{prf:lem:distance-preserve}

To prove this theorem, we introduce two useful lemma about the Fenchel conjugate.

\begin{restatable}{lemma}{lemConjugateOnedim}
\label{lem:conjugate-1dim}
Let $u_1, u_2$ be two convex functions on $X = [\xlow, \xhigh]$. Denote $x_j(t) \in \arg\max_{x \in X} \langle x,t\rangle - u_j(x)$ for $j=1,2$ and $t \in T \coloneqq [\tlow, \thigh]$. Then, if $l_\infty(u_1, u_2) \le \delta$, then $l_1(x_1, x_2) \le \frac{\sqrt{2\delta}\cdot [\xhigh - \xlow + \thigh - \tlow]}{2}$.
\end{restatable}

\begin{restatable}{lemma}{lemConjugateMultidim}
\label{lem:conjugate-multidim}
Let $u_1, u_2$ be two convex functions on $X = \times_\iinn X_i$, where $X_i = [\xlow_i, \xhigh_i]$. Denote $x_j(t) \in \arg\max_{x \in X} \langle x,t\rangle - u_j(x)$ for $j=1,2$ and $t \in T$, where $T = \times_\iinn T_i$ and $T_i = [\tlow_i, \thigh_i]$.
Then, if $l_\infty(u_1, u_2) \le \delta$, then $l_1(x_1, x_2) \le  \frac{\sqrt{2\delta}\cdot [\sum_\iinn (\xhigh_i - \xlow_i)/ (\thigh_i - \tlow_i) + n]\cdot \prod_\iinn (\thigh_i - \tlow_i)}{2}$.
\end{restatable}

These lemmas are proved in \cref{prf:lem:conjugate-1dim} and \cref{prf:lem:conjugate-multidim}, respectively.

Let $M^{m,k} = \{p^{m,k}_i\})_\iinn$ and $M^{d,k} = (\bmx^{d,k}, \bmp^{d,k})$ for $k=1,2$. 
Let $\calX_i$ and $\calT_i$ are $d_i$-dimensional and $\calX_i \subseteq \times_{j=1}^{d_i}[\xlow_{ij}, \xhigh_{ij}]$ and $\calT_i \subseteq \times_{j=1}^{d_i}[\tlow_{ij}, \thigh_{ij}]$.

Assume $l_\infty( p^{m,1}_i,p^{m,2}_i) \le \delta $, then by \cref{lem:conjugate-multidim}, we know that $l_1(x^{d,1}_i, x^{d,2}_i) \le c_i \cdot \sqrt{\delta}$, where $c_i = \frac{\sqrt{2}}{2}(\prod_{j\in[d_i]} (\thigh_{ij} - \tlow_{ij}) \cdot [n + \sum_{j\in[d_i]} (\xhigh_{ij} - \xlow_{ij}) / (\thigh_{ij} - \tlow_{ij})] )$ is a constant. 

We analyze $\dist(M^{d,1}, M^{d,2}) = \sum_\iinn l_1(x^{d,1}_i,x^{d,2}_i) + \sum_\iinn l_1(p^{d,1}_i, p^{d,2}_i)$. Given $\varepsilon > 0$, we only need to find $\delta$ such that $l_1(x^{d,1}_i, x^{d,2}_i) \le \frac{\varepsilon}{2n}$ and $l_1(p^{d,1}_i, p^{d,2}_i) \le \frac{\varepsilon}{2n}$ for a certain $i$.

To satisfy $l_1(x^{d,1}_i, x^{d,2}_i) \le \frac{\varepsilon}{2n}$, we can choose $\delta \le (\frac{\varepsilon}{2n c_i})^2$ for all $i$. We focus on satisfying $l_1(p^{d,1}_i, p^{d,2}_i) \le \frac{\varepsilon}{2n}$.

We know that by equivalence relation, $p^{d,j}_i(\bmt) = p^{m,j}_i(x^{d,j}_i(\bmt);\bmt_{-i})$. By argmax property of $x^{d,j}_i$, we have
\begin{align*}
& \langle x^{d,2}_i(\bmt), t_i \rangle + c_i(x^{d,2}_i(\bmt)) - p^{m,2}_i(x^{d,2}_i(\bmt);\bmt_{-i})
\ge \langle x^{d,1}_i(\bmt), t_i \rangle + c_i(x^{d,1}_i(\bmt)) - p^{m,2}_i(x^{d,1}_i(\bmt);\bmt_{-i})
\\
\Rightarrow& p^{m,2}_i(x^{d,2}_i(\bmt);\bmt_{-i}) \le \langle x^{d,2}_i(\bmt) - x^{d,1}_i(\bmt), t_i\rangle + c_i(x^{d,2}_i(\bmt)) - c_i(x^{d,1}_i(\bmt)) + p^{m,2}_i(x^{d,1}_i(\bmt);\bmt_{-i})
\\
\le& \langle x^{d,2}_i(\bmt) - x^{d,1}_i(\bmt), t_i\rangle + c_i(x^{d,2}_i(\bmt)) - c_i(x^{d,1}_i(\bmt)) + p^{m,1}_i(x^{d,1}_i(\bmt);\bmt_{-i}) + \delta
\end{align*}
where the last inequality is because the $l_\infty(p^{m,1}_i, p^{m,2}_i) \le \delta$.

Then,
\begin{align*}
p^{d,2}_i(\bmt) - p^{d,1}_i(\bmt) =& p^{m,2}_i(x^{d,2}_i(\bmt);\bmt_{-i}) - p^{m,1}_i(x^{d,1}_i(\bmt);\bmt_{-i}) 
\\
\le& \langle x^{d,2}_i(\bmt) - x^{d,1}_i(\bmt), t_i\rangle + c_i(x^{d,2}_i(\bmt)) - c_i(x^{d,1}_i(\bmt)) + \delta
\\
\le& \diam(\calT_i) \cdot \| x^{d,2}_i(\bmt) - x^{d,1}_i(\bmt)\|_1 + c_i(x^{d,2}_i(\bmt)) - c_i(x^{d,1}_i(\bmt)) + \delta
\end{align*}

The inverse direction holds similarly, therefore, 
\begin{align*}
|p^{d,2}_i(\bmt) - p^{d,1}_i(\bmt)|
\le& \diam(\calT_i) \cdot \| x^{d,2}_i(\bmt) - x^{d,1}_i(\bmt)\|_1 + |c_i(x^{d,2}_i(\bmt)) - c_i(x^{d,1}_i(\bmt))| + \delta
\end{align*}

Denote $\calT^{\varepsilon_0} = \{\bmt\in\calT: \| x^{d,1}_i(\bmt) - x^{d,2}_i(\bmt) \|_1 \le \varepsilon_0 \}$ where $\varepsilon_0$ is a pending parameter.
Recall that,
\begin{align*}
l_1(p^{d,1}_i, p^{d,2}_i) =& \int_{\calT} | p^{d,1}_i(\bmt) - p^{d,2}_i(\bmt) | \dd \mu(\bmt)
\\
\le& \diam(\calT_i) \int_{\calT} \| x^{d,2}_i(\bmt) - x^{d,1}_i(\bmt)\|_1 \dd \mu(\bmt) + \delta \mu(\calT) + \int_{\calT} |c_i(x^{d,2}_i(\bmt)) - c_i(x^{d,1}_i(\bmt))| \dd \mu(\bmt)
\\
\le& \diam(\calT_i)\cdot l_1(x^{d,1}_i, x^{d,2}_i) + \delta \mu(\calT) 
\\
+& \int_{\calT^{\varepsilon_0}} |c_i(x^{d,2}_i(\bmt)) - c_i(x^{d,1}_i(\bmt))| \dd \mu(\bmt) + \int_{\calT \backslash\calT^{\varepsilon_0}} |c_i(x^{d,2}_i(\bmt)) - c_i(x^{d,1}_i(\bmt))| \dd \mu(\bmt)
\end{align*}

We let $\delta \le \min\{\frac{\varepsilon}{8n\mu(\calT)}, (\frac{\varepsilon}{8n c_i \diam(\calT_i)})^2\}$, then the first two term $\le \frac{\varepsilon}{8n}$. Next, since $c_i$ is continuous on $\calX_i$ thus uniformly continuous and has bound $\pm B_i$ where $B_i > 0$ is a constant that relies only on problem instance. Then,
\begin{align*}
l_1(p^{d,1}_i, p^{d,2}_i) 
\le& \frac{\varepsilon}{4n} 
+ \int_{\calT^{\varepsilon_0}} |c_i(x^{d,2}_i(\bmt)) - c_i(x^{d,1}_i(\bmt))| \dd \mu(\bmt) 
+ 2B_i \mu(\calT \backslash\calT^{\varepsilon_0})
\end{align*}

By Markov Inequality we have $\mu(\calT \backslash\calT^{\varepsilon_0}) \le \frac{l_1(x^{d,1}_i,x^{d,2}_i)}{\varepsilon_0}$. Thus by choosing $\delta$ so small such that $l_1(x^{d,1}_i,x^{d,2}_i) \le \frac{\varepsilon\varepsilon_0}{16n B_i}$, we can satisfy that $2B_i \mu(\calT \backslash\calT^{\varepsilon_0}) \le \frac{\varepsilon}{8n}$. We can choose $\delta \le (\frac{\varepsilon\varepsilon_0}{16n B_i c_i})^2$ that satisfies this requirement.

Next, by uniformly continuity of $c_i$, we know that there is $\varepsilon_0>0$ such that $\| x^{d,2}_i - x^{d,1}_i \|_1 \le \varepsilon_0$ indicates that $|c_i(x^{d,2}_i) - c_i(x^{d,1}_i)| \le \frac{\varepsilon}{8n \mu(\calT)}$. By choosing such $\varepsilon_0$ we have that
\begin{align*}
& \int_{\calT^{\varepsilon_0}} |c_i(x^{d,2}_i(\bmt)) - c_i(x^{d,1}_i(\bmt))| \dd \mu(\bmt) 
\\
\le& \int_{\calT^{\varepsilon_0}}\frac{\varepsilon}{8n \mu(\calT)} \dd \mu(\bmt) 
\\
\le& \frac{\varepsilon}{8n}
\end{align*}

and then 
\begin{align*}
l_1(p^{d,1}_i, p^{d,2}_i) 
\le& \frac{\varepsilon}{4n} + \frac{\varepsilon}{8n} + \frac{\varepsilon}{8n}
= \frac{\varepsilon}{2n}
\end{align*}

Above all, by choosing $\delta \le \min_i \min \{(\frac{\varepsilon\varepsilon_0}{16n B_i c_i})^2, \frac{\varepsilon}{8n\mu(\calT)}, (\frac{\varepsilon}{8n c_i \diam(\calT_i)})^2, (\frac{\varepsilon}{2n c_i})^2\}$, we have that $\dist(M^{m,1},M^{m,2}) \le \delta$ indicates that $\dist(M^{d,1}, M^{d,2}) \le \varepsilon$. 
Thus we complete the proof.

\end{proof}

\subsection{Proof of \texorpdfstring{\cref{lem:direct-EU}}{}}

\lemDirectEU*


\begin{proof}[Proof of \cref{lem:direct-EU}]
\label{prf:lem:direct-EU}

Let $\varepsilon>0$.

Non-degenerate distribution is equivalent to that $\calF$ is absolutely-continuous on $\mu$, where $\mu$ is Lebesgue measure on $\calT$. By Radon–Nikodym theorem \citep{radon-nikodym:nikodym1930generalisation}, there is a measurable function $g: \calT \to \bbR_+$ such that for all measurable subset $\calT_0 \subseteq \calT$, we have that $\calF(\calT_0) \coloneqq \Pr_{\bmt \in \calF}[\bmt\in\calT_0] = \int_{\bmt\in \calT_0} g(\bmt) \dd \mu(\bmt)$.

Then, denote $M^d = (\bmx^d, \bmp^d)$,
\begin{align*}
\EU(M^d) =& \bbE_{\bmt\sim \calF} \left[ u_0(\bmx^d(\bmt), \bmp^d(\bmt);\bmt \right]
\\
=& \int_{\bmt \sim \calT} u_0(\bmx^d(\bmt), \bmp^d(\bmt);\bmt) g(\bmt) \dd \mu(\bmt)
\end{align*}

Therefore, denote $M^{d,1} = (\bmx^{d,1}, \bmp^{d,1})$ and $M^{d,2} = (\bmx^{d,2}, \bmp^{d,2})$,
\begin{equation}
\label{eq:lem:direct-EU:1}
\begin{aligned}
& |\EU(M^{d,1}) - \EU(M^{d,2})| 
\\
=& | \int_{\bmt \in \calT} u_0(\bmx^{d,1}(\bmt), \bmp^{d,1}(\bmt);\bmt) g(\bmt) \dd \mu(\bmt)
- \int_{\bmt \sim \calT} u_0(\bmx^{d,2}(\bmt), \bmp^{d,2}(\bmt);\bmt) g(\bmt) \dd \mu(\bmt) |
\\
\le& \int_{\bmt \in \calT} |u_0(\bmx^{d,1}(\bmt), \bmp^{d,1}(\bmt);\bmt) - u_0(\bmx^{d,2}(\bmt), \bmp^{d,2}(\bmt);\bmt)| g(\bmt) \dd \mu(\bmt)
\end{aligned}
\end{equation}

Since $u_0$ is continuous on a compact set $\calX \times [-B,B]^n \times \calT$, thus it is uniformly continuous and bounded by $\pm B_0$ for some real number $B_0 > 0$.

Since $\int g(\bmt) \dd \mu(\bmt) = 1$, we know that there is $M_0 > 0$ such that $\int_{g(\bmt) > M_0} g(\bmt) \dd \mu(\bmt) < \frac{\varepsilon}{4 B_0}$. Thus, \cref{eq:lem:direct-EU:1} can be transformed to,
\begin{equation}
\label{eq:lem:direct-EU:2}
\begin{aligned}
& \int_{\bmt \in \calT} |u_0(\bmx^{d,1}(\bmt), \bmp^{d,1}(\bmt);\bmt) - u_0(\bmx^{d,2}(\bmt), \bmp^{d,2}(\bmt);\bmt)| g(\bmt) \dd \mu(\bmt)
\\
=& \int_{g(\bmt) > M_0} |u_0(\bmx^{d,1}(\bmt), \bmp^{d,1}(\bmt);\bmt) - u_0(\bmx^{d,2}(\bmt), \bmp^{d,2}(\bmt);\bmt)| g(\bmt) \dd \mu(\bmt)
\\
+& \int_{g(\bmt) \le M_0} |u_0(\bmx^{d,1}(\bmt), \bmp^{d,1}(\bmt);\bmt) - u_0(\bmx^{d,2}(\bmt), \bmp^{d,2}(\bmt);\bmt)| g(\bmt) \dd \mu(\bmt)
\\
\le& \int_{g(\bmt) > M_0} g(\bmt) \dd \mu(\bmt) \cdot \max |u_0(\bmx^{d,1}(\bmt), \bmp^{d,1}(\bmt);\bmt) - u_0(\bmx^{d,2}(\bmt), \bmp^{d,2}(\bmt);\bmt)|
\\
+& \int_{g(\bmt) \le M_0} |u_0(\bmx^{d,1}(\bmt), \bmp^{d,1}(\bmt);\bmt) - u_0(\bmx^{d,2}(\bmt), \bmp^{d,2}(\bmt);\bmt)| M_0 \dd \mu(\bmt)
\\
\le& 2B_0 \cdot \frac{\varepsilon}{4B_0} 
+ M_0 \int_{g(\bmt) \le M_0} |u_0(\bmx^{d,1}(\bmt), \bmp^{d,1}(\bmt);\bmt) - u_0(\bmx^{d,2}(\bmt), \bmp^{d,2}(\bmt);\bmt)| \dd \mu(\bmt)
\\
=& \frac{\varepsilon}{2} 
+ M_0 \int_{\bmt \in \calT} |u_0(\bmx^{d,1}(\bmt), \bmp^{d,1}(\bmt);\bmt) - u_0(\bmx^{d,2}(\bmt), \bmp^{d,2}(\bmt);\bmt)| \dd \mu(\bmt)
\end{aligned}
\end{equation}

Next we focus on the integral term, $I \coloneqq \int_{\bmt \in \calT} |u_0(\bmx^{d,1}(\bmt), \bmp^{d,1}(\bmt);\bmt) - u_0(\bmx^{d,2}(\bmt), \bmp^{d,2}(\bmt);\bmt)| \dd \mu(\bmt)$.

We define the $\calT^\delta_x, \calT^\delta_p \subseteq \calT$ as follows where $\delta$ is a pending parameter: $\calT^\delta_x = \{\bmt\in\calT: \sum_\iinn \|x^{d,1}_i(\bmt) - x^{d,2}_i(\bmt) \| \le \delta\}$,
$\calT^\delta_p = \{ \bmt\in\calT: \sum_\iinn | p^{d,1}_i(\bmt) - p^{d,2}_i(\bmt) | \le \delta \}$ and $\calT^\delta = \calT^\delta_x \cap \calT^\delta_p$.

Then,
\begin{equation}
\label{eq:lem:direct-EU:3}
\begin{aligned}
I =& \int_{\bmt \in \calT} |u_0(\bmx^{d,1}(\bmt), \bmp^{d,1}(\bmt);\bmt) - u_0(\bmx^{d,2}(\bmt), \bmp^{d,2}(\bmt);\bmt)| \dd \mu(\bmt)
\\
=&\int_{\bmt \in \calT^\delta} |u_0(\bmx^{d,1}(\bmt), \bmp^{d,1}(\bmt);\bmt) - u_0(\bmx^{d,2}(\bmt), \bmp^{d,2}(\bmt);\bmt)| \dd \mu(\bmt)
\\
+& \int_{\bmt \in \calT \backslash \calT^\delta} |u_0(\bmx^{d,1}(\bmt), \bmp^{d,1}(\bmt);\bmt) - u_0(\bmx^{d,2}(\bmt), \bmp^{d,2}(\bmt);\bmt)| \dd \mu(\bmt)
\\
\le& \int_{\bmt \in \calT^\delta} |u_0(\bmx^{d,1}(\bmt), \bmp^{d,1}(\bmt);\bmt) - u_0(\bmx^{d,2}(\bmt), \bmp^{d,2}(\bmt);\bmt)| \dd \mu(\bmt)
+ 2B_0 \mu(\calT \backslash \calT^\delta)
\\
\end{aligned}
\end{equation}

By Markov Inequality, $\mu(\calT \backslash \calT^\delta)$ is upper bounded by $\dist(M^{d,1}, M^{d,2})$. Specifically, 
\begin{align*}
& \int_\calT \sum_\iinn \| x^{d,1}_i(\bmt), x^{d,2}_i(\bmt) \| \dd \mu(\bmt)
\\
\ge& \int_{\calT\backslash \calT^\delta_x} \sum_\iinn \| x^{d,1}_i(\bmt), x^{d,2}_i(\bmt) \| \dd \mu(\bmt)
\\
\ge& \delta \cdot \mu(\calT\backslash \calT^\delta_x)
\end{align*}

Similarly, 
\begin{align*}
\int_\calT \sum_\iinn | p^{d,1}_i(\bmt), p^{d,2}_i(\bmt) | \dd \mu(\bmt) \ge \delta \cdot \mu(\calT\backslash \calT^\delta_p)
\end{align*}

This means that
\begin{align*}
\mu(\calT \backslash \calT^\delta) \le& \mu(\calT \backslash \calT^\delta_x) + \mu(\calT \backslash \calT^\delta_p)
\\
\le& \frac{1}{\delta}
\left[
\int_\calT \sum_\iinn | p^{d,1}_i(\bmt), p^{d,2}_i(\bmt) | \dd \mu(\bmt)
+ \int_\calT \sum_\iinn \| x^{d,1}_i(\bmt), x^{d,2}_i(\bmt) \| \dd \mu(\bmt) 
\right]
\\
=& \frac{\dist(M^{d,1}, M^{d,2})}{\delta}
\end{align*}

and \cref{eq:lem:direct-EU:3} becomes,
\begin{equation}
\label{eq:lem:direct-EU:4}
\begin{aligned}
I \le& \int_{\bmt \in \calT^\delta} |u_0(\bmx^{d,1}(\bmt), \bmp^{d,1}(\bmt);\bmt) - u_0(\bmx^{d,2}(\bmt), \bmp^{d,2}(\bmt);\bmt)| \dd \mu(\bmt)
+ \frac{2B_0 \dist(M^{d,1}, M^{d,2})}{\delta}
\end{aligned}
\end{equation}

By uniform continuity of $u_0$, we know that there is $\delta_0 > 0$ such that for all $\bmt \in \calT^{\delta_0}$ we have $|u_0(\bmx^{d,1}(\bmt), \bmp^{d,1}(\bmt);\bmt) - u_0(\bmx^{d,2}(\bmt), \bmp^{d,2}(\bmt);\bmt)| < \frac{\varepsilon}{4 M_0 \cdot \mu(\calT)}$.
Taking $\delta = \delta_0$ in \cref{eq:lem:direct-EU:4} (since \cref{eq:lem:direct-EU:4} holds for all $\delta$), we get
\begin{equation}
\label{eq:lem:direct-EU:5}
\begin{aligned}
I \le& \int_{\bmt \in \calT^{\delta_0}} |u_0(\bmx^{d,1}(\bmt), \bmp^{d,1}(\bmt);\bmt) - u_0(\bmx^{d,2}(\bmt), \bmp^{d,2}(\bmt);\bmt)| \dd \mu(\bmt)
+ \frac{2B_0 \dist(M^{d,1}, M^{d,2})}{\delta_0}
\\
\le& \int_{\bmt \in \calT^{\delta_0}} \frac{\varepsilon}{4M_0 \mu(\calT)} \dd \mu(\bmt)
+ \frac{2B_0 \dist(M^{d,1}, M^{d,2})}{\delta_0}
\\
\le& \frac{\varepsilon}{4 M_0} + \frac{2B_0 \dist(M^{d,1}, M^{d,2})}{\delta_0}
\end{aligned}
\end{equation}

Taking $I$ into \cref{eq:lem:direct-EU:2}, we achieve
\begin{equation}
\label{eq:lem:direct-EU:6}
\begin{aligned}
& |\EU(M^{d,1}) - \EU(M^{d,2})|
\\
\le& \frac{\varepsilon}{2} + M_0 I
\\
\le& \frac{3\varepsilon}{4} + \frac{2B_0 M_0 \dist(M^{d,1}, M^{d,2})}{\delta_0}
\end{aligned}
\end{equation}

By letting $\delta = \frac{\delta_0 \varepsilon}{8 B_0 M_0}$, we complete the proof.

\end{proof}

\subsection{Proof of \texorpdfstring{\cref{thm:tedi-expressive}}{}}

\thmTediExpressive*

\begin{proof}[Proof of \cref{thm:tedi-expressive}]
\label{prf:thm:tedi-expressive}
Given any $\varepsilon > 0$, we will show that there is $M^m \in \calM^\method$ such that $\EU(M^m) > \SEU - \varepsilon$.
By \cref{lem:nbnp-wlog} we know that $\SEU = \sup_{M^m \in \calM^m_2} \EU(M^m)$. 
Therefore, we can find $M^{m,1} \in \calM^m_2$ such that $\EU(M^{m,1}) > \SEU - \frac{\varepsilon}{2}$.
Also, by definition of $\EU(M^{m,1})$ we can find a $M^{d,1}$ that is equivalent to $M^{m,1}$ and $\EU(M^{d,1}) > \EU(M^{m,1}) - \frac{\varepsilon}{4}$.

By \cref{lem:direct-EU}, we know that there is $\delta_1 > 0$ such that for all $M^{d,2}$ such that $\dist(M^{d,1}, M^{d,2}) < \delta_1$, it holds that $| \EU(M^{d,1}) - \EU(M^{d,2})| < \frac{\varepsilon}{8}$.
Then by \cref{lem:distance-preserve}, we know that there is $\delta_2 > 0$ such that for all $M^{m,2}$ such that $\dist(M^{m,1}, M^{m,2}) < \delta_2$ and all $M^{d,2}$ is equivalent to $M^{m,2}$, we have that $\dist(M^{d,1}, M^{d,2}) < \delta_1$.

Since $M^{m,1} \in \calM^m_2$, then by \cref{prop:method-expressive}, there is a \method-express $M^{m,2}\in \calM^\method$ that $\dist(M^{m,1}, M^{m,2}) < \delta_2$.
Denote $M^{d,2}$ is equivalent to $M^{m,2}$, we have that $\dist(M^{d,1}, M^{d,2}) < \delta_1$ and therefore, $\EU(M^{d,2}) > \EU(M^{d,1}) - \frac{\varepsilon}{8}$, and therefore $\EU(M^{m,2}) > \EU(M^{d,1}) - \frac{\varepsilon}{8} > \EU(M^{m,1}) - \frac{\varepsilon}{4} - \frac{\varepsilon}{8} > \SEU - \frac{\varepsilon}{2} - \frac{\varepsilon}{4} - \frac{\varepsilon}{8} > \SEU - \varepsilon$.
Above all, it shows that $\SEU = \sup_{M^m \in \calM^\method} \EU(M^m)$ and we complete the proof.

\end{proof}

\subsection{Proof of \texorpdfstring{\cref{lem:conjugate-1dim}}{}}

\lemConjugateOnedim*

\begin{proof}
\label{prf:lem:conjugate-1dim}

We can without loss of generality assume $\xlow = \tlow = 0$ and $0\le u'_j(x) \le \thigh$, since a) we can add a translation as well as an affine function to hardcode $\xlow = \tlow = 0$ without weakening the conclusions, and b) the domain in which $u'_j < 0$ or $u'_j > \thigh$ are never used in $x_1(t)$ and $x_2(t)$, thus ironing these domains do not weaken the conclusion (also do not strengthen the condition as well).

We define $X_j(t) = \argmax_{x\in X} \langle x,t \rangle - u_j(x)$.
By definition of $x_j(t)$, we know that $t \in \nabla_x u_j(x_j(t))$ and $x_j(t) \in X_j(t)$. 
We assume that at some point $t_0$ we have $x_0 \in X_1(t_0) \cap X_2(t_0)$, by a little computation we have
\begin{align*}
u_j(x_0) =& \int_0^{x_0} \nabla_x u_j(s) \dd s
\\
=& \int_0^{t_0} \nabla_x u_j(x_j(t)) \dd x_j(t)
\\
=&  \int_0^{t_0} t \dd x_j(t)
\\
=& t x_j(t) |^{t_0}_0 - \int_0^{t_0} x_j(t) \dd t
\\
=& t_0 \cdot x_0 - \int_0^{t_0} x_j(t) \dd t
\end{align*}
It means that
\begin{align*}
u_1(x_0) - u_2(x_0) = \int_0^{t_0} x_2(t) \dd t - \int_0^{t_0} x_1(t) \dd t
\end{align*}

Denote $t_0 = 0, t_1,..., t_{N-1}, t_N = \thigh$ are the points (or maximal closed intervals) such that $X_1(t) \cap X_2(t) \ne \emptyset$ (notice that $0\in X_1(0) \cap X_2(0)$ and $\xhigh \in X_1(\thigh) \cap X_2(\thigh)$). 
By \citet{debreu:debreu1964continuity} and upper-semi-continuity of $X_j(t)$, we know that such points are always countable.
We temporarily abuse the notation and assume that such points are finite ($N$) and have a natural linear order.
One can easily extend the proof to the case of countable points.
By monotonicity and upper-semi-continuity of $X_j(t)$, we know that one of $X_1(t) > X_2(t)$ or $X_2(t) > X_1(t)$ always holds for $t \in (t_{k}, t_{k+1})$.

We without loss of generality assuming $X_2(t) > X_1(t)$ always holds for $t \in (t_{k}, t_{k+1})$, then, denote $x_{k+1} \in X_1(t_{k+1}) \cap X_2(t_{k+1})$ and $x_{k} \in X_1(t_{k}) \cap X_2(t_{k})$, we have
\begin{align*}
u_1(x_{k+1}) - u_2(x_{k+1}) =& \int_0^{t_{k+1}} x_2(t) \dd t - \int_0^{t_{k+1}} x_1(t) \dd t
\\
u_1(x_{k}) - u_2(x_{k}) =& \int_0^{t_{k}} x_2(t) \dd t - \int_0^{t_{k}} x_1(t) \dd t
\end{align*}

Subtracting them, we achieve,
\begin{align*}
& \left[ u_1(x_{k+1}) - u_2(x_{k+1}) \right] - \left[ u_1(x_{k}) - u_2(x_{k}) \right]
\\
=& \int_{t_k}^{t_{k+1}} x_2(t) \dd t - \int_{t_k}^{t_{k+1}} x_1(t) \dd t
= \int_{t_k}^{t_{k+1}} x_2(t) - x_1(t) \dd t
\end{align*}

Since $l_\infty(u_1, u_2) < \delta$, it means that $\left[ u_1(x_{k+1}) - u_2(x_{k+1}) \right] - \left[ u_1(x_{k}) - u_2(x_{k}) \right] \le 2\delta$, then
\begin{align*}
\int_{t_k}^{t_{k+1}} x_2(t) - x_1(t) \dd t \le 2\delta
\end{align*}

If $X_1(t) > X_2(t)$ always holds for $t \in (t_{k}, t_{k+1})$, then similarly $\int_{t_k}^{t_{k+1}} x_1(t) - x_2(t) \dd t \le 2\delta$.

Besides, we have
\begin{align*}
& \int_{t_k}^{t_{k+1}} x_2(t) - x_1(t) \dd t
\\
\le& \int_{t_k}^{t_{k+1}} [x_{k+1} - x_k] \dd t
\\
=& (x_{k+1} - x_k) \cdot (t_{k+1} - t_k)
\end{align*}
where the third equation holds because $x_k \le x_j(t) \le x_{k+1}$ when $t_k \le t \le t_{k+1}$ by definition.

Combining them, we have
\begin{align*}
\int_{t_k}^{t_{k+1}} x_2(t) - x_1(t) \dd t \le \sqrt{2\delta (x_{k+1} - x_k) \cdot (t_{k+1} - t_k)}
\end{align*}

Above all, 
\begin{align*}
l_1(x_1, x_2) =& \int_0^\thigh |x_1(t) - x_2(t)| \dd t
\\
=& \sum_{k=0}^N \int_{t_k}^{t_{k+1}} |x_1(t) - x_2(t)| \dd t
\\
\le& \sum_{k=0}^N \sqrt{2\delta} \cdot \sqrt{(x_{k+1} - x_k) \cdot (t_{k+1} - t_k)}
\\
\le& \sqrt{2\delta} \cdot \sum_{k=0}^N \frac{(x_{k+1} - x_k) + (t_{k+1} - t_k)}{2}
\\
\le& \sqrt{2\delta} \cdot \frac{\xhigh + \thigh}{2}
\end{align*}

Note that the last inequality is generally not equality because when we add through $t_{k+1} - t_k$ ($x_{k+1} - x_k$), we actually do not take the closed interval in $\{t_k\}$ ($\{x_k\}$) into count.
Above all, we complete the proof.

\end{proof}

\subsection{Proof of \texorpdfstring{\cref{lem:conjugate-multidim}}{}}

\lemConjugateMultidim*

\begin{proof}
\label{prf:lem:conjugate-multidim}

Similar with \cref{prf:lem:conjugate-1dim}, we assume $\xlow_i = \tlow_i = 0$ and $0 \le \nabla_{x_i} u_j(x) \le \thigh_i$ for all $x\in X$ without loss of generality.

Define $u^i_j(x_i;t_{-i}) = \min_{x_{-i}\in X_{-i}} u_j(x) - \langle x_{-i}, t_{-i}\rangle$, following statements about $u^i_j(x_i;t_{-i})$ are true for $j=1,2$ and $\iinn$:

\begin{enumerate}
\item $u_j^i(x_i;t_{-i})$ is convex on $x_i$.
\item $0 \le \nabla_{x_i} u^i_j(x_i;t_{-i}) \le \thigh_i$.
\item $l_\infty(u^i_1(\cdot;t_{-i}), u^i_2(\cdot;t_{-i})) \le \delta$.
\item $x^i_j(t_i;t_{-i}) \in \argmax_{x_i \in X_i} \langle x_i, t_i\rangle - u^i_j(x_i;t_{-i})$, where $x^i_j(t)$ is the $i$'th element of $x_j(t)$.
\end{enumerate}

Among them, statement (1) is true because $u_j(x) - \langle x_{-i}, t_{-i} \rangle$ is jointly convex on $(x_i,x_{-i})$, and taking the minimum value on partial variables do not affect the convexity of remaining variables \citep{convex:boyd2004convex}. 

Statement (2) is true because $\nabla_{x_i} (u_j(x) - \langle x_{-i}, t_{-i}\rangle) = \nabla_{x_i} u_j(x)$ is between $[0,\thigh_i]$, then statement (2) is a direct corollary of envelope theorem \citep{envelope-theorem:milgrom2002envelope}.

Statement (3) is true because we have $u^i_1(x_i;t_{-i}) \le u^i_2(x_i;t_{-i}) + \delta$ always holds. To see this, we let $x^2_{-i}$ be the minimum value $u^i_2$ achieve, then, $u^i_2(x_i;t_{-i}) = u_2(x_i, x_{-i}^2) - \langle x_{-i}^2, t_{-i}\rangle$.
We have $u^i_1(x_i;t_{-i}) \le u_1(x_i, x_{-i}^2) - \langle x_{-i}^2, t_{-i}\rangle$ by definition of minimum. Since $l_\infty(u_1, u_2) \le \delta$, we know that $u_1(x_i, x_{-i}^2) \le u_2(x_i, x_{-i}^2) + \delta$, then $u^i_1(x_i;t_{-i}) \le u^i_2(x_i;t_{-i}) + \delta$. The inverse direction is symmetric.

Statement (4) is true because by definition of $x_j(t) \in \argmax_{x\in X} \langle x,t \rangle - u_j(x)$, we have that 
\begin{align*}
x^i_j(t) \in& \argmax_{x_i \in X} \max_{x_{-i} \in X_{-i}} \langle x,t\rangle - u_j(x) 
\\
=& \argmax_{x_i \in X} \max_{x_{-i} \in X_{-i}} \langle x_i, t_i \rangle + \langle x_{-i}, t_{-i} \rangle - u_j(x) 
\\
=& \argmax_{x_i \in X} [\langle x_i, t_i \rangle + \max_{x_{-i} \in X_{-i}} \langle x_{-i}, t_{-i} \rangle - u_j(x)] 
\\
=& \argmax_{x_i \in X} \langle x_i, t_i \rangle - u^i_j(x_i;t_{-i})
\end{align*}

Based on these statements, we can apply \cref{lem:conjugate-1dim} on each $u^i_1$ and $u^i_2$ pair and any $t_{-i}$. We can get that for each $i$ and each $t_{-i}$, $l_1(x^i_1(\cdot;t_{-i}), x^i_2(\cdot;t_{-i})) \le \frac{\sqrt{2\delta}\cdot [\xhigh_i + \thigh_i]}{2}$. Then,
\begin{align*}
l_1(x^i_1, x^i_2) \coloneqq& \int_{t\in T} | x^i_1(t) - x^i_2(t)| \dd t
\\
=& \int_{t_{-i} \in T_{-i}} \dd t_{-i} \int_{t_i \in T_i } | x^i_1(t_i;t_{-i}) - x^i_2(t_i;t_{-i})| \dd t_i
\\
\le& \int_{t_{-i} \in T_{-i}} \dd t_{-i} \frac{\sqrt{2\delta}\cdot [\xhigh_i + \thigh_i]}{2}
\\
=& \frac{\sqrt{2\delta}\cdot [\xhigh_i + \thigh_i]\cdot \prod_\jnei \thigh_j}{2}
\\
=& \frac{\sqrt{2\delta}\cdot [\xhigh_i/ \thigh_i + 1]\cdot \prod_\iinn \thigh_i}{2}
\end{align*}
and
\begin{align*}
l_1(x_1, x_2) =& \sum_\iinn l_1(x^i_1, x^i_2)
\\
\le& \frac{\sqrt{2\delta}\cdot [\sum_\iinn \xhigh_i/ \thigh_i + n]\cdot \prod_\iinn \thigh_i}{2}
\end{align*}
This completes the proof.

\end{proof}

\subsection{A Self-Contained Proof of the Partial Convexity of \network}

\begin{proof}
\label{prf:thm:network-to-convex-self}

Recall the computation flow of \Network (line 10 - line 15) in \cref{alg:network-computation}, where $W^x_i$ for $i \ge 1$ and $w_x$ is non-negative. 
Since all parameters ($W^x_i, W^r_i, b_i, w^x, w^r$) only relies on $y$, therefore, we see them as constant when we focus on the convexity on $x$.
Denote $\theta$ as the set of all these parameters.
The interim variables used in the computation process are denoted as $h^x_i(x;\theta)$ and $x_i(x;\theta)$, in line 12 and line 13 respectively. We denote $z(x;\theta)$ in line 15 as the final output.

To begin with, since $x_0(x;\theta) = x$ so each element in $h^x_1(x;\theta) \in \bbR^{h_x}$ is linear on $x$, thus convex on $x$. 
Next we use mathematical reduction to show that for all $i$, each element in $h^x_i(x;\theta) \in \bbR^{h_x}$ and $x_i(x;\theta) \in \bbR^G$ is convex on $x$.

Assume the statement holds for $k \le i-1$. 
For $k=i$, we have $h^x_i(x;\theta) = W^x_{i-1} x_{i-1}(x;\theta) + W^r_{i-1} x_0(x;\theta) + b_{i-1}$. 
From this expression we know that each element of $h^x_i(x;\theta)$ consists of three part: 
The first part is a positive weighted combination of $x_{i-1}(x;\theta)$, and this part is convex on $x$ since $x_{i-1}(x;\theta)$ is convex on $x$, and convex functions are closed under positive linear combination.
The second part is a linear combination of $x_0(x;\theta)$, and this part is linear on $x$ because $x_0(x;\theta)$ is linear on $x$.
The last part is evidently independent with $x$.
Overall, each element of $h^x_i(x;\theta)$ is convex on $x$.

Next, we know that each element of $x_i(x;\theta)$ is a maximum of some elements in $h^x_i(x;\theta)$. Therefore, each element of $x_i(x;\theta)$ is also convex on $x$. Thus the above statement is recursively correct.

Finally, we claim that $z(x;\theta)$ is convex on $x$, similar with the convexity of each element of $h^x_i(x;\theta)$ on $x$.
Above all, we complete the proof.

\end{proof}

\newpage
\section{Experimental Details}
\label{app:exp}

We provide more experimental details in this section, including implementation details for all approaches in experiments in \cref{app:subsec:exp:implementation-detail}, more experimental results in \cref{app:subsec:exp:more-results}, more analysis on experimental results in \cref{app:subsec:exp:more-analysis}.

\subsection{Implementation Details}
\label{app:subsec:exp:implementation-detail}

\subsubsection{Default Configurations.}
\label{app:subsubsec:exp:detail-experiment}

To make a fair comparison, we choose the default sample size $K = 65,536$ and the batch size $B=4,096$ during training procedures for all approaches. 
The network parameter is updated with $T = 5,000$ iterations for all approaches, except for MD-RegretNet we let $T = 10,000$.
All temperatures in \method, MD-GemNet, MD-LotteryAMA, and MD-MenuNet achieve their maximum value $\beta = 512$ in their last iteration, except the $B_{1,m}$ experiment in which the maximum temperature is $\beta = 4,096$.

We also validate learned models in each 500 network iterations during the training period. 
Therefore, all approaches (except MD-RegretNet) are validated 10 times.
For all approaches, we use $2^{14} = 16,384$ samples for $1$-player settings or $2^{12} = 4,096$ samples for $3$-player settings to validate the model performance and choose the model with the largest mechanism designer's utility (or other measures \Wrt MD-RegretNet) estimated on the validation samples.

When testing a model, we use $2^{18} = 262,114$ samples for $1$-player settings or $2^{16} = 65,536$ samples for $3$-player settings to achieve an estimation of the mechanism designer's value.
All values are reported based on these samples in tables.
Note that the standard deviation of estimation error has magnitude $\approx 10^{-3}$ by Central Limit Theorem, which is negligible among the results for different approaches in most cases.

All experiments are run on a single RTX 4090 graphical card with $24$G memory and same CPU configurations.

These configurations are default used except explicit specifications.

\subsubsection{Details of \method.}
\label{app:subsubsec:exp:detail-method}

\paragraph{Architecture of \network.}
The hidden layer of \Network is selected to be 1 for single-player experiments and 2 for multi-player experiments in all experiments. 
We choose the hidden dimension $h_x = 64 * (m+3)$ and the number of group $G = 2 * (m+3)$ for all cases, except for the $S^{Ber}$ case where we set $h_x = 24 * (m+3)$.
When we implement the GroupMax activation function, we instead use a ``soft'' version of the max function, \ie, $\mathrm{differentialmax}(x_1,...,x_k) = \frac{1}{\beta} \log \sum_\iink \exp(\beta \cdot x_i)$ with the default value $\beta = 4,096$. 
This technique is used for ease of optimization on the \Network parameters because the conventional max operator will pass the gradient to only one of the many inputs. 
It's not hard to verify that, when ``$\max$'' is replaced with ``differentialmax'' in GroupMax activation, the partial convexity as well as the universal approximation property of \Network still hold.

\paragraph{Architecture of \pan.}

The implementation of \Pan differs between $n=1$ player and $n>1$ players.
When the problem instance has $n > 1$ players, the \Pan is implemented with a Multi-Layer Perceptron (MLP) with 2 layers, with hidden dimension fixed as $24 * (m+3)$. 
We choose leaky-relu with negative slope = 0.01 to serve as activation functions on each hidden layer. 
After computing affine transformations $W \in \bbR^{d_{in}}$ and $b\in \bbR^{d_{out}}$, we either post-process an element-wise $\softplus$ function to guarantee that each element in $W$ is positive, or do nothing and output $W$ directly. It depends on whether this network was pre-empted always to output positive affine transformation or (standard) affine transformation.

When the problem instance has only one player, the input $t_{-i}$ makes no sense. Therefore, we hardcode $W \in \bbR^{d_{in}}$ and $b\in \bbR^{d_{out}}$ as the model parameter, and the post $\softplus$ function is also required when \Pan was pre-empted to output positive affine transformation.

\paragraph{Initialization}

The initialization of parameters in \Network is far different from traditional initialization for MLP (\eg, the default ``Kaiming-He Initialization'' in PyTorch).
It is because ``Kaiming-He Initialization'' makes the magnitude of each neuron $O(1/\sqrt{d})$ where $d$ is the input dimension, therefore, leading each element in output dimension the magnitude with input dimension. However, since all elements are hard-coded positive in \Network sometimes, we require the magnitude of the affine transformation $O(1/d)$ to guarantee the output magnitude same with input magnitude. Therefore, we multiply $1/d$ to the output of \Pan when it is preempted to hardcode positive affine transform when $n>1$.
When $n=1$, \Pan directly parameterizes $W, b$, so we additionally scale $W$ with a factor of $1/\sqrt{d}$ when $W$ is required to be positive.

\paragraph{Training Procedures.}

We use an Adam optimizer with an initial learning rate of $5\times 10^{-4}$ to optimize all the network parameters, as well as an SGD optimizer with an initial learning rate of $0.03$ to update auxiliary variables $\{\bmy, \bmz\}$ in continuous sampling steps. Both learning rates are gradually decreased to the minimum value ($1 \times 10^{-5}$ and $0.01$, respectively) through the learning procedure.

Before each iteration on network parameters, we follow with some iterations on continuous sampling on $\{\bmy^k,\bmz^k\}_\kinK$. When there is $n=1$ player, we take a continuous sampling step on all $K$ data for $S=2$ times. When there are $n > 1$ players, we only take a continuous sampling step on the $B$ batch data for $S=16$ times.
This difference arises from the fact that calling \Pan is more costly than calling \network, therefore, by this technique, we can decrease the cost of calling \Pan in multi-player settings.

\paragraph{Inference Procedures.}

In the inference period, we use the SGD optimizer with fine-tuned learning rate and momentum $\beta=0.9$ to find the player utility-maximizing outcomes. We update $5000$ iterations on outcomes optimization for each type profile. The learning rate is initialized with $\gamma=0.2$ and gradually decreases to $\gamma=0.0003$. 
We observe that the accuracies on player utilities are less than $10^{-6}$ and these accuracies pose $ < 10^{-3}$ errors on platform expected utilities, which is even smaller than the standard deviation of errors on reporting values from samples, making these errors negligible and acceptable for estimating platform expected utility of \method.

\paragraph{Complexity Analysis of \network.}

When there is only $n=1$ player, the complexity of one network call of \Network is $O(m^2)$ (one $O(m)$ for the number of groups or the input dimension and another $O(m)$ for hidden dimension).
When there are $n>1$ players, the complexity becomes $O(n^2 m^3)$ (an additional $O(m)$ for the hidden dimension in \pan; one $O(n)$ for independent computations of $n$ players, another $O(n)$ for the input of \pan, $\bmt_{-i}$). 
Therefore, we fine-tune the complexity of all learning-based baselines to make the complexity comparable.
Specifically, the menu size of MD-MenuNet is set as $O(m)$ to make the total computational cost $O(m^2)$. 
The menu size of MD-LotteryAMA is set to $O(m^2) $, resulting in a total computational cost of $ O (n^2 m^3)$.
The menu size and hidden dimension of MD-GemNet are both set to $O(m)$, resulting in a total computational cost of $ O (n^2 m^3)$.
The hidden dimension of RegretNet is set as $O(m^{1.5})$ to make the total computational cost $O(n^2 m^3)$.
See below implementations of learning-based baselines for more details.



\subsubsection{Details of Learning-based Baselines.}
\label{app:subsubsec:exp:detail-baselines}

\paragraph{MD-MenuNet.}
MD-MenuNet \cite{MenuNet:shen2019automated} is specialized to the multi-good monopolist setting in our experiments. MD-MenuNet consists of a menu mechanism network that generates candidate outcomes and corresponding prices, as well as a buyer network that selects the utility-maximizing outcomes and monetary transfers. Note that only the menu mechanism network has learnable parameters. 
Specifically, we set the menu size as $64 * (m + 3)$. The sigmoid activation is applied to the outputs of the menu mechanism network to hardcode each element of outcomes within $[0,1]$. 
In the training of MD-MenuNet, we use the Adam optimizer with an initial learning rate of $0.01$, which decayed to $0.005$ after $1,000$ epochs and $0.002$ after $4,000$ epochs. The temperature is initialized at $16$ and multiplies with $1.15$ every $2\%$ epochs after the first $20\%$ epochs until it reaches the maximum value of $512$. We find that the value does not increase when the maximum temperature is increased to $4,096$ at the Bernoulli distribution setting. Therefore, we fix the maximum temperature at $512$ in all cases.


\paragraph{MD-GemNet.}
MD-GemNet consists of two separate
networks, namely the "outcome network" and "payment network".
Each network is an MLP with two hidden layers and GeLU \cite{hendrycks2023gaussianerrorlinearunits} activation following the original work \citep{GemNet:wang2024gemnet}. An element-wise sigmoid function is applied to the outputs of the outcome network to ensure that each element of the outcomes is within $[0,1]$. The loss function is set as the negative value of the mechanism designer's utility.
Both menu size and hidden dimension of the MLP are set to $64 * (m + 3)$. The temperature is fine-tuned and initialized as $64$ for $3$ players settings. In all experiments, the temperature multiplies with $1.15$ times every $2\%$ epochs after the first $20\%$ epochs, until it reaches the maximum value of $512$.
The network is optimized by Adam optimizer with the learning rate initialized at $5\times 10^{-4}$ and gradually decreased to the minimum value of $1\times 10^{-5}$.

We note that in the settings of $B^{1,0}_{3,10}, U^{0,0.5}_{3,10}, C^{0,0.5}_{3,10}$, MD-GemNet learns a mechanism with $=0$ designer's utility. Therefore, we continuously change different random seeds until MD-GemNet returns a mechanism with positive designer's utility.

\paragraph{MD-LotteryAMA.}

MD-LotteryAMA directly parameterizes an AMA mechanism, which is inherently truthful. AMA mechanism in MD-LotteryAMA is composed of many outcome candidates, also called lotteries, as well as corresponding shifts on each outcome candidate.
We set the lottery size of MD-LotteryAMA as $8 * (m+3)^2$ (we note that increasing the lottery size does not further improve the model performance evidently). 
We choose Adam Optimizer to optimize the parameters of outcome menus and corresponding shifts. The initial learning rate is $0.1$ and gradually decreases to $0.001$.
We set the initial temperature $\beta_{init} = 64$ for $3$ players settings. In all experiments, the temperature multiplies with $1.15$ times every $2\%$ epochs after the first $20\%$ epochs, until it reaches the maximum value of $512$.
In the inference period, the outcome that maximizes the affine social welfare w/o player $i$ is selected.

\paragraph{MD-RegretNet.}

MD-RegretNet is composed of an allocation network $\bmx: \calT \to \calX$ as well as a payment network $\bmp: \calT \to \bbR^n$ to parameterize the direct mechanisms.
We set the allocation network and payment network 
with an MLP with $2$ layers and hidden neurons $16 * (m+3)^{1.5}$ per layer for $1$ player and $8 * (m+3)^{1.5}$ per layer for $3$ players.
MD-RegretNet has the loss function same with \citep{RegretNet-journal:dutting2024optimal}. 
The default first-order penalty is initialized at $\lambda_1=32$ and the default second-order penalty is fixed at $\lambda_2 = 0$.
We find that in many cases a fixed first-order penalty is enough to learn a direct mechanism with large utility and reasonably low regret, but in the settings $U^{1,0}_{3,2}, U^{0,0.5}_{3,10}, C^{0,0.5}_{3,10}$, it fails to learn a direct mechanism with positive utility. 
Therefore, in these setting, we fine-tune the hyper-parameters to $\lambda_1 = 3, \lambda_2 = 5$.
The update rule of first-order penalty is $\lambda_1 \leftarrow \lambda_2 \cdot \min(rgt, 0.1)$ for every $1\%$ epochs after the first $5\%$ epochs.
The learning rate is chosen as $1 \times 10^{-3}$ initially, and gradually decreased to $2 \times 10^{-5}$ during training.
The network is trained with $10,000$ iterations in total.
We note that since MD-RegretNet has two conflicting objectives, the learning of MD-RegretNet takes more computational resources and has lower performance than other truthfulness-guaranteed approaches.

For the computation of types misreport, we use SGD optimizer with learning rate $0.1$ and momentum $\beta = 0.9$. We optimize the misreport of batch data for $32$ iterations before each network parameter updates in training. These misreports are used in regret computation in the penalty term.
In the inference period of RegretNet, we estimate the misreport through direct optimization of players' utilities with SGD optimizer for 1000 iterations with learning rate $0.1$.
We note that this implementation follows the original implementation \citep{RegretNet-journal:dutting2024optimal}.
The true regret is under-estimated, since the true regrets are approximated by non-convex optimization.

After training, we choose the model with the largest \emph{normalized platform utility} on validation data. The normalized platform utility is computed following the work of \citet{ALGNet:rahme2020auction},
and has following form: 
$\tilde{u}(M^d) = \left(\sqrt{u(M^d)} - \sqrt{rgt(M^d)}\right)^2$, where $rgt(M^d)$ is the (estimated) expected regret value of the mechanism $M^d$
and $u(M^d)$ is the expected mechanism designer's utility of the mechanism $M^d$.


\subsubsection{Details of Traditional Baselines.}
\label{app:subsubsec:exp:detail-baseline-2}

\paragraph{SJA.}

SJA is only implemented for $n=1$. We follow the definition of SJA in Definition 4.1 of \citet{optimal_auction-special_case:giannakopoulos2014duality}. 
The probability is taken as experienced probability on $2^{16} = 65,536$ data, to make sure the data used for learning are identical among all approaches. 

Although this practice may bring out over-fitting issues to SJA parameters, we think such an issue does not need to be overcome since all learning-based approaches may also face over-fitting issues. 
Introducing this issue brings a more fair comparison among \Method and baselines and allows us to better understand the over-fitting degree of these approaches.
Besides, to understand at what level does this overfitting issue have affect on SJA, we also take up to $2^{25} = 33,554,432$ data to estimate SJA parameters more accurately. The performance of SJA with variable training data is shown in \cref{app:subsec:exp:more-results}.

Although the optimality of SJA relies on the $\iid$, $U[0,1]$ type distribution \citep{optimal_auction-special_case:giannakopoulos2014duality}, the definition of SJA does not rely on the type distribution, thus the performance of SJA is also evaluated in other settings with non-uniform distributions, though the performance is suboptimal.

\paragraph{Bundle-OPT.}

As long as $\calX_i = [0,1]^m$, we can bundle all $m$ items together, \ie, reduce $\calX_i$ to $\tilde{\calX}_i = \{(x,x,...,x) \in [0,1]^m: x \in [0,1]\}$. In this way, this mechanism design problem becomes a single-parameter problem (possibly with a complex and intractable distribution of the ``single parameter''). We utilize the characterization of the optimal mechanism in single-parameter settings to parameterize the optimal mechanism with one parameter. 



\paragraph{Separable-OPT.}

As long as $\calX_i = [0,1]^m$ and utilities of players and mechanism designer is separable (\ie, $u_i = \sum_\jinm u_{ij}, u_0 = \sum_\jinm u_{0j}$), we can also dissemble the problem to $m$ independent sub-problems and independently solve these sub-problems. 
Each sub-problem $j$ has the outcome space $\calX_{i,j} = [0,1]$, player utilities $u_{ij}$ and mechanism designer utility $u_{0j}$. Sub-problem $j$ is again single-parameter and its optimal mechanism as well as optimal value can be analytically derived. The expected utility of mechanism design in original problems equals the sum of expected utility in all sub-problems.

\paragraph{VCG.}

The conventional VCG mechanism is irrelevant to the mechanism designer's utility.
In this version, if the mechanism designer's utility is quasi-linear on monetary transfers, we add the mechanism designer's utility to the objective function of VCG, which is very like the boost function on outcomes in the conventional AMA mechanism. This practice can boost VCG's performance compared with traditional VCG.

Denote $u_0(\bmx,\bmp;\bmt) = v_0(\bmx;\bmt) + \sum_\jinm p_j$. Given a type profile $\bmt$, VCG choose the outcome that maximizes the social welfare $\bmx^*(\bmt) \in \argmax_{\bmx\in\calX} \SW(\bmx;\bmt) \coloneqq v_0(\bmx;\bmt) + \sum_\iinn v_i(x_i;t_i)$. 
The monetary transfer of player $i$ is determined by the negative externalities of player $i$. Mathematically, define 
$\SW_{-i}(\bmx;\bmt_{-i}) = \min_{\bmt_i \in \calT_i} v_0(\bmx;\bmt) + \sum_\iinn v_i(x_i;t_i)$ as the social welfare without player $i$ when the outcome is $\bmx$ and others' types are $\bmt_{-i}$, 
$\bmx^*_{-i}(\bmt_{-i}) \in \argmax_{\bmx \in \calX} \SW_{-i}(\bmx;\bmt_{-i})$ be the virtual outcome without player $i$ when other types are $\bmt_{-i}$, and 
$\Ext_i(\bmt) = \SW(\bmx_{-i}^*(\bmt_{-i});\bmt) - \SW_{-i}(\bmx^*(\bmt);\bmt)$ be the negative externality brought by player $i$. 
Then, if we let $p^*_i(\bmt) = \Ext_i(\bmt)$, then $(\bmx^*(\bmt), \bmp^*(\bmt))$ forms the result of VCG. 

Not hard to verify that this version of the VCG mechanism is still truthful and IR. The insight arises from two facts: a) Utility of truthful reporting is always non-negative because $\SW(\bmx^*(\bmt);\bmt) \ge \SW(\bmx_{-i}^*(\bmt);\bmt)$, and b) truthful reporting maximizes $\SW(\bmx^*(\bmt);\bmt)$ and therefore maximizes players' utilities.
We note that only \emph{quasi-linear utility} assumption is enough to achieve this result.

\subsubsection{Details of Ablated Approaches.}
\label{app:subsubsec:exp:detail-ablation}

\paragraph{\Method-Discrete.}

This approach ablates TEDI's learning algorithm by replacing the continuous, discretization-free outcome space with a discrete menu of outcomes. The core training procedure is consequently simplified, removing the need for the covariance trick and continuous sampling. Specifically, for each player, an "outcome-rule" network, identical in architecture to the one used in MD-GemNet, takes other players’ types as input and generates a discrete menu of $32(m+3)$ candidate outcomes. The payment for each of these outcomes is then determined by the Partial GroupMax Network, preserving TEDI's core pricing mechanism. During training, the mechanism is optimized by maximizing the expected designer utility directly, using a standard softmax relaxation (with the same temperature schedule as MD-GemNet) to ensure differentiability over the discrete menu choices. The optimizer and learning rate schedules are identical to those used for MD-GemNet.

\paragraph{\Method-PICNN.}

This approach ablates the network architecture by replacing the Partial GroupMax Network with a Partial Input Convex Neural Network (PICNN) \cite{PICNN:amos2017input}. The PICNN is designed to be provably convex with respect to its partial inputs. Our implementation uses a 2-layer architecture with a hidden dimension of $32*(m+3)$. It maintains convexity in the outcome input $x_i$ through the use of positive weight constraints and ReLU activations, while taking other players' types $\bmt_{-i}$ as standard, non-convex inputs. This model is trained using the exact same procedure as TEDI, including the use of the covariance trick and continuous sampling with Langevin dynamics, to assess the effectiveness of the pricing network architecture in isolation.

\paragraph{\Method-PMA.}

This ablation replaces the pricing network with a Parameterized Max-Affine (PMA) network \citep{PMA-universal_approximator:kim2022parameterized}. PMA networks are also partially convex by construction. Our implementation features a 2-layer base MLP that processes other players' types $\bmp_{-i}$ to generate the weights and biases for $16*(m+3)$ affine functions. The final pricing function is the maximum of these affine functions applied to the outcome input $x_i$. As with TEDI-PICNN, this model is trained using the identical algorithm as the full TEDI approach, allowing for a direct comparison of the complete network architecture of Partial GroupMax Network against the PMA architecture.

\paragraph{\Method-MLP.}

In the most direct network ablation, we replace the Partial GroupMax Network with a standard 2-layer Multi-Layer Perceptron (MLP) with a hidden dimension of $32*(m+3)$. The MLP takes the simple concatenation of the outcome $x_i$ and other players' types $\bmt_{-i}$ as its input. Crucially, this architecture provides no structural guarantee of partial convexity, thereby testing the importance of this property for the stability and performance of the overall learning framework. The training algorithm for TEDI-MLP is identical to that of TEDI.

\subsection{More Experimental Results}
\label{app:subsec:exp:more-results}

\subsubsection{Reverse Auctions as Penalized Over-Allocation of Unit-Supply Auctions}

In this section, we consider a new setting, which is economically explained as \emph{reverse auctions}.

In this setting, we consider a buyer (serving as the mechanism designer) buying $m$ goods from $n$ potential sellers.
We let $\calX_i = [0,1]^m$, $\calT_i = [-1,0]^m$, $c_i(x_i) \equiv 0$.
Hence $-t_{ij}$ is interpreted as the cost to seller $i$ of providing one unit good $j$.
We specify that $t_{ij} \iidd \calU([-1,0])$.
Let $v\in \bbR^m_+$ be the buyer's intrinsic valuation to goods.
We specify $v_j \equiv 1$ for all $\jinm$ within this section.

We consider two cases representing either divisible goods or indivisible goods.
a) Goods are indivisible, and the element $x_{ij}$ in outcome is interpreted as the probability of buying the good $j$ from seller $i$. 
In this case, we assume the buyer's valuation is unit-demand, then we have $u_0(\bmx, \bmp;\bmt) = \langle v, k \rangle + \sum_\iinn p_i$ where $k\in[0,1]^m, k_j = \min(1, \sum_\iinn x_{ij})$ is the probability of good $j$ bought from at least one seller.
Note that in this case, $- p_i$ represents the payoff to the buyer $i$.
b) Goods are divisible, and the element $x_{ij}$ in outcome is interpreted as the amount of good $j$ bought from seller $i$.
In this case, we assume the valuation of buyer has constant absolute risk aversion (CARA) \citep{MWG:mas1995microeconomic}, then we let $u_0(\bmx, \bmp, \bmt) = \sum_\jinm v_j \cdot [1 - \exp(- \sum_i x_{ij}/v_j)] + \sum_\iinn p_i$.

We denote $R_{n,m}^{G}$ as the setting involving $n$ seller, $m$ goods, with $G = D$ or $G=I$ representing divisible goods or indivisible goods settings, respectively. 
We note that the indivisible goods settings are similar to unit-supply auctions with penalized over-allocation.

\cref{tab:exp-reverse} presents the results about mechanism designer's utility.

\begin{table}[t]
\centering
\caption{Experimental results for reverse auctions and digital good auctions.}
\begin{tabular}{c|cccc}
\toprule
Methods/Setting & $R_{3,10}^I$ & $R_{3,20}^I$ & $R^D_{3,10}$ & $R^D_{3,20}$ \\
\midrule
\method & \underline{\textbf{5.5096}} & \underline{\textbf{11.150}} & \underline{\textbf{2.8047}} & \underline{\textbf{5.7293}} \\
\midrule
MD-GemNet & 5.1347 & 8.8076 & 2.5080 & 3.9118 \\
MD-LotteryAMA & 5.1501 & 10.012 & 1.2725 & 3.2019 \\
\midrule
MD-RegretNet & 4.7599 & 9.9737 & 1.6378 & 3.8719 \\
(regret) & 0.0334 & 0.0569 & 0.0302 & 0.0639 \\
\midrule
VCG & 5.0000 & 10.000 & 1.8269 & 3.6538 \\
Separable-OPT & 5.3110 & 10.625 & 2.5639 & 5.1275 \\
Bundle-OPT & 5.0028 & 10.001 & 1.2213 & 2.3390 \\
\bottomrule
\end{tabular}
\label{tab:exp-reverse}
\end{table}

\subsubsection{More Precise Value of SJA.}

\begin{table}[H]
\centering
\caption{Performance of SJA in $U_{1,m}$ for $m = 2,10,20$ with different number of samples.}
\begin{tabular}{c|ccc}
\toprule
\# samples & $U_{1,2}$  & $U_{1,10}$ & $U_{1,20}$ \\
\midrule
$2^{16}$ & 0.5492 & 3.4761 & 7.5132 \\
$2^{20}$ & 0.5487 & 3.4786 & 7.5638 \\
$2^{25}$ & 0.5496 & 3.4748 & 7.5629 \\
\bottomrule
\end{tabular}
\label{tab:SJA-more}
\end{table}

We provide the value of the SJA mechanism with variable training data size in \cref{tab:SJA-more}.
We omit the values in the settings with Bernoulli distributions, since SJA performs poorly on these settings. We suspect that the degeneration of Bernoulli distribution mainly causes poor performance in Bernoulli distribution settings.
We find that the approximation of SJA is fairly unstable when the dimension is large (especially the $U_{1,20}$ setting).

\subsubsection{Comparison Between the Learned Mechanisms and Baselines.}

We visualize the mechanism learned by \Method for the setting of $U_{1,10}$ and compare it with SJA and Bundle-OPT.
We observe that the learned pricing rule of \Method is very close to a piece-wise linear function, therefore, the function can be approximated by the values reported on vertices, \ie, the prices of the deterministic bundles. 
Since there are exponential-many deterministic bundles, we only report the prices on some representative bundles.
Let the bundle be $(x_1,...,x_m)$ with $m = 10$. We consider two types of bundles: a) $x_i = 1$ for $i \le k$ and $x_i = 0$ for $i > k$ for some $1\le k \le 10$. The bundles in this form are called forward bundles; b) $x_i = 1$ for $i > m - k$ and $x_i = 0$ for $i \le m - k$ for some $1\le k \le 10$. The bundles in this form are called backward bundles.
We report the prices of these bundles in \cref{tab:compare-method-SJA}.
Note that the prices of bundles learned by \Method are quite close to the prices of SJA, although the mechanism is learned assuming that players quantile-response with temperature $\beta = 512$ instead of the best response.

\begin{table}[H]
\centering
\caption{Prices of bundles with variable goods in SJA, Bundle-OPT and \Method in the setting of $U_{1,10}$. SJA is reported with $2^{25}$ samples. \method-forward represents that the bundle has the form of $(1,...,1,0,...,0) \in \{0,1\}^{10}$, \method-backward represents that the bundle has the form of $(0,...,0,1,...,1) \in \{0,1\}^{10}$.}
\adjustbox{max width=\textwidth}{
\begin{tabular}{c|cccccccccc}
\toprule
\# goods in bundles & $1$  & $2$ & $3$ & $4$ & $5$ & $6$ & $7$ & $8$ & $9$ & $10$ \\
\midrule
\method-forward & 1.2232 & 1.9819 & 2.5356 & 3.0713 & 3.4577 & 3.8033 & 4.0465 & 4.2164 & 4.3968 & 3.9359 \\
\method-backward & 1.1740 & 1.9189 & 2.5008 & 3.0110 & 3.3468 & 3.7761 & 4.1621 & 4.3293 & 4.4604 & 3.9359\\
SJA & 0.9091 & 1.6897 & 2.3561 & 2.9053 & 3.3620 & 3.6915 & 3.9330 & 4.0650 & 4.0939 & 3.9944 \\
Bundle-OPT & - & - & - & - & - & - & - & - & - & 3.9034 \\
\bottomrule
\end{tabular}
}
\label{tab:compare-method-SJA}
\end{table}

\subsection{More Experimental Analysis}
\label{app:subsec:exp:more-analysis}

In this section, we provide a more in-depth analysis of the experimental results provided in this paper.



\subsubsection{Bundling Effect.}

\citet{optimal_auction-special_case:manelli2006bundling} has shown the insight that selling goods with bundles can bring improvements over selling goods independently in multiple goods monopolists setting (\eg, $U_{1,2}$).
Even though there are exponentially deterministic bundles, the results of Bundle-OPT shed light on the fact that selling only one bundle can achieve $>99\%$ optimal value in some cases ($U_{1,2}, U_{1,10}, U_{1,20}$). We call the bundling effect strong \Wrt some problem instance, if a single bundle can achieve a large percent of optimal value; otherwise, we call the bundling effect weak.

The bundling effect is extremely strong in the $U_{1,m}$ setting. In the case of $U_{1,10}$ and $U_{1,20}$, the performance of MD-MenuNet is very similar to the Bundle-OPT, demonstrating the large probability that MD-MenuNet exactly learns the single bundle of Bundle-OPT in these cases.

For \method, our visualization result in \cref{tab:compare-method-SJA} show that in the case of $U_{1,10}$, \Method successfully learns some pricing rule which is more similar to SJA (which is also conjectured to be optimal) than Bundle-OPT. The performance of \Method also outperforms Bundle-OPT in this case. 
Unluckily, \Method does not perform ideally in $U_{1,20}$, we think two factors mainly cause this phenomenon: one is the limited temperature, and the other is the competitive performance of a single bundle, which represents the value of Bundle-OPT.

In the setting of $B_{1,m}$ and all 3-players setting, the bundling effect is fairly weak. It can be evidenced with the fact that Separable-OPT surpasses Bundle-OPT with significant gaps.
MD-MenuNet surpasses the Bundle-OPT in $B_{1,10}$, yet it still fails to approximate the optimal mechanism with a large gap in all cases.
However, the performance of \Method is stable and approximates the optimal performance with $\approx 98\%$ performance in cases of $B_{1,10}$, when the optimal mechanism is known (exactly Separable-OPT).

The multi-players settings again demonstrate such a phenomenon: When Bundle-OPT is no more competitive (has a lower value compared with Separable-OPT), discretization-based baselines are likely to be suboptimal.

From these results, we can roughly conclude below: \Method exhibits stable performance that is irrespective of the problem settings. However, the performance of discretization-based approaches, such as MD-GemNet, MD-LotteryAMA, and MD-MenuNet, depends highly on whether the bundling effect is strong in different settings. If the bundling effect is weak, the performance of discretization-based approaches will likely be suboptimal.

\subsubsection{Further Revealing Dimension Insensitivity.}

A notable phenomenon is that the performance of \Method in $R_{3,20}$ is two times larger than that in $R_{3,10}$. This result is reasonable since the utility is additive among items; thus, in some settings with $2m$ goods, one could always separate $2m$ goods into two sub-settings, with $m$ goods for each, and run the optimal mechanisms in each sub-setting. With inductive bias, one would believe that if some approach is really insensitive to problem size, then the performance on $2m$-size instances should be two times better than the performance on corresponding $m$-size instances. However, both MD-LotteryAMA and MD-GemNet fail this argument in $R_{3,20}$ compared with $R_{3,10}$, while \Method survives. 
Such a contrast is further evidence of the curse-of-dimensionality faced by existing approaches, revealing the learning power of \Method in various cases.

\subsubsection{Initialization.}

\Method also exhibits the advantage of insensitivity on initialization. \citet{lottery-AMA:curry2023differentiable} discovers that the performance of LotteryAMA highly depends on different initializations, and we discover similar phenomena on MD-MenuNet and MD-GemNet.
The instability of initialization appears in the experiment stated below:
The optimal mechanism in $S^U_2$ includes $3$ deterministic bundles, but MD-MenuNet fails to learn two of three bundles and learns only one bundle in Bundle-OPT sometimes when the random seeds vary. As a comparison, \Method can stably learn the pricing rule represented in \cref{fig:intro}, regardless of random seeds. 

MD-GemNet suffers from this issue even more severely. In $B^{1,0}_{3,10}, U^{0,0.5}_{3,10}, C^{0,0.5}_{3,10}$, GemNet preserves $=0$ values at the first random seeds. We reran $B^{1,0}_{3,10}$ with $3$ random seeds and $U^{0,0.5}_{3,10}, C^{0,0.5}_{3,10}$ with $2$ random seeds to achieve positive values of MD-GemNet.

Such phenomena showcase that learning a function is much more stable than learning many discrete points.



\newpage
\section{Further Discussions}
\label{app:sec:discuss}

In this section, we provide more discussions about this research and discuss the most significant open problems inspired.

\subsection{The Algorithmic Boundary of \method.}
\label{app:sec-discuss:subsec-boundary}

In this section, we will discuss the economic insights as well as the algorithmic boundary of \Method in detail.

The core assumptions \Method poses on model structures are: players' \emph{quasi-linear utilities} and \emph{bi-linear valuations}, as well as \emph{outcome-separability} of outcome space structure. 
By this way we say that the ``algorithmic boundary'' of \Method is all mechanism design problems with above three model structures.
Quasi-linear utility is standard in mechanism design theory and we will not clarify it more.
Below we will provide a discussion of 
\emph{bi-linear valuation} and \emph{outcome-separability} at length.

\subsubsection{Bi-linear Utility.}

The property of bi-linear valuations requires that the second-mixed derivative of valuation function is constant between the outcome \Wrt the player, $x_i$, and her type, $t_i$ (possibly not with same dimensionality). Formally, an equivalent form is $\frac{\pp^2 v_i}{\pp x_i \pp t_i}(x_i;t_i)$ is a constant matrix. Furthermore, it's \Wlg to assume $x_i$ and $t_i$ has same dimensionality and $\frac{\pp^2 v_i}{\pp x_i \pp t_i}(x_i;t_i) \equiv I$ where $I$ is the identity matrix. The reason why this step is \Wlg will be discussed at the end of this part.
Consequently, $v_i(x_i;t_i)$ only has the following form
\begin{equation}
\label{eq:model-3}
    v_i(x_i;t_i) = \langle x_i, t_i\rangle + c^x_i(x_i) + c^t_i(t_i),
\end{equation}
\Wrt some $c^x_i : \calX_i \to \bbR$ and $c^t_i: \calT_i \to \bbR$.
It is \Wlg to assume $c^t_i(t_i) \equiv 0$ and $c^x_i(\zeros) = 0$ by subtracting a constant \Wrt each type \citep{AMD-initial:sandholm2003automated}.
This is because by doing so, the truthfulness and IR constraints (recall that the player will receive utility $v(\zeros;t_i)$ if she chooses to realize the outside option ($\zeros$)) as well as the mechanism designer's utility do not change at all, thus these changes have no effect from the mechanism designer's perspective of solving Program \eqref{eq:model-6}.
Therefore, \cref{eq:model-3} can be simplified as,
\begin{equation}
\label{eq:model-4}
    v_i(x_i;t_i) = \langle x_i, t_i\rangle + c_i(x_i),
\end{equation}
with $c_i(\zeros) = 0$.
\cref{eq:model-4} is equivalent with the generalized model we have discussed at \cref{app:sec-preface}. There are also convexity assumptions on type space and outcome space, as well as the separability assumption on type space ($\calT = \times_\iinn \calT_i$). These assumptions are only technical.


There are two questions left: 1) the economic intuition behinds ``constant second-mixed derivative''; 2) the reason of \Wlg $\frac{\pp^2 v_i}{\pp x_i \pp t_i}(x_i;t_i) \equiv I$.

To answer the first question above, one economic intuition arises from ``risk neutrality'' of players. Let us consider a problem with $K$ (finite) deterministic outcomes and arbitrary valuation on these outcomes, denoted as $v = (v_1,...,v_K)$. If the player is risk-neutral and denote $\Delta^K$ as the standard simplex with dimension $K$, and $x\in \Delta^K$ represents the probability of each deterministic outcome. Then, $\langle v, x\rangle$ is the valuation of randomized outcome $x$, and $\Delta^K$ is the randomized outcome space. This valuation expression has exactly ``constant second-mixed derivative''.

Next we will explain the second question above from a reduction perspective. We will show that for any mechanism design problem $\MD = (n, \calT, \calX, \{u_i\}_\iinn, u_0, \calF)$ satisfying constant second-mixed derivative, we can always construct an ``equivalent'' mechanism design problem $\MD'$ satisfying $\frac{\pp^2 v'_i}{\pp x_i \pp t_i}(x_i;t_i) \equiv I$, where $v'_i$ is the valuation function in $\MD$. By saying ``equivalence'', we mean that there is a natural reduction between a direct mechanism $M$ from $\MD$ and a direct mechanism $M'$ from $\MD'$. $M$ and $M'$ derives identical truthfulness requirement and identical utility to mechanism designer, respectively in $\MD$ and $\MD'$.

Let $\calX_i$ and $\calT_i$ is $d^x_i, d^t_i$ dimensional, respectively.
Then we can write $\calX_i \subseteq \mathbb{R}^{d^x_i}$, $\calT_i \subseteq \mathbb{R}^{d^t_i}$.
Recall that by quasi-linear utility and bi-linear valuation, we can write $u_i(x_i, p_i;t_i) = v_i(x_i; t_i) - p_i$ and $v_i(x_i; t_i) = x_i^T A_i t_i + c_i(x_i)$. We have $A_i \in \bbR^{d^x_i \times d^t_i}$, and $x_i, t_i$ are regarded as column vectors. The superscript $T$ means the transpose operator.

We next make the following transformation:

\begin{itemize}
\item $t'_i = ((A_i t_i)^T, t_i^T)^T$
\item $x_i' = (x_i^T, 0^{d^t_i})^T$
\item $c'_i(x'_i) = c_i(x'_i[1:d^x_i]) = c_i(x_i)$
\item $v'_i(x'_i,t'_i) = \langle x'_i, t'_i \rangle + c'_i(x'_i) = x_i^T A_i t_i + c_i(x_i) = v_i(x_i,t_i)$
\item $u'_i(x'_i,p'_i;t'_i) = v'_i(x'_i,t'_i) - p'_i$
\item $u'(\bmx',\bmp';\bmt') = u(\{x'_{i}[1:d^x_i]\}_\iinn,\bmp';\{t'_{i}[d^x_i+1:d^x_i+d^t_i]\}_\iinn)$
\item $F'$ is the distribution of $\bmt'$ as $\bmt \sim F$
\end{itemize}

By writing $y = x[k:l]$, we mean that $y$ is the sub-vector of $x$, from the $k$'th element to the $l$'th element.

We then construct the mechanism design problem $\MD' = (n, \calX', \calT', \{u'_i\}_\iinn, u', \calF')$ with type space $\calT'_i = \{t'_i = ((A_i t_i)^T, t_i^T)^T | t_i \in \calT_i\}$, outcome space $\calX'_i = \calX \times \{0^{d^t_i}\}$, player $i$'s utility function $u'_i(x'_i,p'_i;t'_i): \calX'_i\times \bbR \times \calT'_i \to \mathbb{R}$, and mechanism designer's utility function as defined above. We can ensure that $\calX'_i$ and $\calT'_i$ are convex and compact sets as long as $\calX$ and $\calT$ are convex and compact.
Note that from our transformations, we have $t_i = t'_i[d^x_i+1:d^x_i+d^t_i]$ and $x_i = x'_i[1:d^x_i]$, thus the transformation is a bijection between ($\bmt,\bmx$) and ($\bmt', \bmx'$).

Let $M'^d = (x'^d,p'^d)$ be a mechanism in $\MD'$, where $x'^d: \calT' \to \calX', p'^d: \calT' \to \mathbb{R}^n$. Similarly, let $M^d = (x^d, p^d)$ be a mechanism in $\MD$. Consider the following equivalence relation between $M^d$ and $M'^d$: $(x^d_i(\bmt), 0^{d^t_i}) = x'^d_i(\bmt')$, $p^d_i(\bmt) = p'^d_i(\bmt')$.
Under this equivalence constraint, there is a one-to-one, polynomial-time computable reduction from $M^d$ to $M'^d$ or from $M'^d$ to $M^d$. Moreover, $M^d$ is truthful in $\MD$ if and only if $M'^d$ is truthful in $\MD'$. In addition, mechanism designer's utility in $M^d$ w.r.t. $\MD$ equals to the utility in $M'^d$ w.r.t. $\MD'$.

Therefore, by solving the optimal mechanism in $\MD'$, we can obtain the optimal mechanism in $\MD$ within polynomial time.
Note that the second-order mixed derivative in $\MD'$ is given by $\frac{\partial^2 v'_i}{\partial x'_i \partial t'_i}(x'_i;t'_i) \equiv I$, where $I$ is the $d^x_i + d^t_i$-dimensional identity matrix.



\subsubsection{Outcome-Separability.}

Outcome-separability means the global outcome space is a cartesian product of local outcome space, \ie, $\calX = \times_\iinn \calX_i$. This structure exhibits arguably strong economic insight: There is no strong restriction on any possible combination of local outcomes, while weak restriction is permissible. 
More specifically, denote $\bmx = (x_1,...,x_n)$ as a combination of local outcomes. 
Recall that any mechanism design problem should economically require the mechanism designer have commitment power over some outcome spaces before the mechanism is carried out and players then rationally act based on mechanism designer's commitment power \cite{borgers-mechanism-design}.
Let $\calX_0$ be the ``desirable'' outcome space of some mechanism design problem.
By saying a mechanism design problem has strong (\Resp weak) restriction on outcome space $\calX_0 \subseteq \calX \coloneqq \times_\iinn \calX_i$, we mean that the mechanism designer does not have (\Resp does have) commitment power on outcomes $\bmx \in  \calX \backslash\calX_0$.
Specifically, when a problem has weak restriction on outcome space, and the local outcome $x_i$ is realized for all player $i$, the mechanism designer should then realize the global outcome $\bmx = (x_1,...,x_n)$, even if $\bmx \in  \calX \backslash\calX_0$. The mechanism designer may have extremely low preference on such $\bmx \in \calX \backslash\calX_0$, but since she have committed to it before the mechanism is carried out, she have no choice but to realize $\bmx$, possibly with extremely high cost. It can be mathematically described by a penalization on ``undesired outcome space'' $\calX \backslash\calX_0$, reflected in mechanism designer's utility function.
Let us take an example of auctions with reproducible goods considered in main body. 
As $\calX_i = [0,1]^m$ represents the allocation of goods to player $i$, \emph{\structure} requires that the mechanism designer commit to any allocation profile $\bmx \in [0,1]^{n\times m}$ by reproducing goods such that the goods meet the demand, even if the reproducing costs are high.

We then explain why \emph{\structure} is necessary for \method. Recall that \Method parameterizes menu mechanism. In a menu mechanism, the local outcome $x^*_i(\bmt)$ that maximizes player's utility function within local outcome space $\calX_i$ is realized. There is no hard constraint among $x^*_i(\bmt)$ for different player $i$, therefore, only by guaranteeing that any combination $(x_i)_\iinn$ is a feasible global outcome as long as $x_i$ is a feasible local outcome, we can guarantee that \Method outputs a feasible menu mechanism (consequently, a truthful direct mechanism).
The above requirement holds as long as $\calX = \times_\iinn \calX_i$.

We note that there is an implicit economical contradiction within our arguments about \emph{\structure} below: Our argument suggests that global outcome space is a consequence of local outcome space. However, in many cases including auctions, local outcome is determined by global outcome. This is because \Structure is only a simplification of some more ``intrinsic'' structure, which does not relate to the concept of local outcome space. 
Actually, local outcome space is something abstracted from preference structure of players.
Let $\calX$ be the (global) outcome space of some mechanism design problem.
Define $\succsim_i(t_i)$ be the preference relation of player $i$ on outcome space $\calX$, when player $i$ has type $t_i$. Then, $\succsim_i(t_i)$ naturally induces an equivalence relation on outcome space as below: $x \sim_i y \Leftrightarrow (x,p) \succsim_i(t_i) (y,q)$ consistently for all $t_i, p, q$, or equivalently, $u_i(x,p;t_i) - u_i(y,q;t_i)$ is non-negative for all $t_i$. This means that although the specific type of player $i$ is unknown, the mechanism designer can infer from type space to derive the conclusion that $x$ and $y$ is invariant to player $i$, and thus $x$ and $y$ can be seen as same outcome. Therefore, $\sim_i$ naturally defines an equivalent class on $\calX$.
Based on above, \emph{\structure} can be stated as follows: Given any two outcomes $x,y \in \calX$ and two different players $i,j$, we can always find an outcome $z \in \calX$ such that $x \sim_i z, y\sim_j z$. It is not hard for one to see that, by defining local outcome space $\calX_i$ as the equivalent class induced by $\sim_i$ on $\calX$, there is a homomorphism from $\calX$ to $\times_\iinn \calX_i$.

For general settings of mechanism design problems where \Structure is no longer satisfied, it is also possible to construct an ``equivalent'' mechanism design problem which satisfies \structure, where the equivalence is defined in the sense that the optimal truthful mechanisms are identical in these two problems.
Chapter 13.1 by \citeauthor{AGT-Ch13:hartline2007profit} in \citet{AGT:Nisan2007AlgorithmicGameTheory} shows that there is always an approach to reduce any mechanism design problem to another equivalent problem with \emph{\structure}.
The reduction can be done by penalizing infeasible outcomes onto the designer's objective.

\subsection{Open Problems}

\subsubsection{Sampling in Stream.}
\label{app:subsec:discuss:sampling}

One costly step of \Method lies in the continuous sampling step on each sample. In this step, the cost linearly scales with the number of samples. Conversely, if the number of samples is not large enough then the learning is likely to fall into over-fitting issue.
The cost is inevitable because the samples in the next time depend on the samples in the last time.
An open problem is how to quickly generate samples that do not rely on the past samples \Wrt a general distribution $q_i(a_i;\bmt_{-i},\theta)$, which is further dependent on the learnable parameter $\theta$. 
With this technique, the training cost of \Method will no longer depend on the training samples and allows for sampling training data in stream. 
We suspect that generative models--for their capabilities to approximate any distribution--would help for this technique.

\subsubsection{Problems Beyond Mechanism Design.}
\label{app:subsec:discuss:problem}

From the menu mechanism perspective, one could see that truthful mechanism design shares some similarities to Stackelberg games \citep{Stackelberg:von2010market}, in which the players' actions share sequential structures.
Algorithms that exhibit first-order gradients have been designed to solve this kind of economic problem \citep{learning:Stackelberg:kroer2015faster}. The techniques proposed in this paper are likely to scale to larger problems without suffering from curse-of-dimensionality.
Moreover, our techniques are even more general and are likely to be performed in economic scenarios where players' actions share sequential structures, even with multi-layer structures.

\subsubsection{Permutation Equivariance Pricing Rule.}
As is pointed out by \citet{permutation-equivariance:qin2022benefits}, a permutation-equivariant network architecture might help for learning if the economic problems hold ``permutation equivariance property''. 
As is known to all, the Transformer is a universal approximator of the permutation-equivariant functions, while the \Network proposed in this paper is a universal approximator of the partial convex functions.
However, it seems unclear how to combine the Transformer and GroupMax architecture to create the universal approximator of the permutation equivariant and convex functions.
Also, note that both convexity and permutation equivariance are significant properties in economic problems, thus we suppose the design of such a universal approximator is a significant future question.
With such a universal approximator, \Method can incorporate it as the pricing rule and is supposed to perform better than \Method with non-permutation-equivariant architecture.

\subsubsection{Computational Complexity of \network.}

In a conventional neural network, if the hidden dimension scales linearly with input dimension $d$, then the computation complexity of one network call is $O(d^2)$. However, the complexity of one network call is $O(nm^3)$ in our \Network implementations, though the hidden dimension is also $O(m)$.
While \Network shares the ideal property of universal approximation property, the cubic dependency on $m$ further decreases the scalability. We hope a new network architecture with universal approximation property on partial convex functions would be proposed with lower computational complexity (possibly within $O(m^2)$) and competitive performance compared with \network.

\end{document}